\documentclass[twocolumn,preprint]{aastex62}

\usepackage{amsmath,amstext}
\usepackage[T1]{fontenc}
\usepackage{apjfonts} 
\usepackage{natbib}
\usepackage{microtype}
\usepackage{longtable}
\usepackage{verbatim}

\newcommand{\AAA}{\hbox{\AA}}
\newcommand{\zone}{\hbox{z $\sim$ 1.1}}
\newcommand{\ztwo}{\hbox{z $\sim$ 1.2}}
\newcommand{\zthree}{\hbox{z $\sim$ 1.3}}
\newcommand{\zsix}{\hbox{z $\sim$ 1.6}}
\newcommand{\hst}{\textit{HST}}
\newcommand{\HST}{\textit{HST}}
\newcommand{\wfcy}{\hbox{$Y_{105}$}}
\newcommand{\wfcj}{\hbox{$J_{125}$}}

\newcommand{\acsz}{\hbox{$z_{850}$}}
\newcommand{\wfcym}{\hbox{$Y_{098}$}}
\newcommand{\lsim}{\lesssim}
\newcommand{\gsim}{\gtrsim}

\newcommand{\hb}{\hbox{H$\beta$}}

\newcommand{\hd}{\hbox{H$\delta$}}

\newcommand{\jwst}{\textit{JWST}}
\newcommand{\threedhst}{\hbox{3D-HST}}
\newcommand{\wfirst}{\textit{WFIRST}}
\newcommand{\grizli}{\textit{Grizli}}
\newcommand{\msol}{\hbox{$M_\odot$}}

\newcommand{\zsol}{\hbox{$Z_\odot$}}
\newcommand{\Zsol}{\hbox{$Z_\odot$}}

\newcommand{\editone}[1]{{\textcolor{black}{#1}}}

\begin{document}

\title{\large \bf CLEAR I:  Ages and Metallicities of Quiescent Galaxies at $\mathbf{1.0 < z < 1.8}$ Derived from Deep \\ Hubble Space Telescope Grism Data}
 
\author[0000-0001-8489-2349]{Vicente Estrada-Carpenter}
\affiliation{Department of Physics and Astronomy, Texas A\&M University, College
Station, TX, 77843-4242 USA}
\affiliation{George P.\ and Cynthia Woods Mitchell Institute for
 Fundamental Physics and Astronomy, Texas A\&M University, College
 Station, TX, 77843-4242 USA}
 
\author[0000-0001-7503-8482]{Casey Papovich}
\affiliation{Department of Physics and Astronomy, Texas A\&M University, College
Station, TX, 77843-4242 USA}
\affiliation{George P.\ and Cynthia Woods Mitchell Institute for
 Fundamental Physics and Astronomy, Texas A\&M University, College
 Station, TX, 77843-4242 USA}
\author[0000-0003-1665-2073]{Ivelina Momcheva}
\affil{Space Telescope Science Institute, 3700 San Martin Drive,
  Baltimore, MD, 21218 USA}
\author{Gabriel Brammer}
\affil{Space Telescope Science Institute, 3700 San Martin Drive,
  Baltimore, MD, 21218 USA}
\author{James Long}
\affiliation{Department of Statistics, Texas A\&M University, College
Station, TX, 77843-3143 USA}
\author[0000-0003-0341-8827]{Ryan F. Quadri}
\affiliation{Department of Physics and Astronomy, Texas A\&M University, College
Station, TX, 77843-4242 USA}
\affiliation{George P.\ and Cynthia Woods Mitchell Institute for
  Fundamental Physics and Astronomy, Texas A\&M University, College
  Station, TX, 77843-4242 USA}

\author[0000-0002-8584-1903]{Joanna Bridge}
\affil{Department of Physics and Astronomy, 102 Natural Science Building, University of Louisville, Louisville, KY, 40292 USA}
\author[0000-0001-5414-5131]{Mark Dickinson}
\affil{National Optical Astronomy Observatory, 950 North Cherry Ave, Tucson, AZ, 85719, USA}
\author{Henry Ferguson}
\affil{Space Telescope Science Institute, 3700 San Martin Drive,
  Baltimore, MD, 21218 USA}
\author[0000-0001-8519-1130]{Steven Finkelstein}
\affil{Department of Astronomy, The University of Texas at Austin, Austin,
TX 78712, USA}
\author{Mauro Giavalisco}
\affil{Astronomy Department, University of Massachusetts,
Amherst, MA 01003, USA}
\author[0000-0001-8023-6002]{Catherine M. Gosmeyer}
\affil{NASA Goddard Space Flight Center, Greenbelt, MD 20771, USA}
%
\author{Jennifer Lotz}
\affil{Space Telescope Science Institute, 3700 San Martin Drive,
  Baltimore, MD, 21218 USA}
\author{Brett Salmon}
\affil{Space Telescope Science Institute, 3700 San Martin Drive,
  Baltimore, MD, 21218 USA}
\author[0000-0001-7393-3336]{Rosalind E. Skelton}
\affil{South African Astronomical Observatory, PO Box 9, Observatory, Cape Town, 7935, South Africa}
\author[0000-0002-1410-0470]{Jonathan R. Trump}
\affil{Department of Physics, University of Connecticut, Storrs, CT 06269, USA}
\author[0000-0001-6065-7483]{Benjamin Weiner}
\affil{MMT/Steward Observatory, 933 N. Cherry St., University of Arizona, Tucson,
AZ 85721, USA}
%


\begin{abstract}
We use deep \textit{Hubble Space Telescope} spectroscopy to  constrain
the metallicities and (\editone{light-weighted}) ages of massive ($\log M_\ast/M_\odot\gtrsim10$)
galaxies selected to have quiescent stellar populations at
$1.0<z<1.8$. The data include 12--orbit depth coverage with the
WFC3/G102 grism covering  $\sim$ $8,000<\lambda<11,500$~\AA\, at a
spectral resolution of $R\sim 210$ taken as part of the CANDELS
Lyman-$\alpha$ Emission at Reionization (CLEAR) survey.  At
$1.0<z<1.8$, the spectra cover important stellar population features
in the rest-frame optical. We simulate a suite of stellar
population models at the grism resolution, fit these to the data
for each galaxy, and derive posterior likelihood distributions for
metallicity and age.  We stack the posteriors for subgroups of
galaxies in different redshift ranges that include different
combinations of stellar absorption features. Our results give \editone{light-weighted
ages of $t_{z \sim 1.1}= 3.2\pm 0.7$~Gyr,
$t_{z \sim 1.2}= 2.2\pm 0.6$~Gyr,
$t_{z\sim1.3}= 3.1\pm 0.6$~Gyr, and
$t_{z\sim1.6}= 2.0 \pm 0.6$~Gyr, \editone{for galaxies at $z\sim 1.1$, 1.2,
1.3, and 1.6.  This} implies that most of the massive
quiescent galaxies at $1<z<1.8$ had formed $>68$\% of their stellar
mass by a redshift of $z>2$}. The posteriors give metallicities of
\editone{$Z_{z\sim1.1}=1.16 \pm 0.29$~$Z_\odot$,
$Z_{z\sim1.2}=1.05 \pm 0.34$~$Z_\odot$,
$Z_{z\sim1.3}=1.00 \pm 0.31$~$Z_\odot$, and
$Z_{z\sim1.6}=0.95 \pm 0.39$~$Z_\odot$}. This is evidence that
massive galaxies had enriched rapidly to approximately Solar
metallicities as early as $z\sim3$. 

\end{abstract}

\section{Introduction}

Astronomers have recently developed a general picture of the
formation, chemical enrichment, and evolution of massive galaxies
($\log M_\ast / M_\odot \gsim 10-11$) from $z \gsim$ 3--4 to the
present day. \editone{These galaxies formed their stellar populations
early, reaching a peak star-formation rate (SFR) at $z > 2$
\citep[e.g.,][]{labb05,papo06,mada14,schr16}, and quenching as early
as $z\sim 3-4$ \citep{spli14,stra14,glaz17,schr18b,schr18}}.
%
%
%
\editone{Massive galaxies continue to quench and remain in phases of
quiescent evolution to lower redshift, where the fraction of quiescent
galaxies with $\log M_\ast/M_\odot > 10.5$  rises  from $\sim$ 30\% at
$z\sim 2$ to $\approx$60\% by $z=0.5$ \citep{muzz13,tomc14,kawi17}.}
Because these massive galaxies formed their stars and quenched so
early, they are testing \editone{grounds for physical processes
associated with gas accretion, star-formation efficiency, feedback,
metal production and retention \citep{oppe12,some15,torr17}. }

%
\editone{The next step is to understand in greater detail the
  galaxies' star-formation and chemical enrichment histories.}
%
%
\editone{These effects are all encoded in the colors and spectral
features of the galaxies' stellar populations, }
%
%
which allows one to constrain the ages and metallicities using
features that are sensitive to independent changes in these
parameters.  \editone{For example, line indexes (such as e.g., $D_n4000$,
\hd$_A$+H$\gamma_A$, \hb, Mg$b$, [MgFe]$^\prime$, [Mg$_2$Fe])
have long been used to study these properties in galaxies, and to
study their evolution with mass and redshift
\citep[e.g.,][]{matt94,trag00,thom05,gall05,gall14,renz06,choi14,wu18}. }
%
%
%
%
%

\editone{Regarding the evolution of metals in galaxies, most studies
have focused on the relation between the gas-phase metallicity
(usually measured from emission line ratios in active galaxies) and
galaxy stellar-mass, and their evolution with redshift
\citep[e.g.,][]{trem04, erb06,zahi14,onod16,sand17}.}
%
%
\editone{It has been more difficult to measure evolution of
metallicity of the stellar populations in galaxies.  The spectral
absorption and continuum features} are sensitive to both
$\alpha$-elements and Iron-peak elements  (which dominate the opacity
in stellar photospheres).  
%
%
%
\editone{Analyzing the spectra of quiescent galaxies at low-redshift
  ($z\lsim 0.1$), many studies have found that massive galaxies have
  Solar (or super-Solar) metallicities and abundance ratios, with
  formation epochs at $z >  2$ \citep[e.g.][]{thom05,thom10,gall05,renz06,conr14}.} 
%
%
\editone{\citet[hereafter Gal14]{gall14} conducted one of the first
studies at higher redshift ($z \sim 0.7$), and showed that the spectra
of quiescent galaxies with $\log M_\ast/M_\odot > 10.5$ are already
consistent with high metallicity ($Z~\sim Z_\odot$), while
star-forming galaxies have sub-Solar metallicities \citep[and require
additional enrichment from $z\sim 0.7$ to the present-day, see
also][]{ferr18}}.   The implication is that some fraction of the
massive, quiescent galaxy population must have enriched to
approximately Solar metallicities at even higher redshift, presumably
at $z \gtrsim 2$.

To understand the evolution of massive quiescent galaxies therefore
requires that we push spectroscopic studies to higher redshift, closer
to the quenching epoch of these galaxies.  \editone{The reason for
this is that evolutionary changes in stellar populations occur on
timescales $d\log t \propto dt / t$ \cite[see e.g.,][]{weis14}}.
Therefore, galaxy properties change more rapidly at higher redshift
\citep{tran07,eise08,blak09,papo10}, and more accurate (relative) measures of
galaxy ages can be made by observing quiescent galaxies at $z > 1$.
\editone{The difficulty is that all the rest-frame optical spectral
features used to measure both stellar population metallicities and
ages shift above $\sim$1~\micron\ for galaxies at $z > 1$, where
studies from ground-based telescopes are subject to higher backgrounds
and a high density of telluric (night-sky) emission lines
\citep[see,][]{sull12}.   Several studies have now measured galaxy
metallicities at high redshift with near-IR spectrographs on
ground-based telescopes \citep[e.g.,][]{lono15,onod15,krie16,toft17},
but are often limited to one or a few objects or require stacking.
An attractive alternative solution is to take advantage of space-based
spectroscopy using, for example, the \hst\ grisms, which have been
used to study the properties of stellar populations in distant
galaxies \citep[e.g.,][]{dadd05,vand10, vand11, whit14, momc16,
fuma16, leeb17, ferr18, mori18}.}
%
%
%

\editone{Here, we use deep grism spectroscopy taken with \hst\
spectroscopy  with WFC3 using the G102 grism, covering wavelengths
$\approx$8000~\AA\ to 11,500~\AA.  We study the stellar populations of
a sample of massive ($\log M/M_\odot \geq 10$) quiescent galaxies at
$1 < z < 1.8$.    At these redshifts, the G102 spectra cover crucial
rest-frame optical absorption features at spectral resolution of
$R\sim 210$,   
%
%
%
including the strength of the redshifted 4000~\AA\ break, Balmer lines
(H$\delta$, H$\gamma$,H$\beta$), Ca HK, and Mg$b$ absorption features)
sensitive to stellar population ages and metallicities of galaxies
down to $J \simeq 23-24$~AB mag.  The G102 data also provide much
improved measurements of the galaxy redshifts.  We use a Bayesian
method with} a forward-modeling approach, whereby we simulate stellar
population models of the two--dimensional (2D) WFC3/G102 spectra using
the galaxies' accurate morphologies at the appropriate roll-angle for
\hst\ combined with the G102 spectroscopic resolution.    
%

One of the goals of this project is to determine the
accuracy of stellar population properties measured with
lower-resolution grism data. This has important implications for the
next  generation of space telescopes (\jwst\ and \wfirst), and will
allow the study of stellar  populations of galaxies in much larger
samples and at higher redshifts.  As we push to higher redshifts, all
important age- and metallicity-sensitive spectroscopic  features shift
into the IR.   As we show here, even
relatively low resolution ($R\sim 100-200$) spectroscopy can recover
metallicity and age diagnostics of higher redshift galaxies (provided
the spectra cover the age-- and metallicity--sensitive features).
This will have a wide range of application for future space missions. 

The remainder of this paper is organized as follows: In Section 2 we
describe the parent sample selection.  In Section 3 we summarize the
\hst\ observations, data reduction, and spectral extraction
methods. In Section 4 we describe the G102 spectral modeling
procedure, including tests to demonstrate our ability to recover
accurate stellar population parameters.  We then discuss the model
fits to the G102 data for the galaxies in our sample.   In Section 5
we discuss the  constraints on metallicities and
\editone{(light-weighted)} age for these galaxies, as well as
implications for the formation, quenching and evolution of these
galaxies.  In Section 6 we present our conclusions and describe future
work using this method.   In the Appendices \editone{we show the model
fits to the full galaxy sample, discuss our method to derive errors on
model templates, } discuss the Bayesian evidence
in support of different stellar populations derived from the modeling,
and we test our posterior stacking technique.

Throughout we use a cosmology with $\Omega_M=0.3$,
$\Omega_\Lambda=0.7$, and $h$ = 0.7, where $H_0 = 100$
$h$ km s$^{-1}$ Mpc$^{-1}$ consistent with the recent constraints from
Planck \citep{plan15} and the local distance scale of
\cite{ries16}.   All magnitudes are in ``absolute bolometric'' (AB)
units \citep{oke83}. 
%
%

\section{Parent Sample and Sample Selection}\label{section:sample}

The \hst\ WFC3 G102 grism wavelength coverage (8000 $< \lambda <$
11500 \AA) provides a wealth of information for distant galaxies as it
probes the \editone{rest-frame UV and  optical range} for  redshifts
$z\sim 1$ (rest-frame coverage: 4000 $< \lambda <$ 5750 \AA)  to
$z\sim 1.8$ (rest-frame coverage: 2850 $< \lambda <$ 4100 \AA).  These
wavelengths cover many age-- and metallicity--sensitive spectral
features, including Ca HK, the 4000 \AA/Balmer break, Balmer-series
lines, several Lick indices, Mg$b$, and some \ion{Fe}{1} lines
\editone{(though not all features are covered at all redshifts by the
G102 grism, as we will discuss below)}.  With this in mind, we focused
on a sample of quiescent galaxies selected to have low levels of star
formation compared to their past averages.   By modeling the spectra
of such galaxies, we are able to constrain the galaxies'
star-formation histories (SFHs, when did they form their stars and when did
they quench?), and enrichment histories (when did they form their
metals?). 

The CLEAR survey includes G102 observations within the CANDELS 
coverage of the GOODS-North deep
(GND) and GOODS-S deep (GSD) fields \citep{grog11,koek11}.   We
selected galaxies from an augmented version of the photometric
catalog from \threedhst\ \citep{skel14,momc16} that
includes photometry from $Y$--band imaging (using the WFC3 F098M
and/or F105W imaging), which exists over the
majority of the GND and GSD CANDELS fields
\citep{koek11}\footnote{All of the GND fields and most of the GSD
fields are covered by F105W from CANDELS \citep{grog11,koek11} and
from the direct imaging associated with our grism program, see Section~3. The
northern portion of the GSD overlaps with the WFC3 ERS field
\citep{wind11} for which F098M is
available.\label{footnote:candels}}.   More importantly, the WFC3
F098M and F105W filters provide imaging at comparable wavelengths
covered by the WFC3 G102 grism, and is important for object
extraction and identifying the locations of nearby objects that may cause
spectral contamination.   

\editone{The original \threedhst\ data release did not include the
F098M nor F105W data in the GOODS fields (because these have less
uniform coverage, and are only available in two of the five CANDELS
fields).  Because we require them for our program, we have added them to
the \threedhst\ photometric catalog following the identical approach
as \cite{skel14}.}  We first matched the PSFs of the F098M and F105W
images to the detection image (using a weighted combination of the
WFC3 F125W + F140W + F160W images).  We then reran SExtractor
\citep[v2.8.6,][]{bert96} in dual-image mode, using the detection
image to photometer sources in the F098M and F105W images.  We then
inserted these fluxes into the existing \threedhst\ catalogs along
with rederived photometric redshifts and rest-frame $U-V$ and $V-J$
colors  using EAZY \citep{bram08}, and stellar masses from FAST
\citep{krie09}  (using a \cite{chab03} IMF). For details of the
procedures, see \cite{skel14}.

\begin{figure}[t]
\epsscale{1.1}
\plotone{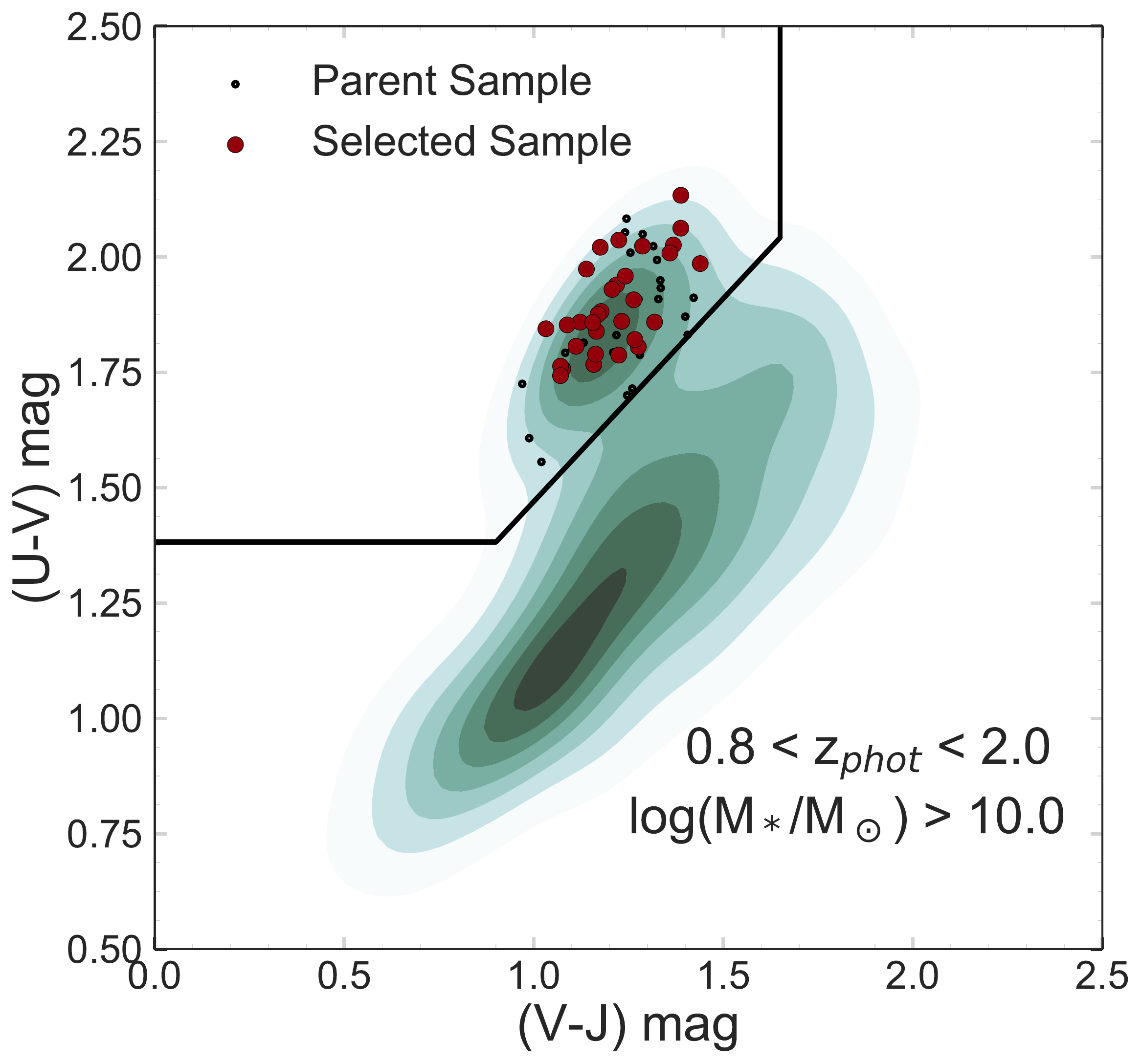}
\caption{Distribution of rest-frame $V-J$ and $U-V$ colors of the
galaxies in the GND and GSD galaxy samples. The green--shaded regions
show the distribution of all galaxies in the parent catalog
with $0.8 \leq z_\mathrm{phot} \leq 2.0$, \texttt{CLASS\_STAR} $< 0.8$
from SExtractor,  and $\log(M_\ast/M_\odot) \geq 10.0$ (the
two-dimensional shows all galaxies in this redshift and mass range,
smoothed using an \citet{epan69} kernel density estimation).  
The polygon delineates ``quiescent'' galaxies (upper
left region) from ``star-forming'' (everywhere else) using the
definition of \cite{will09}.   Red points mark the sample of quiescent
galaxies used here (these galaxies satisfy our selection requirements
and fall in fields covered by the deep \hst/WFC3
G102 spectroscopy). Unfilled points mark the sample of galaxies
which were covered in the data and met most of the selection 
requirements, but were rejected from the final sample.\label{uvj} }
\end{figure} 

\begin{deluxetable}{lcccc}
\small
\tablecaption{Properties of the $1 < z < 1.8$ Quiescent Galaxy Sample\label{iptable}}
\tablehead{
\colhead{ID} &
\colhead{$z_{phot}$} &
\colhead{$J_{125}$ (Mag)}&
\colhead{Log $(M_\ast/M_\odot)$} &
\colhead{SNR}\\
\colhead{(1)} & 
\colhead{(2)} & 
\colhead{(3)} &
\colhead{(4)} &
\colhead{(5)} }
\startdata
\hline
GND23758 & $0.94 _ {- 0.04} ^ {+ 0.04}$ & 20.5 & 11.0 & 12.3 \\
GND37955 & $0.97 _ {- 0.03} ^ {+ 0.03}$ & 21.3 & 10.8 & 7.3 \\
GND16758 & $0.99 _ {- 0.03} ^ {+ 0.04}$ & 21.2 & 10.8 & 10.1 \\
GSD43615 & $1.04 _ {- 0.04} ^ {+ 0.04}$ & 21.7 & 10.7 & 6.9 \\
GSD42221 & $1.04 _ {- 0.04} ^ {+ 0.04}$ & 21.7 & 10.5 & 7.7 \\
GSD39241 & $1.04 _ {- 0.04} ^ {+ 0.05}$ & 20.9 & 10.9 & 12.4 \\
GSD45972 & $1.05 _ {- 0.02} ^ {+ 0.03}$ & 21.1 & 10.9 & 12.3 \\
GSD44620 & $1.08 _ {- 0.03} ^ {+ 0.03}$ & 22.1 & 10.5 & 4.8 \\
GSD39631 & $1.08 _ {- 0.03} ^ {+ 0.03}$ & 21.3 & 10.7 & 9.2 \\
GSD39170 & $1.08 _ {- 0.02} ^ {+ 0.03}$ & 20.3 & 11.1 & 20.8 \\
GND34694 & $1.09 _ {- 0.03} ^ {+ 0.04}$ & 21.0 & 10.9 & 9.9 \\
GND23435 & $1.09 _ {- 0.06} ^ {+ 0.06}$ & 22.5 & 10.3 & 4.4 \\
GSD47677 & $1.10 _ {- 0.04} ^ {+ 0.04}$ & 22.5 & 10.1 & 4.2 \\
GSD39805 & $1.16 _ {- 0.03} ^ {+ 0.03}$ & 22.5 & 10.6 & 3.9 \\
GSD38785 & $1.18 _ {- 0.03} ^ {+ 0.04}$ & 21.5 & 10.9 & 7.5 \\
GND32566 & $1.18 _ {- 0.05} ^ {+ 0.05}$ & 21.7 & 10.6 & 6.6 \\
GSD40476 & $1.19 _ {- 0.03} ^ {+ 0.03}$ & 21.9 & 10.6 & 8.6 \\
GND21156 & $1.20 _ {- 0.04} ^ {+ 0.04}$ & 20.9 & 11.2 & 15.8 \\
GND17070 & $1.22 _ {- 0.04} ^ {+ 0.04}$ & 21.2 & 10.9 & 5.3 \\
GSD35774 & $1.23 _ {- 0.03} ^ {+ 0.03}$ & 21.0 & 10.9 & 10.0 \\
GSD40597 & $1.24 _ {- 0.03} ^ {+ 0.03}$ & 20.9 & 11.0 & 16.3 \\
GND37686 & $1.27 _ {- 0.04} ^ {+ 0.04}$ & 21.3 & 10.9 & 8.4 \\
GSD46066 & $1.32 _ {- 0.03} ^ {+ 0.03}$ & 21.7 & 10.8 & 3.4 \\
GSD40862 & $1.33 _ {- 0.04} ^ {+ 0.04}$ & 21.7 & 10.9 & 4.9 \\
GSD39804 & $1.36 _ {- 0.03} ^ {+ 0.04}$ & 21.6 & 10.9 & 4.5 \\
GND21427 & $1.48 _ {- 0.05} ^ {+ 0.05}$ & 22.0 & 10.7 & 2.5 \\
GSD40623 & $1.49 _ {- 0.09} ^ {+ 0.11}$ & 22.3 & 10.8 & 4.7 \\
GSD41520 & $1.64 _ {- 0.04} ^ {+ 0.04}$ & 22.2 & 10.9 & 3.4 \\
GSD40223 & $1.65 _ {- 0.05} ^ {+ 0.05}$ & 22.7 & 10.7 & 2.3 \\
GSD39012 & $1.66 _ {- 0.06} ^ {+ 0.06}$ & 22.6 & 11.1 & 1.8 \\
GSD44042 & $1.67 _ {- 0.05} ^ {+ 0.05}$ & 21.8 & 11.0 & 4.0 \\
\enddata
\tablecomments{ (1) Galaxy ID number in the GND or GSD \threedhst\ catalog. 
(2) photometric redshift from the \threedhst\ catalog; (3) observed 
\wfcj\ magnitude from the \threedhst\ catalog; (4) stellar mass derived 
using FAST; (5) SNR per pixel measured at 8500-10500~\AA\ in the
stacked G102 spectrum.} 
\end{deluxetable}

\editone{It is our goal to select galaxies that are mostly devoid of
star-formation, with a redshift that places the 4000~\AA/Balmer break
in the G102 grism wavelength coverage).   We first selected a parent
sample of galaxies to have photometric redshifts ($z_{phot}$) in the
range $0.8 < z_\mathrm{phot} < 2.0$ and stellar mass $\log (M_\ast /
\msol) > 10.0$.  This $z_{phot}$ redshift range ensures that we select
\textit{all} galaxies that possibly place the
4000\AA/Balmer break in the G102 wavelength coverage,
when accounting for errors on the photometric redshifts.
Our stellar-mass constraint ensures that the spectra
have sufficient signal-to-noise for accurate modeling (see below).} We
filter stars from our parent sample, using objects that have
stellarity values from the \threedhst\  catalog (defined by
SExtractor) with \texttt{CLASS\_STAR}  $< 0.8$.  We remove any source
with an X-ray detection (within a search  radius of 0$\farcs$5) in the
2 Ms Chandra Deep Field-North Survey  \citep{xue16} and 7 Ms Chandra
Deep Field-South Survey catalogs \citep{luo17}.

Based on the rest-frame $U-V$ and $V-J$ colors, we selected galaxies
that are actively star-forming from those in quiescent phases of
evolution \editone{\citep[compared to star-forming/active
galaxies,][]{wuyt07,will09,whit11}}. We selected a parent sample of
quiescent galaxies from our augmented \threedhst\ catalog using the
definition in \citet{will09},

\begin{eqnarray}
(U-V)\ &\ge\ & 1.382\ \mathrm{mag},\ \nonumber\\
(V-J)\ &\leq\ & 1.65\ \mathrm{mag},\ \mathrm{and}\ \\
(U-V)\ &\ge\ & 0.88 \times (V-J) + 0.59 \nonumber
\end{eqnarray}  
Figure~\ref{uvj} shows the $U-V$ versus $V-J$ color distribution for the full galaxy catalog and for those galaxies in our parent sample. 
Galaxies in the quiescent UVJ region typically have specific SFRs (sSFRs)
$< 10^{-2}$~Gyr$^{-1}$ \citep{papo12}, indicative of galaxies
with lower current SFRs compared to their past averages (e.g., for a
galaxy with a current stellar mass $10^{11}$~\msol\ and
sSFR=10$^{-2}$~\msol\ yr$^{-1}$, the current SFR is $\Psi$=1~\msol\
yr$^{-1}$, whereas the SFR averaged over the past Hubble time is
$\langle \Psi \rangle$ $\sim$ 20~\msol\ yr$^{-1}$).  Galaxies with
such low sSFR qualify as having ``suppressed'' SFRs \citep{krie06}.
Past studies of the evolution of these galaxies show they follow
``passive'' evolution of their mass-to-light ratios from redshifts as high as
$z\sim 2$ (e.g., Fum16). 

\editone{From the parent sample, we further require that all galaxies
fall within the CLEAR \hst/WFC3 G102 coverage (which includes 12
WFC3 pointings divided evenly between the GND and GSD fields, see
Section~3).  We then refit the galaxy redshifts using the G102 grism
spectra themselves (as described in Section~\ref{rshift} below), and
kept only those galaxies with $1.0 < z_{grism} < 1.8$ to ensure the
G102 data cover the redshifted 4000~\AA/Balmer break for all galaxies
(this last step does exclude some galaxies with low SNR in the G102
data, typically $\langle SNR \rangle \lesssim 1.5$, see below).   This
yields our final sample of 31 galaxies.   Table~\ref{iptable} shows
the physical details of these galaxies.}

\begin{figure}[t] 
\epsscale{1.05} 
\plotone{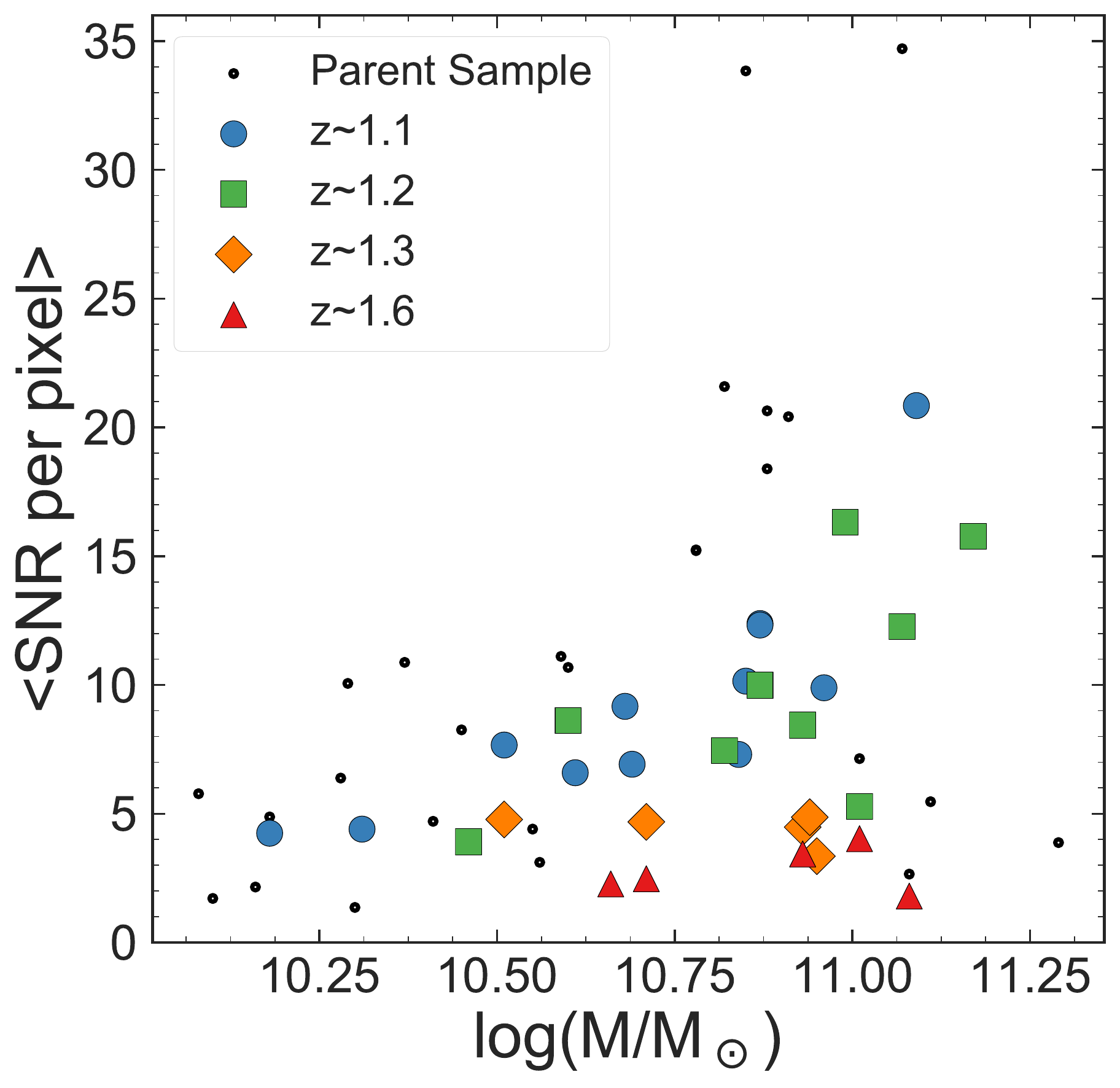} 
\epsscale{1.0}
\caption{The average Signal-to-Noise ratio (SNR) measured in the \hst/WFC3
  G102 grism data for the galaxies in our samples as a function of
  galaxy stellar mass. For each galaxy, we measured the \editone{average SNR per pixel in the 1D spectrum over wavelengths $8500 < \lambda < 10500$ \AA}.   For the majority of
  galaxies the average SNR is $>$3 per spectral pixel, where our tests
  show we are able to derive ``good'' physical constraints on the
  galaxy stellar population parameters. \label{snvm}}
\end{figure}

For quiescent galaxies, the stellar mass limit of $\log M_\ast / \msol
> 10.0$ corresponds roughly to a magnitude limit of $J_{125} \leq
22.7$ mag at $z < 1.8$.   For such galaxies we measure a mean
signal-to-noise ratio (SNR) in the G102 data of $\langle$SNR$\rangle$
$\gsim$3 per pixel, averaged over
$\lambda$=0.85--1.05~\micron\ (see Figure~\ref{snvm} and below). Most
of the quiescent galaxies in our sample lie at stellar masses well
above this limit, where the median mass is $\log(M_\ast / M_\odot) =
10.87$ with an interquartile range (spanning the 25th to 75th
percentile) of $\log(M_\ast / M_\odot) = 10.67$ to 10.96. 

\section{\hst\ WFC3/G102 Observations and Data Reduction} 

\subsection{\hst\  Observing Strategy}\label{sumofo}

\begin{figure*}[ht!]
\epsscale{1.0}
\plotone{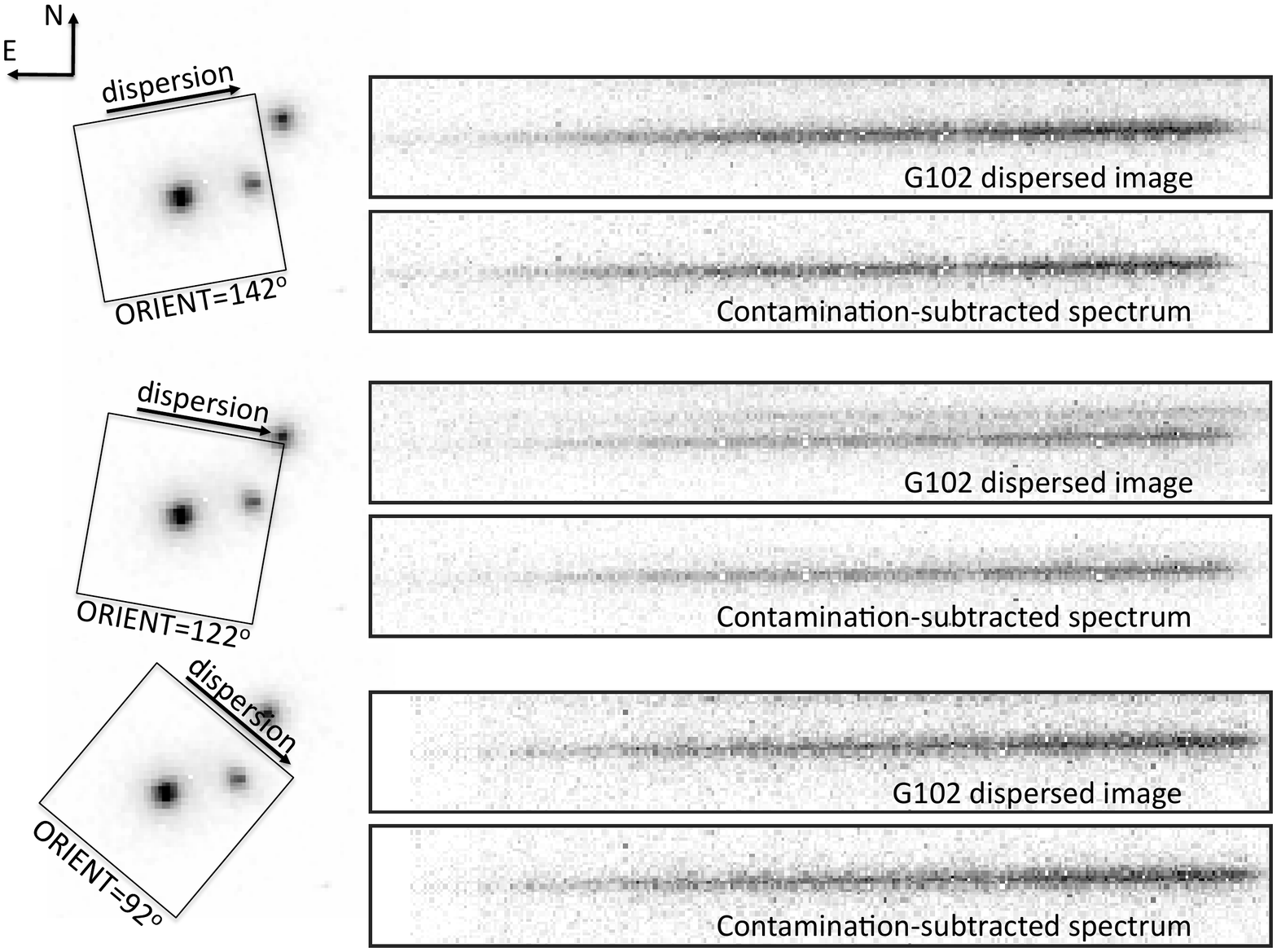}
\caption{Example of \HST\ imaging and spectroscopy for one object in
our sample. The panels show the extracted F105W direct image and 2D
G102 grism spectra for object GSD39170. The data are taken at three
different PAs (ORIENT=142, 122, and 92 deg, as labeled). In each
direct-imaging panel, the arrow shows the direction of the
spectroscopic dispersion in the associated G102 image. Each panel
contains a stack from four orbits at each PA, and shows the spectrum
both before and after the subtraction from nearby sources (modeled
during the the extraction process). \label{diagram}}
\end{figure*} 

We use \hst\ slitless spectroscopic data taken  from the CANDELS
Lyman-$\alpha$ Emission at Reionization (CLEAR; \hst\ GO
14227, PI: Papovich).  CLEAR covers 12 fields in the
GND and GSD portions of CANDELS, with deep WFC3/G102 grism observations
(12 orbits, when combined with some existing data).  
\editone{For unresolved sources the G102 dispersion is 24.5~\AA/pixel$^{-1}$ (for the native WFC3 pixel scale, $\approx$0\farcs13/pixel), yielding a resolution of $R \sim 210$ at 1.0~\micron}. 

The observations for CLEAR were taken over dates ranging from 2015 Nov
14 to 2017 Feb 19.  Each pointing was observed for 10 or 12 orbits,
with direct imaging in the WFC3 F105W (\wfcy) paired with the G102
grism exposures. We followed the sub-pixel dither pattern used by \threedhst\ \citep{bram12}. The observations for each pointing were
divided into 3 pointings of 4-orbits each (except in GND and the ERS
field) separated by $\pm$10 degrees in roll angle to mitigate collisions of
spectra from nearby objects and to avoid detector
defects.    In addition, we combined the CLEAR data with existing
2--orbit--depth data  from other programs.   The observations
in the GND were only 10 orbits, and were combined with existing
2-orbit-depth data from the G102 from program GO~13420 (PI:
Barro). The observations in the ERS were combined with existing
2-orbit-depth G102 data from the WFC3 ERS program \citep[PI:
O'Connell; see][]{wind11,stra11}. 

During the scheduling of the CLEAR observations, care was taken to
protect the G102 grism data from a known time-variable
background. This background is due to \ion{He}{1} emission from the
Earth's atmosphere at 10830~\AA, which contributes to the background
in the $Y$--band (and G102 grism) when \hst\ observes at low limb angles
\citep[see][]{bram14,tilv16,lotz17}. Following the strategy of the
\textit{Hubble} Frontier Fields \citep{lotz17}, we monitored the
predicted observational ephemeris of each CLEAR observation.   We then
structured the observing sequence to place the direct imaging (F105W)
observation at the end (or start) of the orbit when the \ion{He}{1}
emission was expected to have the largest impact, and to place the
grism (G102) observation at the start (or end) of the orbit to take
advantage of the lower background levels. 

\subsection{\hst\ Spectroscopic Data Reduction \label{data_red}}

\begin{figure*}[t]
\epsscale{1.0}
\plotone{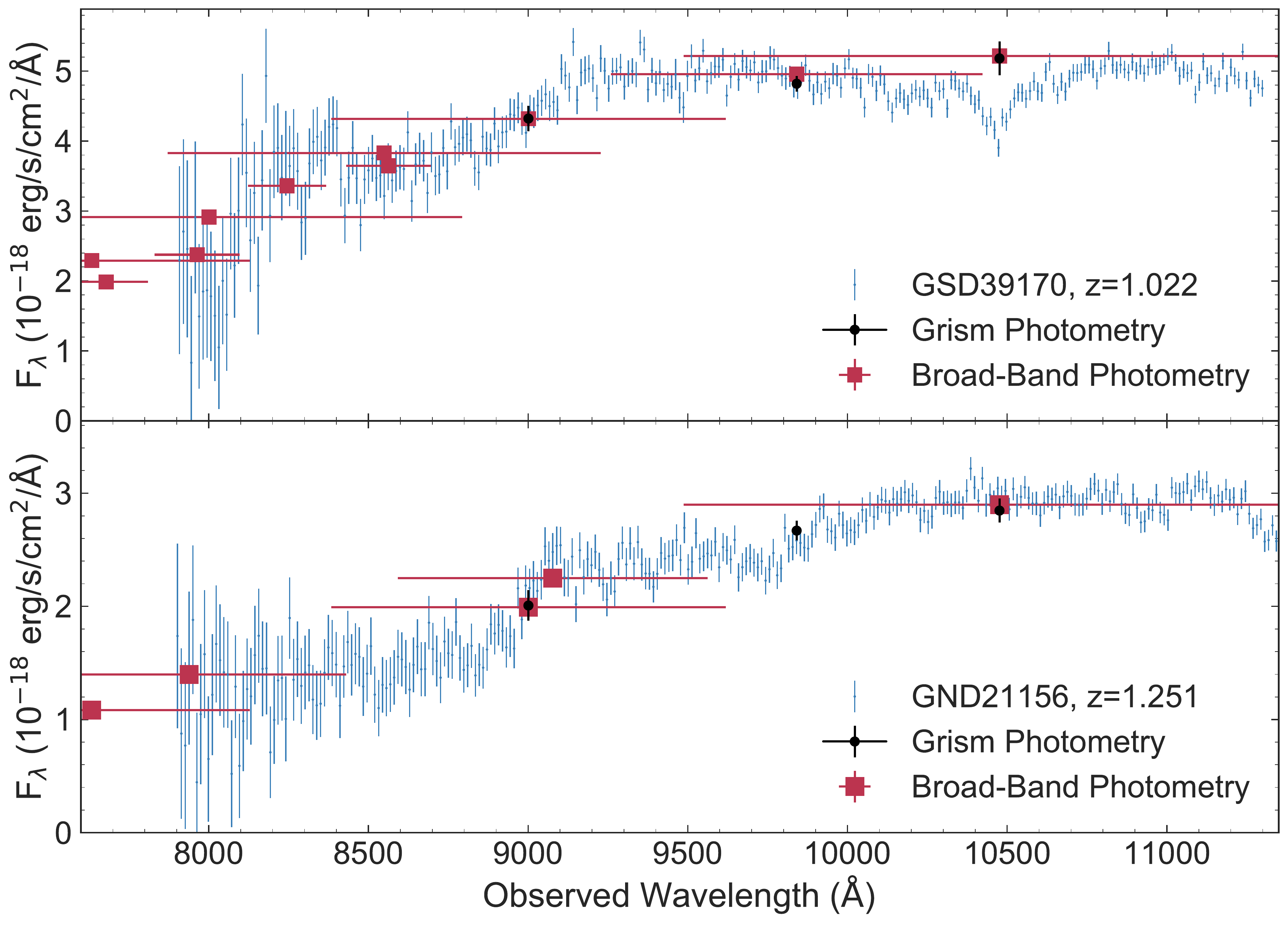}
\caption{Example comparison of the flux-calibration of our
12-orbit depth 1D G102 grism spectra (blue) to the broad-band
photometry (red) from our \threedhst\ catalog (that includes available
\acsz, \wfcy, or \wfcym\ data) for two objects in our sample. Black
points show the synthesized \acsz, \wfcy, and \wfcym\ measurements
from the grism data. The horizontal bars on the photometric points
show the FWHM of the broad-band filter transmission curves. Vertical
error bars show the 1 sigma errors. Filters included in the top panel are the F775W, IA767, 
IA797, F814W, IA827, I, IA856, \acsz, \wfcym, and \wfcy\ bandpasses.
The filters included in the bottom panel are the F775W, I, \acsz, 
Z, and \wfcy\ bandpasses.\label{spec_phot}}
\end{figure*} 

 The data reduction of the G102 spectra follow the procedures from the
\threedhst\ pipeline\footnote{https://github.com/gbrammer/threedhst}$^,$\footnote{https://github.com/gbrammer/unicorn } \citep{bram12,momc16} and custom
scripts\footnote{https://github.com/ivastar/clear}.  
This  includes interlacing of
the data to reduce pixel-to-pixel correlations (c.f.,
drizzling). Following the \texttt{calwf3} processing, we inspect the
images for artifacts (satellite trails) and any regions of elevated
background. Satellites typically affect a single WFC3 grism read, and in
those instances, we remove the read (or reads) containing the
satellite trail.   If they persisted we masked them. 
We then rejected cosmic rays by using
\texttt{AstroDrizzle}.  

To calibrate the G102 images we first divided them by the 
WFC3/F105W imaging flat field \citep[see][]{bram12} and then 
subtracted the background following the methods in \citet{bram14}. We
then subtracted the masked column average in each image to create the
final background--subtracted images and to remove low-level residuals
not accounted for in the backgrounds. Lastly, we combined the
exposures taken at the same ORIENT using the interlacing discussed in
\cite{momc16}. This provided stacked grism images containing
between two and four orbits of data. 

From these stacked grism images we extracted 2D and 1D spectra for
individual objects in our sample defined in
Section~\ref{section:sample} using the procedures in \citep{momc16}. We required a
reference direct image to provide the
positions and morphologies of each object for spectroscopic
extraction. \editone{As described by \cite{momc16}, the pipeline uses the direct image to identify and
model the expected location of spectral traces associated with
contaminating sources, using the positions and redshifts of sources in
the input catalog (see Section~\ref{section:sample}).}

We used the WFC3 F105W mosaics for all extractions, and for the
modeling and subtraction of contamination from nearby
sources (we found that using the F105W provided the best contamination
modeling and subtraction compared to other WFC3 bands, likely because
the F105W matches the wavelength coverage of the G102 data). 

Prior to stacking spectra for each object, we visually inspected the
data to ensure they were not affected by severe contamination or by
other cosmetic issues.  In most cases, the pipeline removed most of
the contamination from nearby sources.  In rare cases we identified
residual contamination present in one of the stacks (in all but one
case for our sample this affected only one ORIENT).  In this case we either
discarded the contaminated stack, or if the contamination was minor,
we masked affected pixels before stacking the extracted spectra. 

Figure~\ref{diagram} shows the CLEAR data for one of the galaxies in
our sample (GSD 39170 with a redshift derived from the grism spectrum of
$z_{grism}=1.02$, see below), which shows the direct imaging and grism data
extracted from the 4-orbit depth stacks at each of the three different
roll angles (ORIENTs).   This target highlights the
contamination subtraction capabilities of the CLEAR pipeline, as a
contaminating source lies near our target. In one orient its spectrum
falls near the target, and in another it lies directly over our
spectrum. Using multiple orients we can characterize the contaminating
source spectrum, and subtract it from the raw data. 

\subsection{Tests of the  \hst\ G102 Grism Flux Calibration}

Our study requires that the relative flux calibration of
the \hst\ WFC3/G102 grism data be accurate as we use the continuum in
our analysis of the galaxy stellar populations. In general, the \editone{absolute flux
calibration of \hst\ is very stable with average temporal variations
constrained to be less than 1\% \citep{lee14},}
and allows us to use the continuum for this purpose. This is a
significant advantage of space-based slitless spectroscopy compared to
slit-fed spectroscopy from the ground, which suffers from systematics
associated with terrestrial, astronomical, and instrumental
backgrounds \citep{sull12}, and partly \editone{explains} why 
\editone{slitless spectroscopy with \hst\ is superior in 
some aspects compared to ground-based 10~m-class telescopes
\citep{momc16,tilv16}, even though the current \hst\ IR instruments operate at lower spectral resolution.} 

\begin{figure}[t]
\epsscale{1.}
\plotone{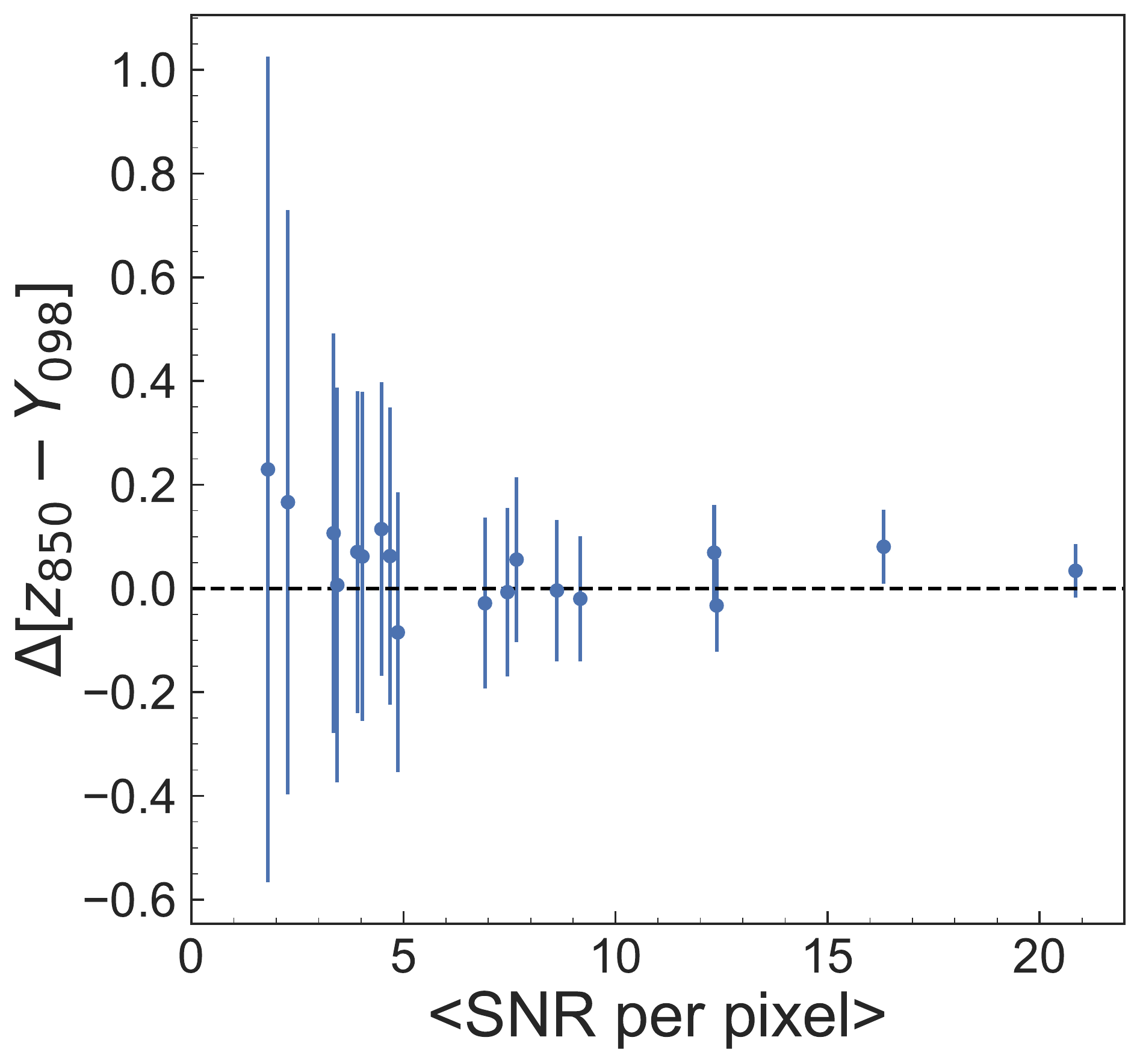}
\caption{Difference between the $\acsz - \wfcym$ color derived from
broad-band photometry and the color in those bands synthesized from
the G102 spectra, plotted as a function of the G102 SNR measured at
8500-10500~\AA. Here we can see the residuals are consistent 
with zero. \label{dcolor}}
\end{figure} 


We tested the flux calibration of the extracted grism data for objects
in our sample by comparing the spectra with the broad-band flux
measurements from \hst\ broad-band imaging. We synthesized
photometry from the grism spectra by integrating them with the \hst\
ACS F850LP and WFC3 F098M filters (\acsz\ and \wfcym), as these cover
wavelengths also covered by the G102 grism. Figure~\ref{spec_phot}
shows a comparison of the (flux-calibrated) extracted G102 spectra for
two objects in our sample (GSD~39170 and GND~21156) to their
broad-band photometry from our augmented \threedhst\ catalog (see
Section~2). The overall agreement between the broad--band and
synthesized photometry is better than 3\%. 

Next, we investigated the accuracy of the relative flux as a
function of wavelength (i.e., the accuracy of the ``color'' of the
spectra as these would impact our ability to measure 
stellar population parameters, such as the ages and metallicity).  The
G102 grism covers wavelengths also covered entirely by the ACS F850LP
and WFC3 F098M filters, and we therefore focus only on the subset of
objects in the GSD (ERS) that have coverage in both band-passes.
Figure~\ref{dcolor} compares the $\acsz - \wfcym$ color synthesized
from the G102 data to the color measured directly by these bands in
our photometric catalog, plotted as a function of SNR  in the G102
grism spectrum measured at 8500-10500\AA.   Taking all objects, the
median color difference is $\Delta(\acsz - \wfcym) = 0.05$~mag,
this is consistent with a measurement of 0.0 mag as our measured errors
are larger than the offset.
Therefore we conclude that the grism flux calibration is accurate
(compared to the broad-band colors), and we consider any systematic
uncertainty in flux calibration to be negligible in our analysis.

\begin{figure*}[t]
\epsscale{1.}
\plotone{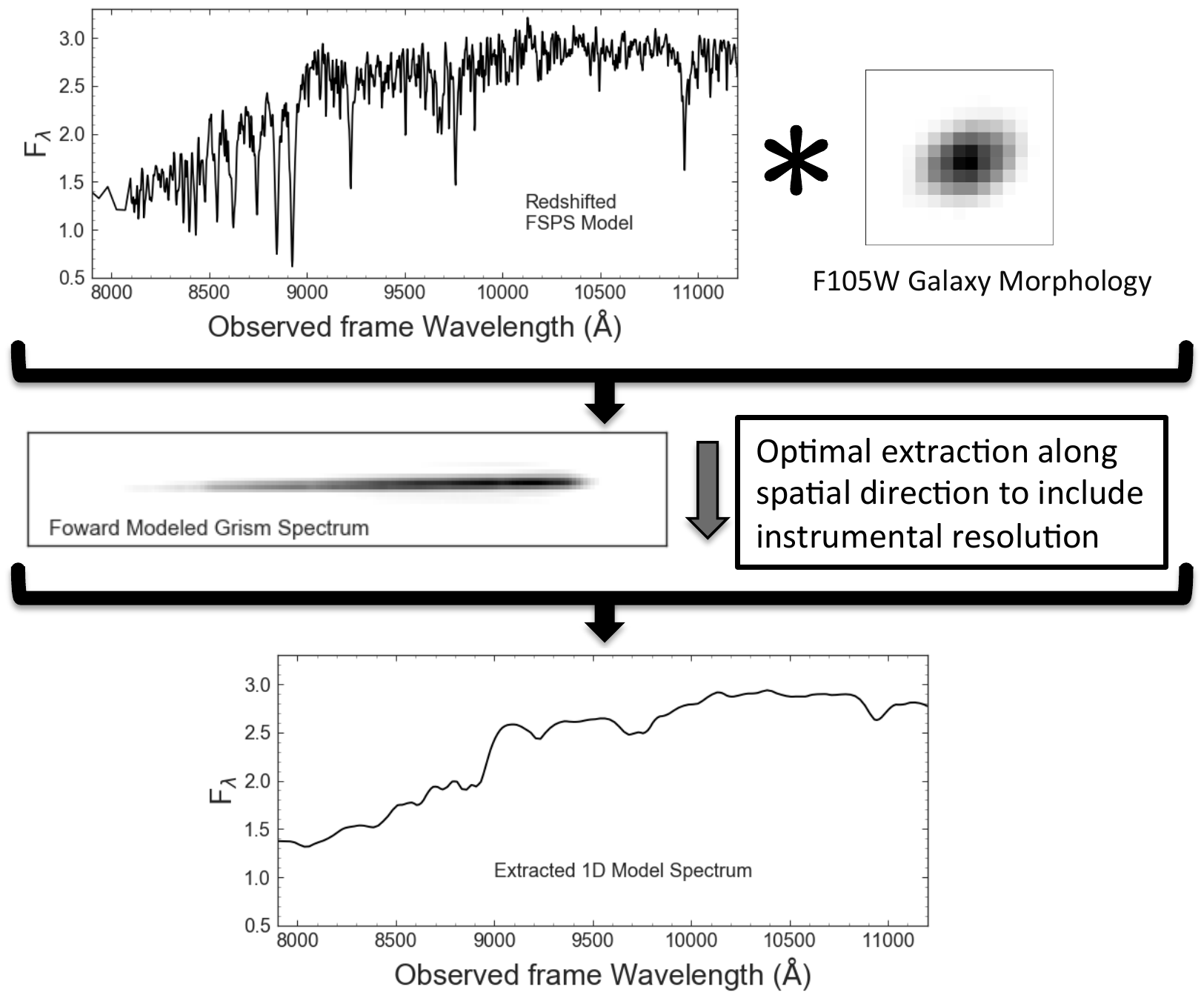}
\caption{Flow chart of the forward modeling process for the G102
spectra.   Top row shows the input, observed frame spectrum with the
native resolution of the stellar population synthesis model (in this
case FSPS; top left panel).  This is then  convolved with the 
$Y$--band (F105W or F098M) imaging (second right panel) and G102 grism
resolution to produce an accurate model of the 2D
G102 spectrum for this galaxy for this stellar population model (illustrated in the
middle row).   We then optimally extract a 1D spectrum for this model, which
now includes an accurate  G102 spectroscopic resolution (that includes
the morphological broadening appropriate for each galaxy).  For each
galaxy we build a suite of these models over a range of stellar
population parameters, which we then fit to the observed spectra for
each galaxy. \label{simdata}}
\end{figure*} 

\section{Methods and Tests of Model Fitting}

Our primary goal is to constrain the
ages and metallicities  of the quiescent galaxies in our sample. 
Here, we describe our method to
model the spectra, to fit them to the data, and to extract posteriors of
the stellar population parameters. 

\subsection{Forward Modeling of \hst\ Grism Data Using Stellar
Population Models}\label{section:models}

We use composite stellar population models in order to
investigate the SFHs, ages, and metallicities of
the quiescent galaxies at $1.0 < z_{grism} < 1.8$ in our sample. To generate
model spectra, we tested two sets of models:  the Flexible Stellar Population Synthesis
(FSPS) models \citep{conr09,conr10};  and the set of models from
\citet[BC03]{bruz03}.   The FSPS models use isochrones from the Padova
stellar evolution tracks, with the MiLeS empirical stellar library for
wavelengths $3500 < \lambda/\AAA < 7500$ (and BaSeL theoretical
spectra elsewhere).   The BC03 models use the Padova tracks and
isochrones combined with the STELIB empirical stellar library for
wavelengths $3200 < \lambda/\AAA < 9500$ and BaSeL spectra elsewhere.
For both the BC03 and FSPS models we assume a \citet{salp55} IMF
(although this has minimal impact on the age or metallicity
constraints we infer). 

While we consider both BC03 and FSPS for our model fitting. In 
Appendix~\ref{fsps-v-bc03}, we find Bayesian evidence against the 
BC03 models in favor of the FSPS
models. Fum16 obtain similar results, where they show that FSPS models
provide better agreement (lower $\chi^2$ values) for fits to stacked
quiescent galaxy spectra of similar resolution and rest-frame
wavelength.   We therefore report results from the FSPS for the
remainder of this work in light of the Bayesian evidence against the
BC03 models. 

We generated a large range of spectra from the FSPS models,  spanning
a range in age ($t = 0.5-6.0$ Gyr, in steps of \editone{0.1} Gyr),
metallicity ($Z = 0.1-1.6$~\zsol\ in steps of 0.05 \zsol), and
SFH ($\tau = 0.0-3.0$~Gyr in steps of 0.1 Gyr).
\editone{Regarding the metallicity, as we fit to the full spectra, our
measurements of the metallicities are affected both by the individual
elements in the spectral absorption features and those that affect the
stellar opacities that make up the continua.  Since the latter are
dominated primarily by Iron peak elements \citep{conr14},  we expect
that the ``metallicity'' ($Z$) here is a proxy for [Fe/H].
Nevertheless, there is evidence for higher abundances of [$\alpha$/Fe]
in massive galaxies at low redshift \citep[see, e.g.,][and references
therein]{matt94,thom05,choi14,conr14} and some evidence for this in
galaxies at high redshift \citep{krie16,stei16}.     Future modeling
of the abundance of individual elements would provide more detailed
insight into chemical enrichment histories of galaxies.   }

We assume "delayed $\tau$" SFHs that increase linearly
with time followed by an exponential decay characterized by an e-folding 
timescale $\tau$, such that the star-formation rate (SFR, $\Psi(t)$) at a given
age, $t$, is parameterized by $\Psi(t) \propto t \times \exp(-t/\tau)$.
These best represent the average SFHs of
quiescent galaxies at early and late times
\citep[e.g.,][]{papo11,smha14,paci16,iyer17}.  
 In our analysis we marginalize over $\tau$ as a nuisance parameter. 

For the remainder of this work, we interpret the age using the
``light--weighted'' ages, $\langle t \rangle_L$, which have been
averaged over the past SFH weighted by the luminosity of the stellar
population.  This is because $\langle t \rangle_L$ best corresponds to
the age of the stellar populations that dominate the galaxy
light. This is more robustly quantifiable than the SFH itself \citep[see e.g.,][]{papo01,salm15}, and best represents
the age of the stars that dominate the observed spectrum (see also
Fum16).  We derive the light-weighted ages using the SFHs and
(instantaneous) ages as 
\begin{equation}
\langle t(t^\ast, \tau) \rangle_L = \frac {\int _0 ^{t^*}
  \Psi(t,\tau)\ L(t^*-t,\tau)\  (t^*-t)\ dt} {\int _0 ^{t^*} \Psi(t,\tau)\ L(t^*-t,\tau)\ dt}
\label{avgage}
\end{equation}
where  $t^\ast$ is the instantaneous age of the model. The quantity
$L(t^* - t,\tau)$ is the luminosity at an age of $t^* - t$, 
for a given SFH, $\Psi(t,\tau)$, measured in a given filter band (we use
the SDSS $g$ band in the rest frame).

To include the effects of dust, we attenuate the model spectra by the value $A(V)$ using the \cite{calz00}
dust law \citep[to be consistent with the dust measurements from \threedhst][]{skel14}. 
\begin{equation}\label{eqn:dust}
F(\lambda)_\mathrm{observed} = F(\lambda)_\mathrm{unattenuated} \times
10^{-0.4\ k(\lambda)\ A(V) / R(V)}
\end{equation}
where $k(\lambda)$ is the starburst reddening curve, 
and $R(V)$ is the total--to--selected
attenuation in the $V$-band, with $R(V)=4.05$.

We then simulated the 2D and 1D grism spectra from using the stellar population models with the software package
\grizli\ (the grism redshift and line analysis software),
developed by CLEAR team member
(G.~Brammer)\footnote{https://github.com/gbrammer/grizli}. \grizli\ uses
as input the stellar population model spectrum, the galaxy redshift,
and the galaxy image to correctly model the 2D grism data.   For this
latter step, we use imaging from F105W (or F098M) from CANDELS.  
Using the correct morphology is highly important, as the
galaxy morphology effectively ``smooths'' the resolution of the
(slitless) grism spectrum \citep[referred to as ``morphological
broadening'',][]{vand11}, which is caused by the image of the galaxy
dispersed over the range of wavelengths covered by the grism.  The
broadening of features is therefore correlated to the morphology of
the galaxy (and at the resolution of our data this is the dominant effect; 
we can neglect the contribution from dynamical motions within the galaxy). 
The correlation with morphology means the spectral resolution decreases 
with increasing galaxy size (where narrow lines can be lost or smoothed 
over). More compact galaxies, due to their smaller size, will produce 
the least morphologically broadened spectra (i.e., they will have higher
spectral resolution). This modeling process of the data is illustrated in
Figure~\ref{simdata}.   

\editone{
We further adopt an additional ``uncertainty'' to account for
systematic errors, which could either arise from incompleteness or
inaccuracies in the stellar population models, or from systematic
errors in the data \citep[see e.g., discussion in][]{papo01,bram08}.   To
counter this ``model error'' we add an additional 
systematic error term (a template
error function) following \citet{bram08} (detailed in 
Appendix \ref{app-tmperr}) .}

\subsection{Fitting Grism Models to Grism Data}\label{section:fitData}

In what follows we describe fitting the 1D spectral models derived
from \grizli\ and the stellar populations to the 1D G102 data for each
galaxy.  Because we are considering only a small number of parameters
at present, we generate models over a grid of parameter values.  We
derive a $\chi^2$ goodness-of-fit measurement for each
combination of model for the G102 data, and we then calculate a likelihood
distribution using the following, 
\begin{equation}\label{eqn:p_chisq}
P(D| \Theta) \propto \exp(- \chi^2/2)
\end{equation}
where $D$ is the data and $\Theta$ is the set of parameters
we consider.   We derive the joint probability density function
using Bayes' theorem
\begin{equation}
P(\Theta | D) \propto {P(D | \Theta) P(\Theta)}
\end{equation}
where $P(\Theta)$ represents prior information. Here we assume
``flat'' priors over the parameter range \citep[see
discussion in][]{salm15}; for metallicity, \editone{(light-weighted)
  age}, SFH, and redshift. \editone{For the dust attenuation, we use a
  prior derived by fitting a skewed Gaussian to the distribution of
  $A(V)$ values derived from broad-band photometry from \threedhst\ for galaxies in our redshift  and mass range.} We then marginalize to get posteriors on individual
parameters.  For example, given $\Theta$ = ($X$, $Y$, $Z$), we would
derive the posterior on parameter $X$ as, 
\begin{equation}\label{eqn:p_margin}
P(X) = \int_Y \int_Z\ P(X,Y,Z|D)\ dZ\ dY
\end{equation}

In the sections that follow, we test the ability of our method to
recover model parameters using simulated data.   We first test the
results for model stellar populations in different redshift ranges
(Section~ \ref{fit_tech}). Next, we test the accuracy of recovered
stellar population parameters for models fit to simulated data with
and without the spectral continua
(Section~\ref{section:sim_continua}).  As a result of these tests we
concluded a best practice to (1) sub-divide galaxies by redshift so
that the G102 spectra cover different spectral features which allows
us to study systematics resulting from differences in rest-frame
wavelength coverage, and (2) fit the full spectrum including the
continua as this provides the most accurate constraints on the stellar
population parameters.  Following these tests we apply these methods
to (re-)measure galaxy redshifts from the G102 grism data
(Section~\ref{rshift}).  We then fit the grism data to derive stellar
population parameters for the galaxies in our
sample. (Section~\ref{section:fit_data}).  

\subsubsection{Tests using Simulated Data in Different Redshift
  Ranges } \label{fit_tech}

We tested our ability to measure meaningful constraints on model
parameters of galaxies at different redshifts. We selected (direct)
images and redshifts from four real galaxies in our sample, and input
models with 6 different combinations of parameter values spaced
throughout a plausible range of values; $t = 1.9$ to $4.5$~Gyr, $Z=$
0.42 to 1.32~$Z_\odot$, and a delayed SFH  with $\tau = 0.5$ Gyr.   In
all cases, we simulated model spectra, as illustrated in
Figure~\ref{simdata}.     For the simulated ``data'' we added real
noise measured from extracted the G102 spectral data.

We then fit the simulated model data using the method described in
Equations~\ref{eqn:p_chisq}--\ref{eqn:p_margin}, fixing redshift to
the true value and setting A($V$)$ = 0$.  We repeated these
simulations 1000 times (for each set of ``truth'' parameters) to
generate likelihoods for the accuracy of the recovered parameters. One
limitation of this simulation is that model spectra exactly match the
(simulated) ``data'', but this allows us to determine the limitations
of the fitting procedure in the idealized case before applying it to
real data.   We then derived posteriors on the model parameters. 

\begin{figure}[t]
\epsscale{1.}
\plotone{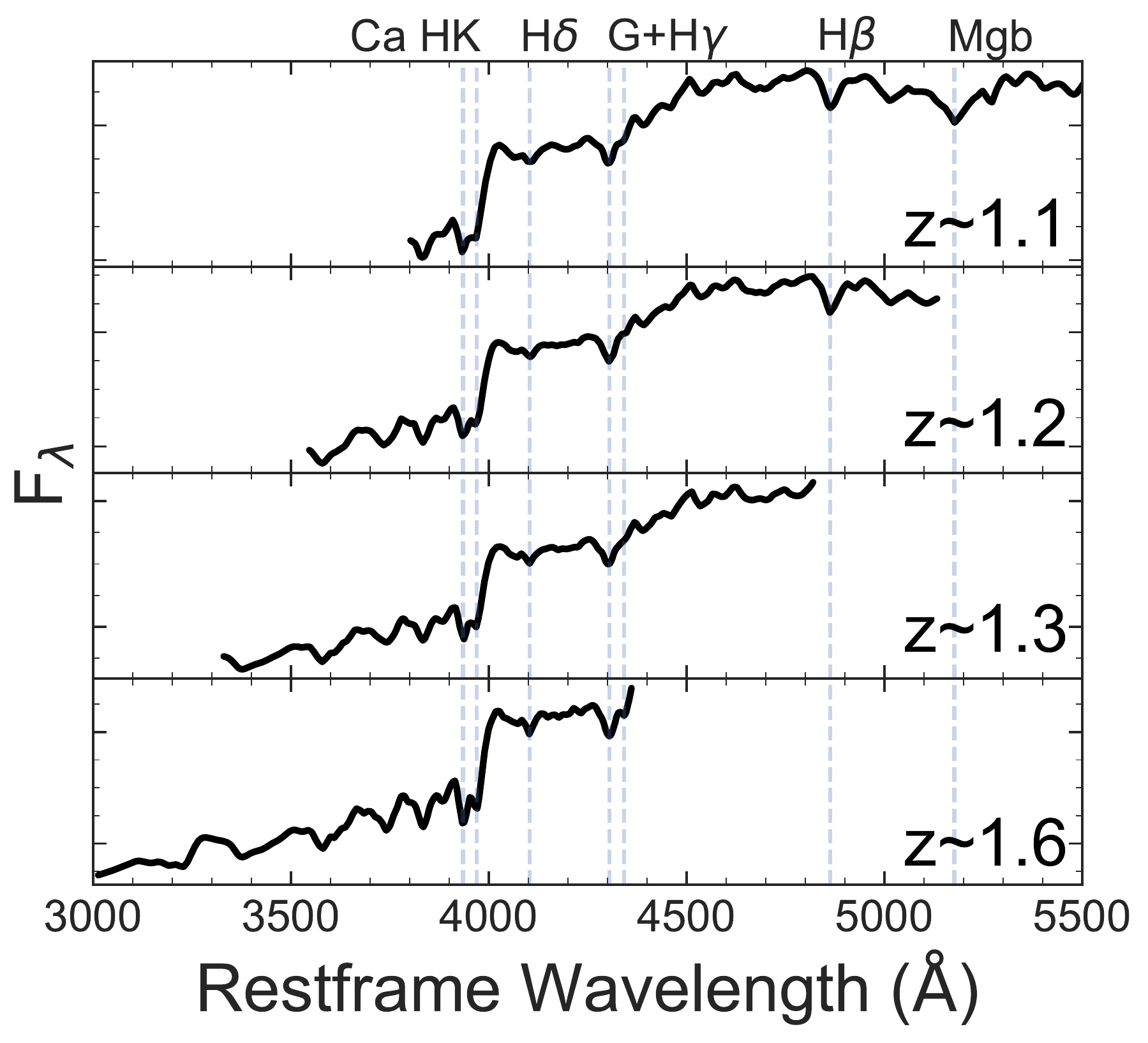}
\caption{Comparison of redshift subgroups. Each panel shows the portion of the rest-frame spectrum covered by the G102 in each redshift subgroup. Important age-- and metallicity--sensitive spectral features are labeled.\label{spec_diff}}
\end{figure} 

\begin{figure}[t]
\epsscale{0.85}
\plotone{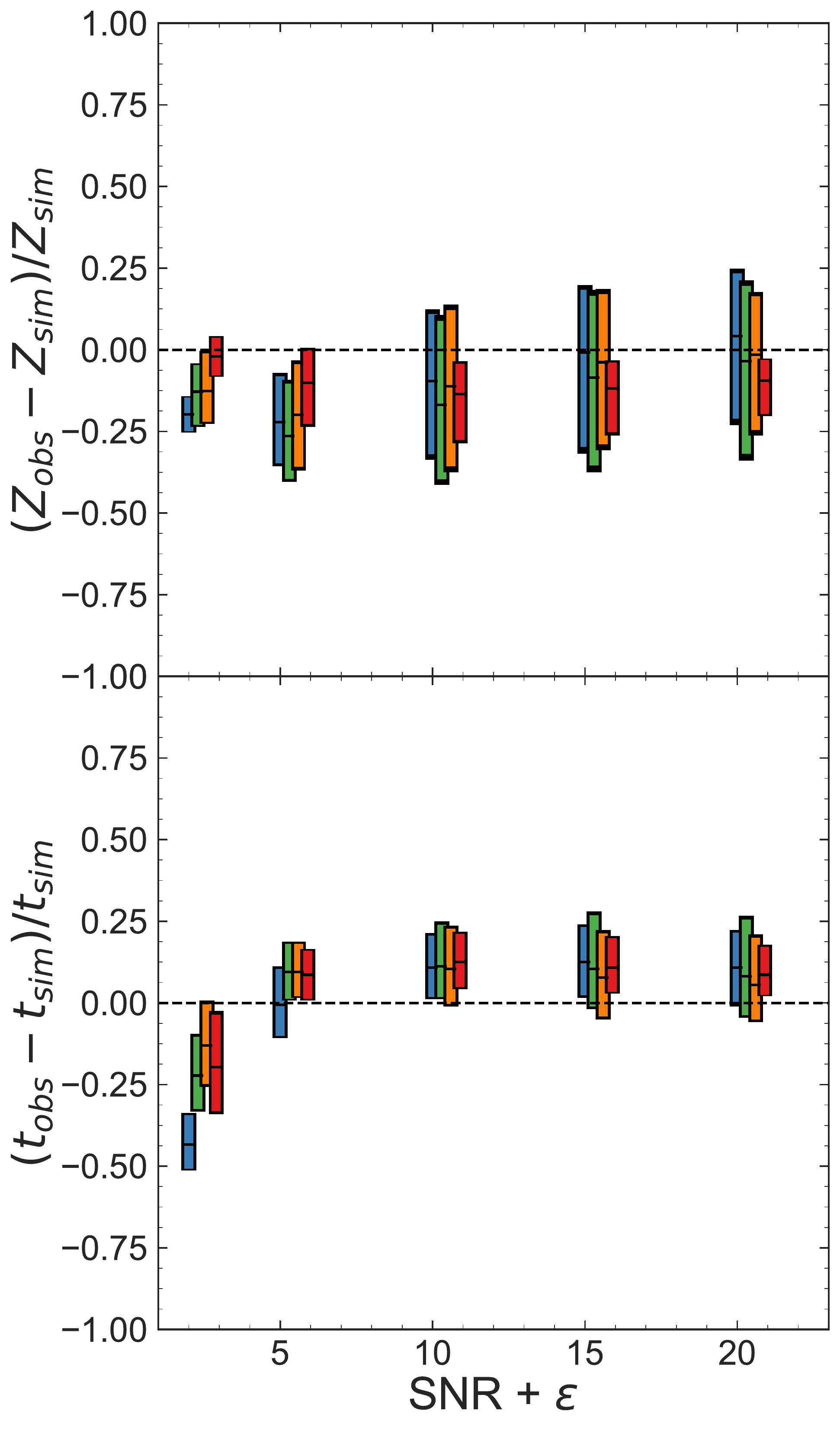}
\caption{Illustration of the distribution of fits on stellar population 
parameters derived  from fits to simulated data. This was done for 
each redshift subgroup as a function of SNR (shown with a slight offset ($\epsilon$)
for readability). Fits shown here are in relation
to simulated spectra with ``true'' values $t=$2.5~Gyr and $Z=$1~$Z_\odot$ .
Because the G102 spectrum probes different portions of the rest-frame spectrum,
the accuracy of the derived parameters changes as the important
features shift across the grism coverage.  \label{rshift-sim}}
\end{figure} 

Inspecting our results, we identified four natural redshift ranges which
probe different rest-frame spectroscopic features, and therefore have
slightly different systematics in the results from the fitting. These
redshift ranges have medians of approximately $z=1.1$, 1.2, 1.3, and
1.6, and correspond to where the different age-- and
metallicity--sensitive spectral features shift in and out of the grism
wavelength coverage, as illustrated in Figure~\ref{spec_diff}.  At all
redshifts ($1.0 < z_{grism} < 1.8$) the G102 spectra contain the 4000~\AA
/Balmer break, which provides constraining power on the ages and
metallicities of the stellar populations. 

Figure~\ref{rshift-sim} shows the measured distributions on
\editone{(light-weighted)} age ($t$) and metallicity ($Z$) for
simulated models, for one combination of ``true''
\editone{(light-weighted)} age $t=$2.5~Gyr and metallicity
$Z=$1~$Z_\odot$, in each of the four redshift bins (we obtain similar
results for other combinations of  age and metallicity, but show these
as they are close to the values we derive for the real data for the
galaxies in our sample). 

In what follows we qualitatively describe the fits in each subgroup 
for one set of parameters. This set was chosen as it represents a 
region of the parameter space we expect to be well populated. We quote
values for fits of spectra with SNR = 10 (as an example of how well 
``good'' spectra can constrain parameters), though we find that we 
are able to recover the parameters accurately  down to SNR	$\sim$ 3. 

For the first redshift subgroup,  \zone\ ($ 1.00 < z_{grism} < 1.16$),
the G102 data cover wavelengths that probe features out through
Mg$b$. Unlike the Balmer lines, which are mostly age dependent,
including the Mg$b$ feature provides improved constraints on
metallicity. The age constraints are also reliable, but
suffer from the lack of spectral coverage the rest-frame $U$-band
data. 

The second redshift subgroup, \ztwo\ ($ 1.16 < z_{grism} < 1.30$),
includes features out through H$\beta$. The lack of Mg$b$ could make
constraining metallicity more difficult, but with a better defined
4000~\AA\ break and coverage of the rest-frame $U$ band these data
probe the shape of the rest-frame $U$--$B$ continuum and yields
relatively accurate constraints on the age and metallicity.
%

The third redshift subgroup, \zthree\ ($ 1.30 < z_{grism} < 1.45$),
contains galaxy spectra that lack coverage of Mg$b$ and H$\beta$, but still
contain the G+H$\gamma$ features. The constraints on the age are aided
due to stronger presence of the rest-frame $U$--$B$ continuum.   
%

The fourth subgroup is our highest redshift group, \zsix\ ($ 1.45 <
z_{grism} < 1.70 $).  The most prominent features of the \zsix\ group are
the 4000 \AA\ break and the large amount of  $U$--$B$ continuum. This group
differs from the \zthree\ group by
having the $H_{\gamma}$ feature in a noisy region of the spectra. 
%

\editone{For the all groups, we are able to recover the metallicities
and \editone{light-weighted} ages   with typical uncertainties of
$\sigma(Z)$$\approx$0.30 Z$_\odot$ and $\sigma(t)$$\approx$0.3 Gyr
($\sigma(t)$$\approx$0.2 Gyr for \zsix) respectively. The tighter
constraint seen in the \zsix\ age measurement is an artifact of the
definition of SNR. We use the SNR per pixel, averaged over
8500--11,500~\AA.  For $z\sim 1.6$, this observed wavelength range
mostly covers the $U$ band, and the portion around (and above) the
rest-frame 4000~\AA/Balmer break has much higher S/N. } 
%

\editone{When taking all redshift subgroups, and parameters into account,
Figure~\ref{rshift-sim} indicates there may be a slight bias in
light-weighted age  and metallicity on the order of $\approx +10 \%$ 
and $\approx -10 \%$, respectively (for SNR $>$ 5).  However,
these biases are small relative to the constraints we derive on these
parameters for each galaxy. We therefore make no attempt to correct these.}

\begin{figure}[t]
\epsscale{0.85}
\plotone{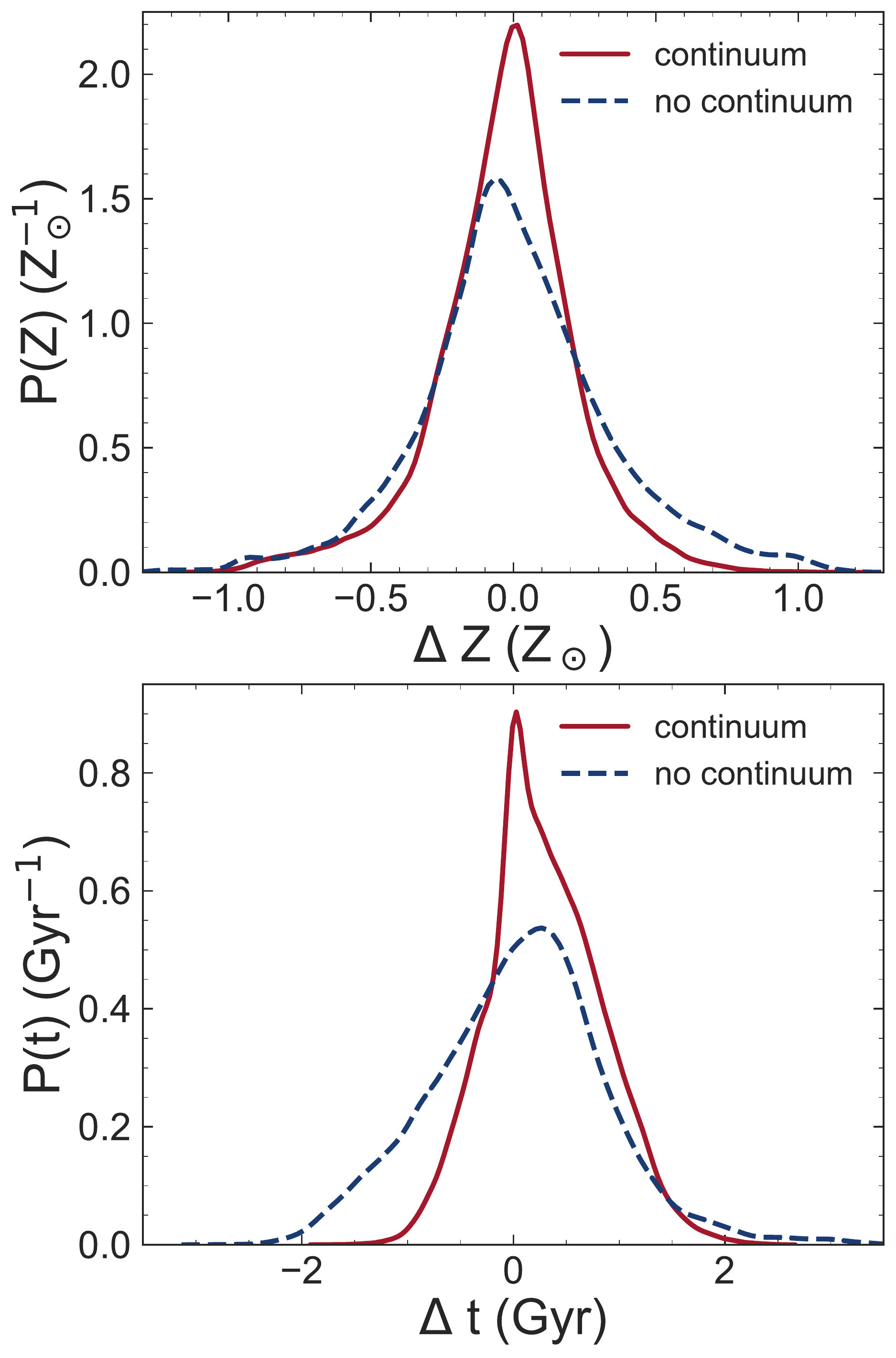}
\caption{Comparison of parameter distributions from fitting with and without the continuum. 
The top panel shows the posteriors on the metallicity distributions. The bottom panel shows 
the light-weighted age distributions. 
\label{mc-cont}}
\end{figure} 

\subsubsection{Tests Using Simulated Data with and without Continua}\label{section:sim_continua}

We performed an additional test to determine the importance of fitting
to the galaxy spectra including the continuum and fitting with the
continuum divided out.  Although we show in Section~\ref{sumofo} that
the flux calibration is not a significant source of error, past
studies frequently remove the continuum, allowing for fits directly to
the stellar population features, and mitigating against uncertainties
in flux calibration \citep[Fum16]{conr10}.  However, there is
information in the continuum as it is the superposition of the
photospheres of the composite stellar populations, whose
characteristics depend strongly on age, metallicity and SFH.  
As we have determined that the \hst\ flux calibration is
both accurate and stable, including the continuum provides important
information for the model fits. 

For our test, we removed the continua (on the models and simulated
``data'') by dividing the spectra with a third order polynomial fit, 
masking out regions with possible emission or absorption
features except for the 4000 \AA\ break.  
We then fit the models to the data, derived
likelihoods, and marginalized over parameters to derive posteriors. 

Figure~\ref{mc-cont} compares the recovered parameters derived by
fitting simulated spectra with and without the continuum. The abscissa
in both the top and bottom panels show the difference between the true
value and the derived median value. We smooth the distribution of
points with a kernel density estimator to derive the
likelihood. While we recover, on average, the true value for both
cases, using the continuum provides a tighter and more symmetric
distribution, with improved results.  Formally, using the continuum, 
we derive median and 68\% confidence intervals for
the offset in metallicity and \editone{light-weighted} age as $\Delta Z_{c} =
-0.013_{-0.22}^{+0.19}$ \zsol\ and  $\Delta t_{c} =
-0.26_{-0.46}^{+0.59}$ Gyr.   
Using fits to data where the continuum
has been divided out, we derive
$\Delta Z_{nc} = -0.013_{-0.27}^{+0.32}$ \zsol\ and $\Delta t_{nc} =
-0.08_{-0.86}^{+0.71}$ Gyr.    For the case that includes fits to the
full spectra (i.e., including the continua and all absorption
features), the offsets and uncertainties are smaller (and the
posterior is more symmetric). We therefore fit to the full spectra in
our analysis of the galaxies in our sample=.  

\begin{figure}[t]
\epsscale{1.2}
\plotone{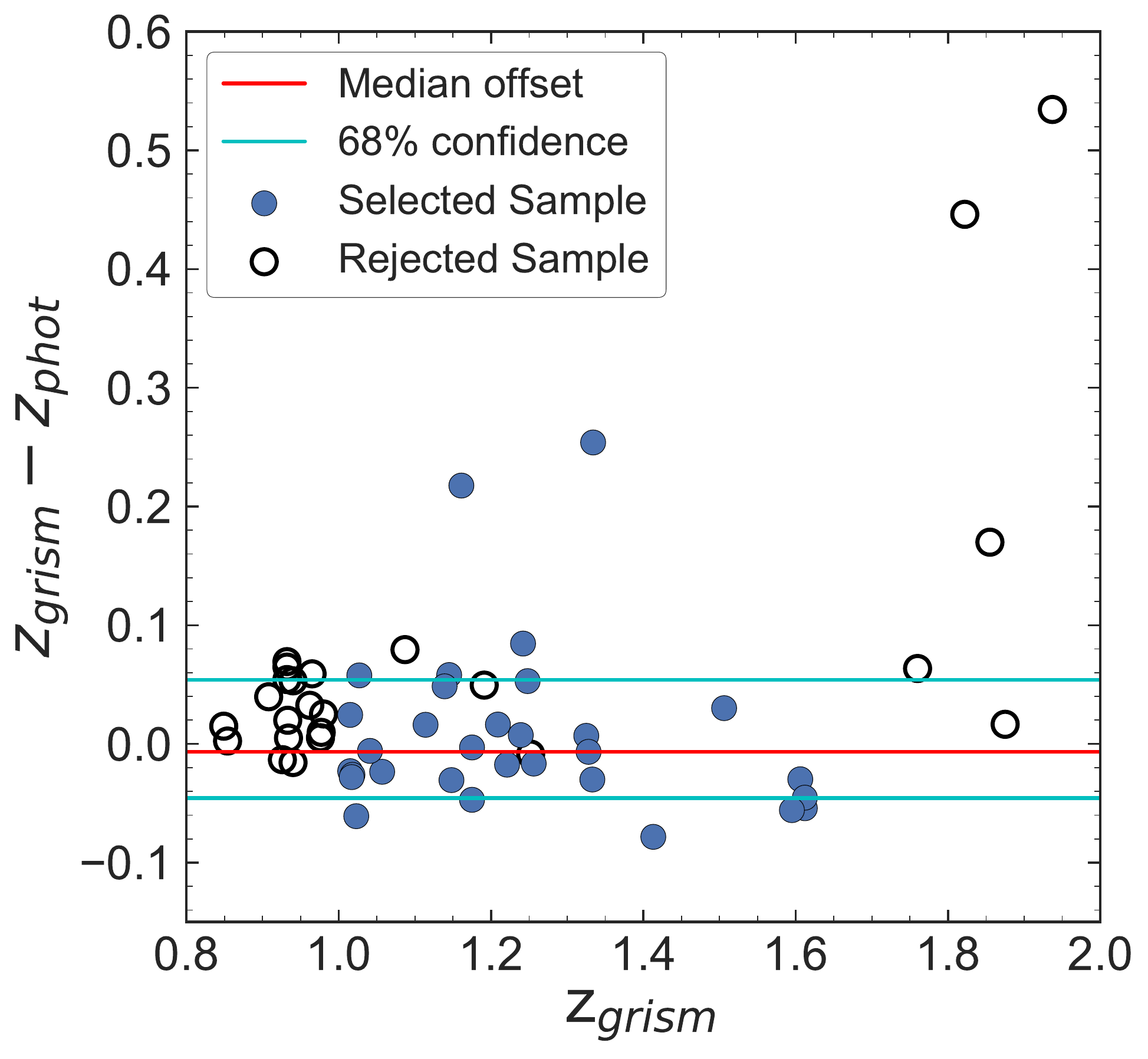}
\caption{Comparison of our photometric redshifts derived from the
deep  CLEAR G102 spectra ($z_{grism}$) to those derived from
broad--band photometry from \threedhst\ ($z_{phot}$).
The median offset between the  two redshifts ($\Delta z
\sim$ 0.01) is marked by a red line and the 68\% scatter
intervals ($-0.07 < \Delta z < 0.04$) are bounded by cyan lines.
Empty circles mark the rejected sample; this includes the 
sample of galaxies rejected after fitting $z_{grism}$. 
\label{zdiff}}
\end{figure}

\section{Results} 

\subsection{Measuring Redshifts from the Grism G102 Data\label{rshift}}

It is important to have accurate redshifts when modeling the galaxy
stellar populations. We therefore re-derived galaxy redshifts,
$z_{grism}$, by fitting the model grism spectrum to the data, \editone{using an
iterative method.  We firstly fit $z_{grism}$ over a coarse grid
  of model parameters to estimate its value.  We secondly fit
  $z_{grism}$ over the full set of parameters but limiting the range
  of redshifts.  These two steps saved computation time by a factor of
$\approx$25. }

For the first $z_{grism}$ iteration, we generated models with fixed $\tau$ = 100
Myr over a coarser grid of metallicities and ages, with a very fine
grid of redshift ($\Delta z = 0.001$) over a range of $0.8 < z < 2.0$.  For the galaxies in
our sample, the choice of $\tau$ does not affect the measurement of
$z_{grism}$ for $\tau < 500$~My for the quiescent galaxies in our
sample. 
We fit using the set of parameters $\Theta= (Z,t, z_{grism})$,  and then
marginalize to obtain a posterior on $z_{grism}$,
\begin{equation}
P(z_{grism}) = \int_Z \int_t P(Z,t,z_{grism})\ dt\ dZ
\end{equation}

From the $P(z)$ we derive median values and a 68\%-tile range on
$z_{grism}$ for each galaxy. \editone{These median values are used in
our full model fitting (below) to set the redshift range used (where
we fit over a range $z_{median} \pm dz$ with $dz=0.02$ (which spans
the peak and majority of the probability mass in redshift space for
galaxies in our sample).   Table~\ref{optable} reports the median and
68\% confidence interval on $z_{grism}$ from these fits.  The
grism-derived redshifts have typical uncertainties $\sigma_z \lsim
0.004$, and are significantly improved compared to the
broad-band-derived photometric redshifts with $\sigma_z \approx
0.02-0.11$ (see Table~\ref{iptable} and Figure~\ref{zdiff}). Similar
results are seen in \cite{momc16}}

\begin{figure*}[t]
\epsscale{1.17}
\plotone{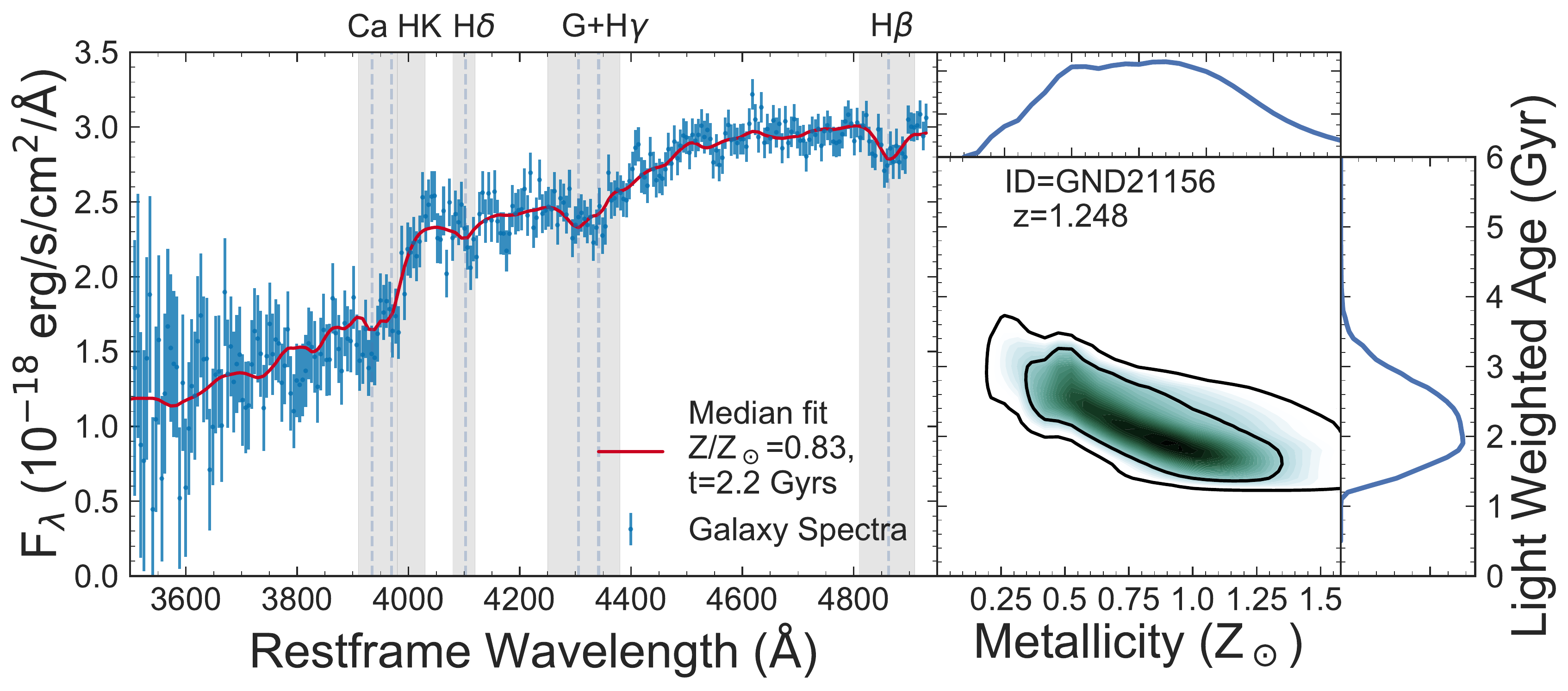}
\caption{\textbf{Left}: The 1D G102 grism 12-orbit data for the galaxy
GND21156 at $z=1.251$ as a function of rest-frame wavelength.   The
shaded regions show the locations of prominent spectral features.  The
red solid line shows a model with parameters given by the median
metallicity ($Z$) and light-weight age ($t$) derived from the
posteriors derived on each parameter.  \textbf{Right}:The posteriors
on the stellar population parameters of metallicity and light-weighted
\editone{age} for the galaxy in the left panel.  The main panel shows
the joint likelihood (with the 68$\%$ and 95$\%$ confidence intervals
outlined in black) derived on both parameters jointly using
Equation~\ref{eqn:p_margin}. The sub-panels to the right and above the
main panel show the individual posteriors on \editone{light-weighted}
age and metallicity.  From these we derived median and 68\%-tile
ranges for each parameter for each galaxy in our sample.
\label{galspec}}
\end{figure*} 

\subsection{Measuring Stellar Population Parameters from the CLEAR G102 Data }\label{section:fit_data}

To constrain the stellar population parameters for the galaxies in our
sample, we generated a large range of spectra from the FSPS models.
\editone{Our parameter space consists of metallicity, age, and SFH 
(ranges defined in Section \ref{section:models}), the redshift range described
in Section~\ref{rshift}, and a range of dust attenuation of $ 0.0 < $ A$(V) /
\mathrm{mag} < 1.0$.} 

\editone{As discussed above, the use of delayed-$\tau$ SFHs is
justified by the fact that our sample consists of quiescent galaxies,
where the \editone{light-weighted} ages are typically many times the
$e$--folding timescale. Such SFHs are motivated by
studies that find galaxy SFRs rise at early times  and decline at
later times \citep[e.g.][]{papo11,salm15,paci16,carn17}.  We note that
using other parameterizations of the SFH would
change the stellar--population ages and metallicities by $<$0.1 dex
(see e.g., Gal14).  Our tests using purely exponentially declining
SFHs (e.g., SFR($t$) $\propto exp(-t/\tau)$)  results in (light-weighted) ages younger by
$\approx$15\% compared to the results here using delayed-$\tau$
models.  This would push the formation redshifts of the galaxies in
our sample to later epochs, but the effect is systematic and would not
effect our overall conclusions.}

We fit the models to each galaxy individually. For each galaxy, we
generate the suite of 2D model spectra using the galaxy's F105W (or
F098M) image and extract a 1D spectrum as illustrated in
Figure~\ref{simdata}.  We then fit the models to the data for each
galaxy using Equations~\ref{eqn:p_chisq} to \ref{eqn:p_margin},
\editone{where we marginalize to generate posteriors on each
parameter.   We then derive median and 68\% confidence intervals on
each parameter.   Table \ref{optable} lists these values for each
galaxy in our sample;  including fits to the (lighted-weighted) age,
metallicity, $\tau$, $z_{grism}$, and dust attenuation.}
 
Figure~\ref{galspec} shows an example fit for one galaxy in our
sample.  The model has the median values derived for the stellar
population parameter posteriors (i.e., this is not the \textit{best
fit} model, but instead is the model with metallicity and
light-weighted age matching the median of the posterior
distributions). The right panel  of Figure~\ref{galspec} shows the
corresponding joint posterior on \editone{(light-weighted)} age and
metallicity, $P(t, Z)$ and individual parameter posteriors, $P(t)$ and
$P(Z)$. This figure also shows the benefits of using the light-weighed
age rather than the instantaneous age.  Based on our tests, the
light-weighed age is best at breaking the metallicity-- age degeneracy
and we see an anti-correlation in the joint $P(t, Z)$ posteriors.   As
a result, the distributions on the individual parameters ($Z$ and $t$)
are more symmetric and well constrained.  \editone{In Appendix~\ref{section:appendix_allplots}
  we show 
%
%
similar plots for all the galaxies in our sample,
along with their respective median fit models, and metallicity  and
age joint likelihoods.}

%
%

We observe possible \hb\ emission in 3 of the galaxies ($<$10\%) in
our sample.  When this residual \hb\ emission  is present we mask out
this region when fitting. \editone{We have also refit all the galaxies in our sample,
removing the central region of \hb, and find no measurable impact on
the parameter fits (in 11 out of 18 galaxies), with random (i.e., not
systematic) changes of $<$10\% in the median fits on parameters in the
other cases.  We therefore make no correction for \hb\ (or other nebular)
emission.  }  

\editone{Although it is our goal to derive model fits solely from the
G102 grism data for each galaxy in this study, we have compared the
best-fit model fits from our analysis to available broad-band
photometry \citep[from][]{skel14}.   A quantitative comparison yields
limited information  as the best-fit models do not capture the full
range of allowable parameter values, but they do offer guidance to the
fidelity of the model parameters.  We find that the broad-band
photometry and best-fit model fits span the same range of color across
out to longward of (rest-frame) 1~\micron\ in 70\% of cases, and this
is improved to nearly 100\% when we fix the dust content in the models
to be $A(V) = 0$~mag.   Because we marginalize over dust attenuation
here, our uncertainties include this information.  We plan to explore this
more fully in a future work, using all available grism and broad-band
data to constrain the stellar populations of galaxies. } 

\subsection{Stacked Results for Galaxies in Redshift Subgroups}

To derive parameter constraints for all galaxies in each redshift
subgroup, we adapt the ``stack-smooth-iterate'' technique discussed in
\cite{bak00}.   This allows us to combine the posterior likelihoods,
$P(Z)$ and $P(t)$, derived from each individual galaxy, placing
constraints on the subgroups \textit{as a population}. 

The steps in our ``stack--smooth--iterate'' method are as follows. We explain the steps in  greater detail below. 
\begin{enumerate}
\item  Sum the posteriors using weights to remove large peaks from the summed distribution.
\item Apply a prior derived from the previous iteration  (the prior is flat on the first iteration).
\item Smooth the distribution to remove smaller residual peaks.

\item Set this smoothed distribution as the new prior.
\item Iterate until asymptotically reaching the parent distribution.
\end{enumerate}

The process begins by deriving the weights, $w_i$. To do this we 
take the inverse variance derived by a jackknife process.   The
jackknife begins by summing
the posteriors to create a distribution $Y(\Theta)$
\begin{equation}
Y(\Theta) = \frac{1}{n}  \sum_i  P(\Theta | D)_i
\end{equation}
where $n$ is the total number of posteriors. We then quantify the effect a single
posterior has on $Y(\Theta)$ by leaving it out of a newly summed distribution $\bar{Y}(\Theta)_i$
\begin{equation}
\bar{Y}(\Theta)_i =   \frac{1}{n-1} \sum_j^{ j \neq i}  P(\Theta | D)_j
\end{equation}
We then calculate  the weights $w_i$ to be 
\begin{equation}
w_i = \frac{1}{ \int_\Theta (\bar{Y}(\Theta)_{i} - Y(\Theta))^2 d\Theta}
\end{equation}
Here we find the variance between $Y(\Theta)$ and $\bar{Y}(\Theta)_i$, where we excluded the $i$th
posterior and integrated over $\Theta$. The advantage of weighting in this way is that the weights naturally
handle individual $P(\Theta | D)_i$ that are sharply peaked (e.g., a
distribution approximately that of a $\delta$-function), as this would
otherwise dominate an average of the subgroup distributions.

We then stack our posteriors using a weighted sum
\begin{equation}
P^*(\Theta) =\frac {\sum_i w_i P(\Theta | D)_i P(\Theta)}{\int_\Theta \sum_i w_i P(\Theta | D)_i P(\Theta) d\Theta} \; .
\label{stack}
\end{equation}
$P(\Theta)$ is a prior on $\Theta$, which is derived from the data
itself (i.e., $P(\Theta)$ is flat in the 1st iteration, then taken as
$P(\Theta) = P^\ast(\Theta)$ on successive iterations, see below). 

We then iterate to calculate $P^*(\Theta)$ from Equation~\ref{stack}.
On each iteration (including the first) we smooth the $P^*(\Theta)$
distribution using local linear regression \citep{clev79} to remove any
residual peaks. While
the choice of smoothing algorithm is not as important, the smoothing
step is important
because it will remove any peaks on the distribution that the
weights did not. During the iteration process these residual peaks (because 
it is a multiplicative process) will begin grow and shift the distribution.

Once smoothed we set $P(\Theta)=P^*(\Theta)$ , and iterate.  In this
way we derive the prior $P(\Theta)$ from the data
itself. \editone{(Note that we apply this ``prior'' only to derive the
stacked likelihood here, and not to alter the likelihoods for
individual galaxies derived above.)}  After several iterations
$P^*(\Theta)$ will converge to reach a distribution, which is an
estimate of the parent distribution of the sample. We emphasize that
these are approximations of the parent distributions, and larger
sample sizes would more reliably recover these distributions. Appendix
\ref{stacking} describes our tests to check the accuracy of this
method to recover a true known parent distribution.

\begin{figure*}[t]
\epsscale{1.1}
\plotone{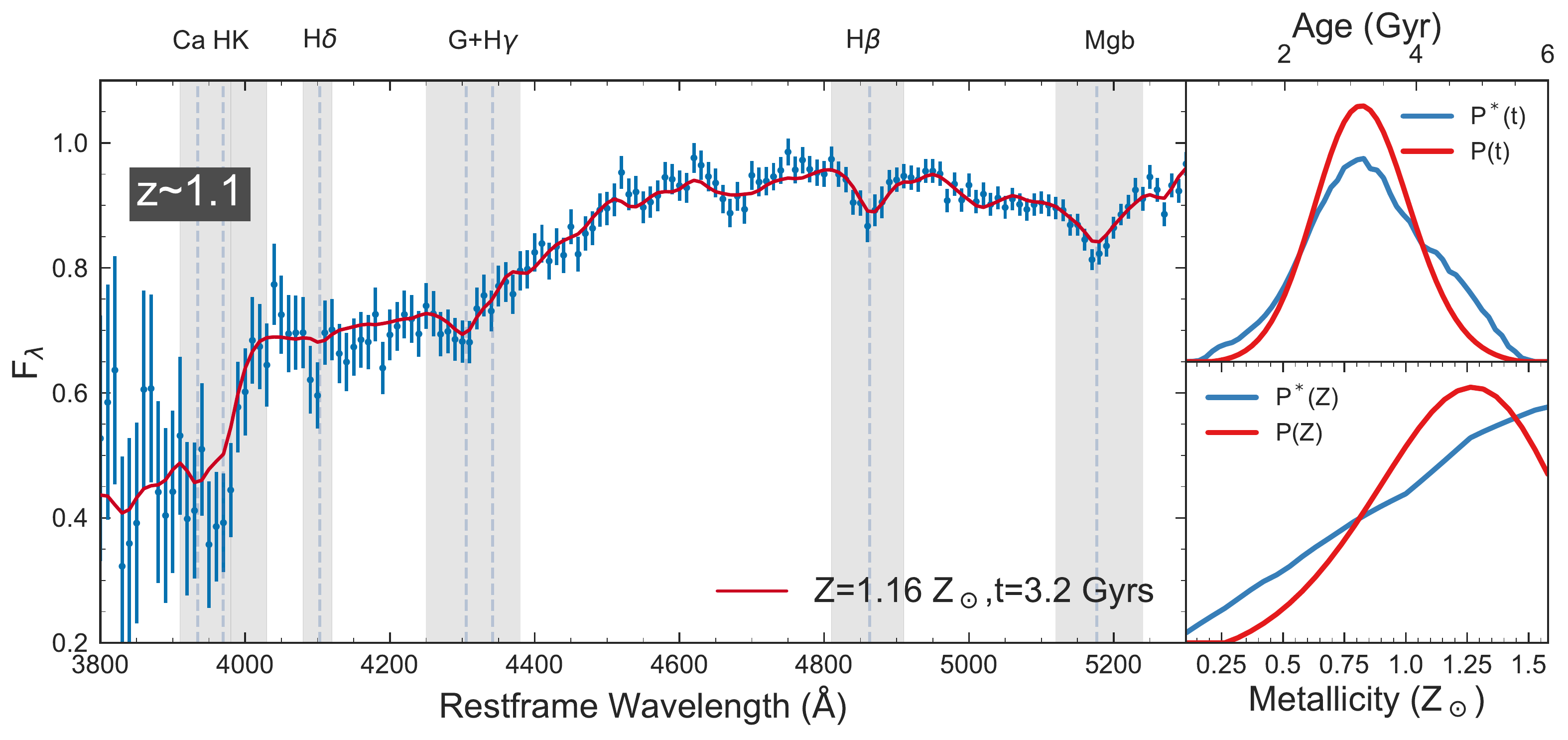}
\plotone{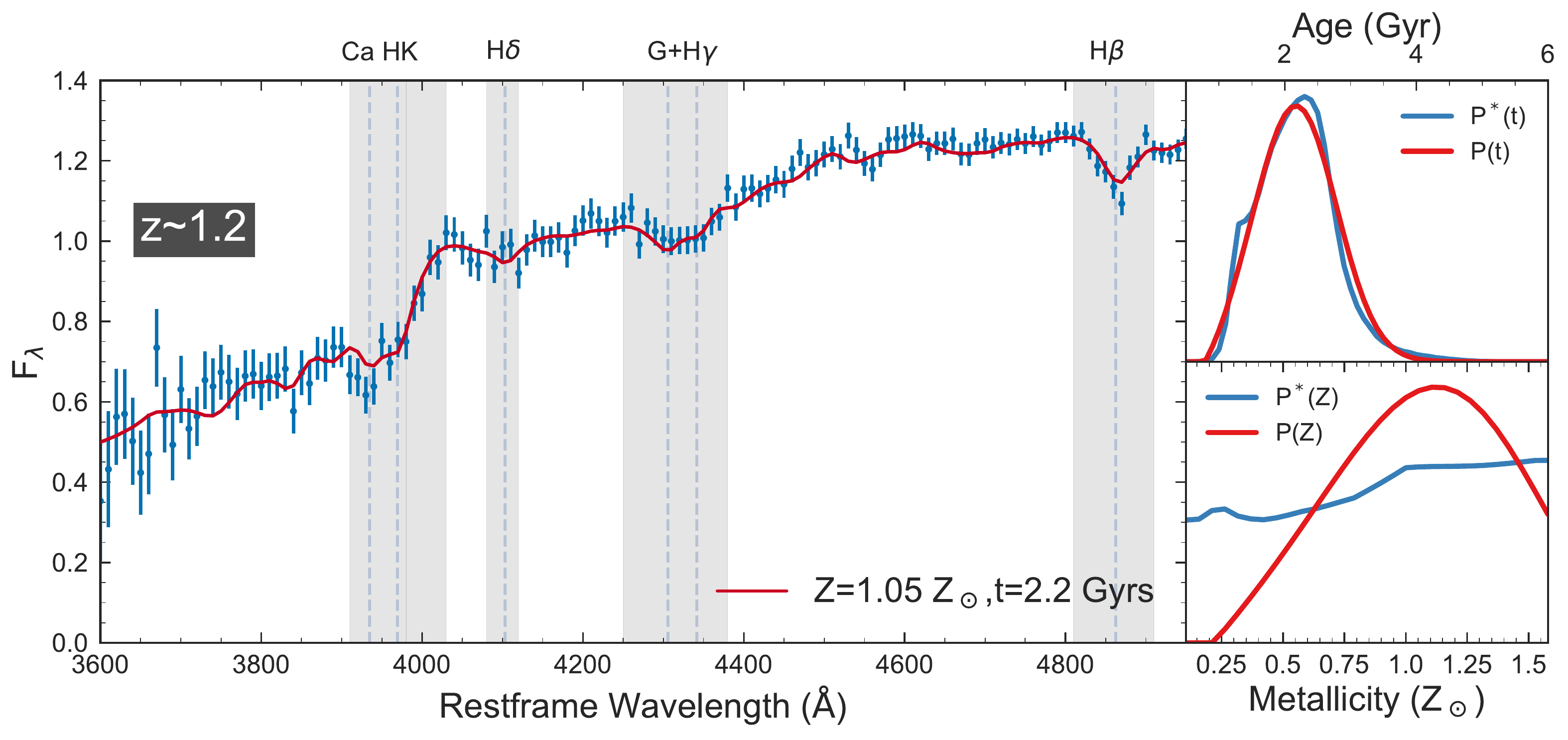}
\caption{Top:  Stacked spectra and posteriors for galaxies in the
\zone\ redshift subgroup.   The main panel shows the stacked 1D G102
grism data against rest-frame wavelength.   The sub-panels on the right
show the stacked posteriors (blue:weighted stacking, 
red:``stack-smooth-iterate'' method ) on \editone{light-weighted} age ($t$) and metallicity ($Z$) derived
using the method described in Section \ref{section:fit_data}.   The
red-solid line in the main panel shows a model with the median $Z$ and
$t$ taken from these individual-parameter posteriors (i.e., these are
\textit{not} best-fit models to the stack).    Bottom: Same
plots for the \ztwo\ redshift subgroup of galaxies.  \label{lowz}}
\end{figure*}

\begin{figure*}[t]
\epsscale{1.1}
\plotone{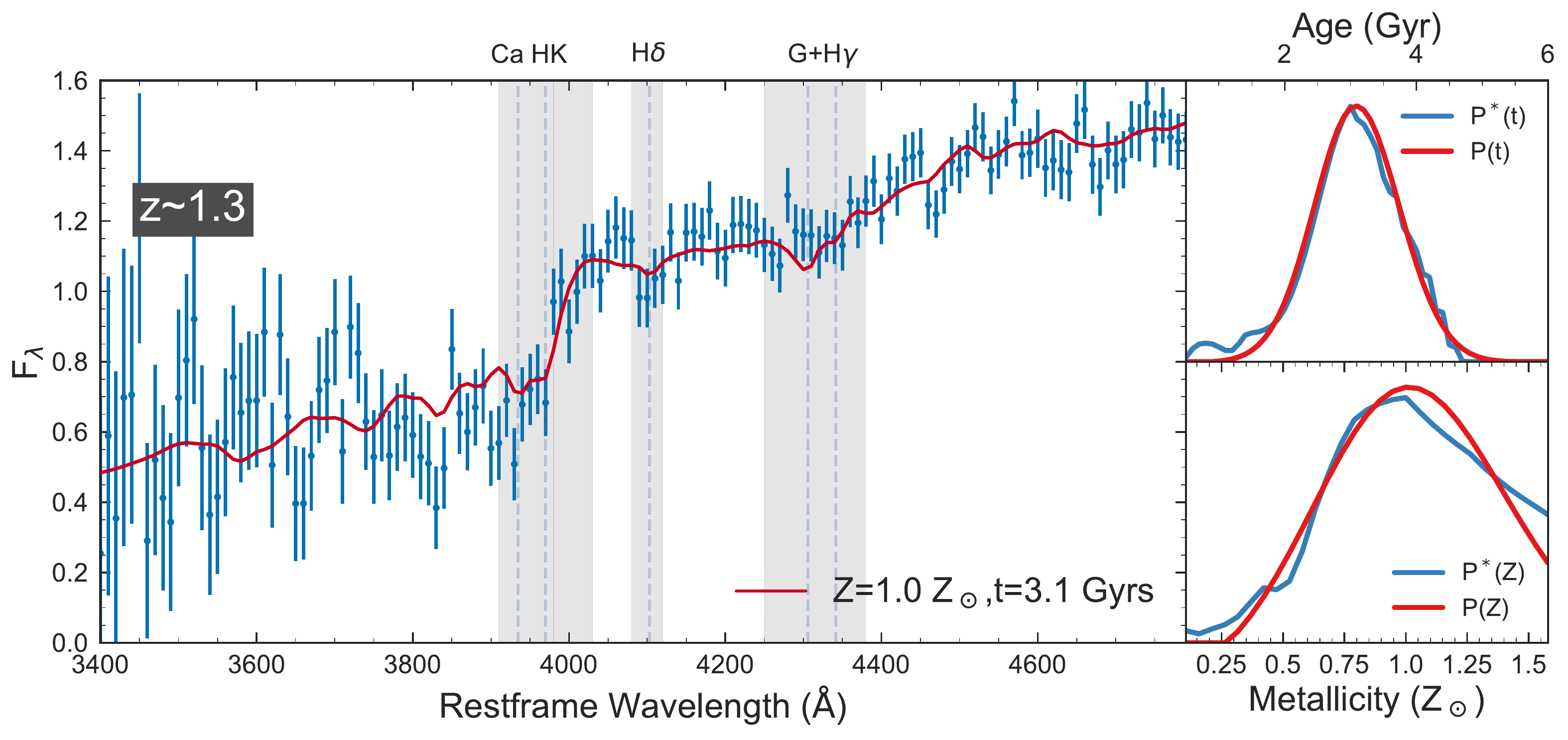}
\plotone{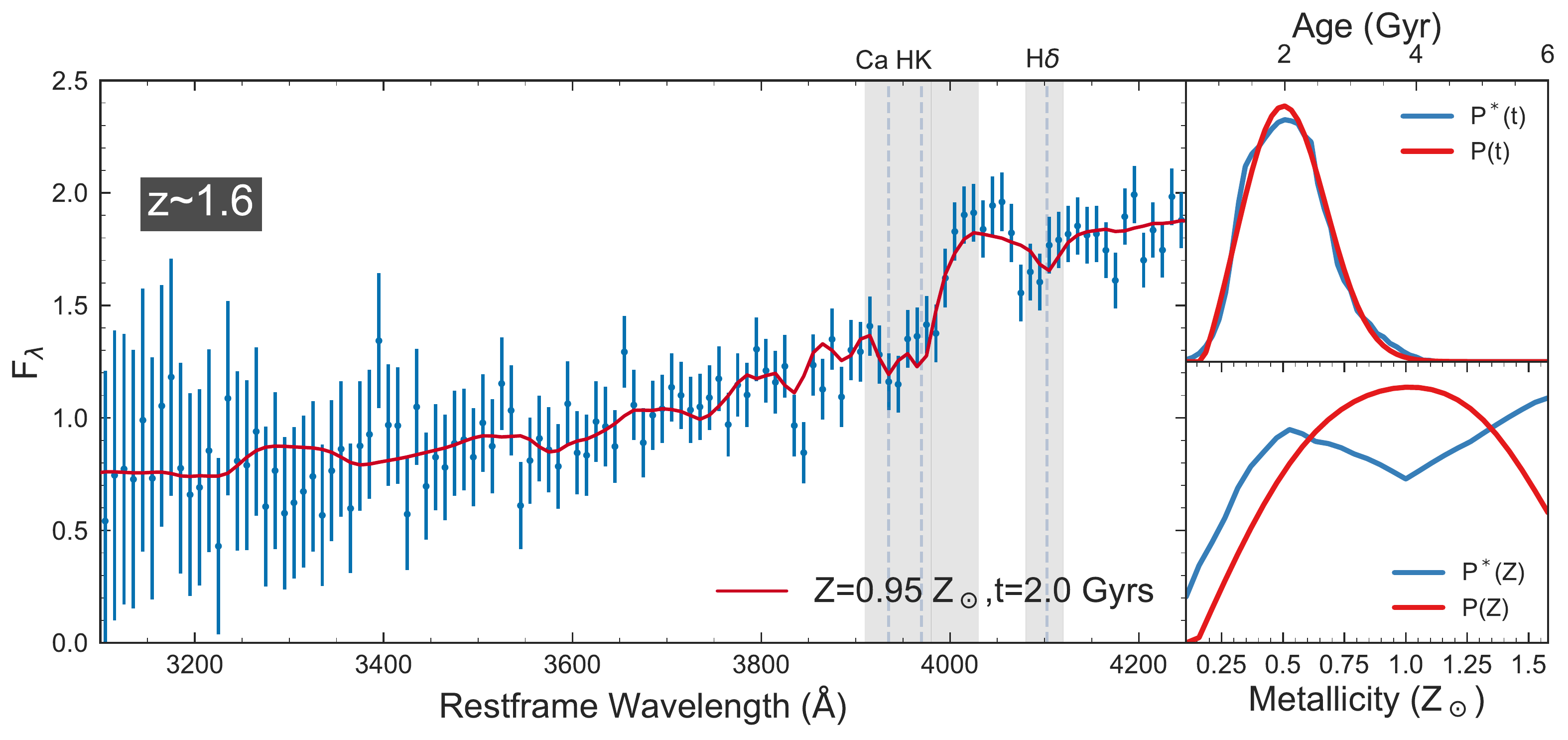}
\caption{Same as Figure~\ref{lowz} for galaxies in the \zthree\ (top)
and \zsix\ (bottom) galaxy subgroups. \label{hiz}}
\end{figure*}

Figures \ref{lowz} and \ref{hiz} show the stacked spectra and models
with the median parameters from the stacked posteriors for the
galaxies in each of our redshift subgroups.  The stacked spectra  are
the weighted average of the spectra for each galaxy in each subgroup:
we first shift these all to the rest-frame, then average, weighting
by the inverse variance as a function of wavelength for each
spectrum. Each figure shows the ``median'' model, which is the
stellar population with parameters equal to the median \editone{light-weighted} age and
metallicity from the stacked parameter
posteriors (i.e., these are not best fits). Nevertheless, the
agreement between the model spectra and the data is high.  This gives us
confidence that the models reliably represent the data and therefore
inform us about the stellar population parameters for these
galaxies.

From the stacked posteriors we derive median and 68\%-tile ranges on the
light weighted ages and metallicities 
for the galaxies in the different redshift subgroups. Based on our tests 
(in Appendix~\ref{stacking}) we interpret the 68\%-tile distributions as an estimate of the intrinsic scatter in parent distribution of the population.  Table
\ref{STKTBL} lists these values as well as the mass range and number
of galaxies for each redshift subgroup.

\begin{deluxetable}{lccc@{\extracolsep{4pt}}cc}
\tabletypesize{\footnotesize}
\tablecaption{Measured parameters for individual galaxies from this work \label{optable}}
\tablehead{
\colhead{ID} &
\colhead{$z_{grism}$} &
\colhead{Metallicity} &
\colhead{Age} &
\colhead{$\tau$} &
\colhead{A(V)} \\
\colhead{} & 
\colhead{} & 
\colhead{(\zsol)} & 
\colhead{(Gyr)} & 
\colhead{(Gyr)} & 
\colhead{(Mag)}\\
\colhead{(1)} & 
\colhead{(2)} & 
\colhead{(3)} & 
\colhead{(4)} & 
\colhead{(5)} & 
\colhead{(6)} }
\startdata
\hline
GND16758 & $1.015_{-0.002}^{+0.002}$ & $1.23_{-0.32}^{+0.25}$ & $3.56_{-0.50}^{+0.76}$ & $0.31_{-0.21}^{+0.24}$ & $0.06_{-0.05}^{+0.08}$ \\
GSD39241 & $1.017_{-0.002}^{+0.002}$ & $1.24_{-0.27}^{+0.23}$ & $3.74_{-0.55}^{+0.75}$ & $0.27_{-0.18}^{+0.22}$ & $0.17_{-0.09}^{+0.11}$ \\
GSD42221 & $1.018_{-0.005}^{+0.005}$ & $1.20_{-0.50}^{+0.27}$ & $1.24_{-0.31}^{+0.85}$ & $0.26_{-0.20}^{+0.67}$ & $0.20_{-0.12}^{+0.18}$ \\
GSD43615 & $1.020_{-0.005}^{+0.002}$ & $1.12_{-0.38}^{+0.31}$ & $3.86_{-0.78}^{+0.90}$ & $0.27_{-0.19}^{+0.24}$ & $0.48_{-0.16}^{+0.20}$ \\
GSD39170 & $1.023_{-0.002}^{+0.002}$ & $1.40_{-0.24}^{+0.13}$ & $3.42_{-0.34}^{+0.31}$ & $0.48_{-0.28}^{+0.14}$ & $0.04_{-0.03}^{+0.04}$ \\
GND37955 & $1.027_{-0.002}^{+0.005}$ & $1.24_{-0.54}^{+0.25}$ & $2.67_{-0.36}^{+0.45}$ & $0.68_{-0.33}^{+0.22}$ & $0.63_{-0.25}^{+0.23}$ \\
GSD45972 & $1.041_{-0.002}^{+0.002}$ & $0.73_{-0.23}^{+0.42}$ & $3.48_{-0.77}^{+1.05}$ & $0.27_{-0.18}^{+0.25}$ & $0.08_{-0.06}^{+0.09}$ \\
GSD39631 & $1.057_{-0.010}^{+0.002}$ & $0.64_{-0.29}^{+0.50}$ & $3.56_{-0.90}^{+1.00}$ & $0.25_{-0.18}^{+0.24}$ & $0.19_{-0.11}^{+0.13}$ \\
GSD47677 & $1.117_{-0.005}^{+0.002}$ & $1.05_{-0.47}^{+0.38}$ & $3.11_{-0.67}^{+0.90}$ & $0.30_{-0.21}^{+0.26}$ & $0.30_{-0.16}^{+0.22}$ \\
GND23435 & $1.139_{-0.007}^{+0.005}$ & $0.73_{-0.43}^{+0.57}$ & $2.23_{-0.73}^{+0.65}$ & $0.34_{-0.24}^{+0.34}$ & $0.38_{-0.20}^{+0.27}$ \\
GND34694 & $1.145_{-0.002}^{+0.002}$ & $1.12_{-0.39}^{+0.29}$ & $2.21_{-0.47}^{+0.71}$ & $0.18_{-0.12}^{+0.21}$ & $0.25_{-0.13}^{+0.14}$ \\
GND32566 & $1.148_{-0.002}^{+0.005}$ & $1.15_{-0.48}^{+0.31}$ & $2.11_{-0.63}^{+0.39}$ & $0.54_{-0.35}^{+0.31}$ & $0.36_{-0.19}^{+0.24}$ \\
\hline
GND23758 & $1.161_{-0.002}^{+0.005}$ & $0.60_{-0.45}^{+0.61}$ & $2.10_{-0.25}^{+0.28}$ & $1.02_{-0.24}^{+0.21}$ & $0.90_{-0.14}^{+0.07}$ \\
GSD38785 & $1.169_{-0.002}^{+0.005}$ & $0.80_{-0.53}^{+0.50}$ & $2.47_{-0.56}^{+0.63}$ & $0.43_{-0.29}^{+0.28}$ & $0.29_{-0.16}^{+0.22}$ \\
GND17070 & $1.175_{-0.014}^{+0.005}$ & $1.09_{-0.43}^{+0.34}$ & $2.10_{-0.67}^{+0.50}$ & $0.34_{-0.25}^{+0.33}$ & $0.15_{-0.10}^{+0.14}$ \\
GSD40476 & $1.209_{-0.002}^{+0.002}$ & $0.51_{-0.27}^{+0.45}$ & $1.83_{-0.57}^{+0.68}$ & $0.22_{-0.15}^{+0.28}$ & $0.44_{-0.19}^{+0.24}$ \\
GSD40597 & $1.221_{-0.002}^{+0.002}$ & $1.22_{-0.40}^{+0.25}$ & $2.49_{-0.34}^{+0.21}$ & $0.63_{-0.25}^{+0.16}$ & $0.20_{-0.12}^{+0.14}$ \\
GSD35774 & $1.227_{-0.005}^{+0.002}$ & $1.28_{-0.27}^{+0.21}$ & $1.65_{-0.34}^{+0.52}$ & $0.12_{-0.08}^{+0.11}$ & $0.57_{-0.24}^{+0.26}$ \\
GSD39805 & $1.243_{-0.019}^{+0.051}$ & $1.06_{-0.59}^{+0.38}$ & $2.10_{-0.47}^{+0.36}$ & $0.59_{-0.31}^{+0.27}$ & $0.40_{-0.22}^{+0.29}$ \\
GND21156 & $1.249_{-0.002}^{+0.002}$ & $0.83_{-0.34}^{+0.36}$ & $2.20_{-0.52}^{+0.58}$ & $0.26_{-0.17}^{+0.22}$ & $0.31_{-0.14}^{+0.16}$ \\
GND37686 & $1.256_{-0.002}^{+0.001}$ & $0.58_{-0.24}^{+0.43}$ & $3.10_{-0.63}^{+0.78}$ & $0.23_{-0.16}^{+0.24}$ & $0.49_{-0.18}^{+0.20}$ \\
\hline
GSD46066 & $1.326_{-0.002}^{+0.012}$ & $1.04_{-0.28}^{+0.31}$ & $3.48_{-0.58}^{+0.57}$ & $0.18_{-0.12}^{+0.17}$ & $0.44_{-0.22}^{+0.28}$ \\
GSD40862 & $1.328_{-0.005}^{+0.002}$ & $1.15_{-0.45}^{+0.30}$ & $2.52_{-0.69}^{+0.72}$ & $0.19_{-0.13}^{+0.19}$ & $0.45_{-0.23}^{+0.27}$ \\
GSD39804 & $1.333_{-0.001}^{+0.005}$ & $0.94_{-0.28}^{+0.37}$ & $3.17_{-0.54}^{+0.67}$ & $0.21_{-0.15}^{+0.20}$ & $0.28_{-0.16}^{+0.21}$ \\
GSD44620 & $1.334_{-0.007}^{+0.005}$ & $0.80_{-0.48}^{+0.54}$ & $0.96_{-0.26}^{+0.53}$ & $0.22_{-0.14}^{+0.33}$ & $0.62_{-0.28}^{+0.25}$ \\
GSD40623 & $1.413_{-0.005}^{+0.019}$ & $1.02_{-0.23}^{+0.29}$ & $3.08_{-0.50}^{+0.56}$ & $0.16_{-0.11}^{+0.16}$ & $0.27_{-0.15}^{+0.19}$ \\
\hline
GND21427 & $1.506_{-0.022}^{+0.072}$ & $0.91_{-0.46}^{+0.44}$ & $2.43_{-0.60}^{+0.64}$ & $0.25_{-0.18}^{+0.24}$ & $0.40_{-0.21}^{+0.27}$ \\
GSD40223 & $1.595_{-0.005}^{+0.005}$ & $0.81_{-0.42}^{+0.49}$ & $1.82_{-0.48}^{+0.44}$ & $0.47_{-0.22}^{+0.22}$ & $0.48_{-0.24}^{+0.29}$ \\
GSD41520 & $1.605_{-0.002}^{+0.002}$ & $0.97_{-0.49}^{+0.46}$ & $1.98_{-0.55}^{+0.66}$ & $0.18_{-0.12}^{+0.18}$ & $0.73_{-0.20}^{+0.17}$ \\
GSD39012 & $1.612_{-0.024}^{+0.010}$ & $0.75_{-0.43}^{+0.55}$ & $1.87_{-0.53}^{+0.58}$ & $0.45_{-0.27}^{+0.27}$ & $0.42_{-0.23}^{+0.30}$ \\
GSD44042 & $1.612_{-0.002}^{+0.002}$ & $1.28_{-0.85}^{+0.22}$ & $2.33_{-0.28}^{+0.33}$ & $0.47_{-0.20}^{+0.11}$ & $0.22_{-0.13}^{+0.18}$ \\
\enddata
\tablecomments{ (1) Galaxy ID number in the GND or GSD \threedhst\ catalog; (2) measured grism redshift; 
(3) median metallicity; (4) median \editone{light-weighted} age; (5) median e-folding time for delayed $\tau$ SFH; 
(6) median A(V) for Calzetti dust law; all measurements provided with 68\% confidence intervals; 
horizontal lines show the galaxies in each of the separate redshift subgroups}
\end{deluxetable}



\begin{deluxetable}{ccccc}
\small
\tablecaption{Measured parameters for each redshift subgroup \label{STKTBL}}
\tablehead{
\colhead{Redshift group} &
\colhead{Metallicity} &
\colhead{Age} &
\colhead{Mass Range} &
\colhead{Sample Size} \\
\colhead{} &
\colhead{(\zsol)} &
\colhead{(Gyr)} &
\colhead{(log(M$_*$/M$_\odot$))} &
\colhead{(N)} \\
\colhead{(1)} & 
\colhead{(2)} & 
\colhead{(3)} &
\colhead{(4)} &
\colhead{(5)}}
\startdata
\hline
\zone &$1.16_{-0.33}^{+0.26}$ & $3.2 \pm 0.7$ & 10.1--11.1  & 12\\
\ztwo &$1.05_{-0.36}^{+0.32}$&$2.2 \pm 0.6$   & 10.6--11.2 & 9\\
\zthree &$1.00_{-0.31}^{+0.31}$&$3.1 \pm 0.6$ & 10.5--11.0 & 5 \\
\zsix &$0.95_{-0.40}^{+0.37}$&$2.0 \pm 0.6$  & 10.7--11.1 & 5\\
\enddata
\tablecomments{Values on (light-weighted) age and metallicity correspond to the median and
  68\% confidence range for each parameter, which we interpret as an
  estimate of the intrinsic scatter in the subgroup (see text);   The
  columns show (1) Redshift of the sub-group of the stacked posteriors; (2)
metallicity; (3) \editone{light-weighted} age, (4) Mass range of the galaxies in each
of the samples; (5) number of galaxies in each redshift sub-group.} 
\end{deluxetable}

\section{Discussion}\label{section:discussion}

Our dataset and analysis  allow us to explore the correlations between
stellar mass, \editone{(light-weighted)} age, and metallicity in
galaxies at $z > 1$, much closer in time to the galaxies' quenching
time.  We can use these to constrain both the SFHs
and enrichment histories of these galaxies.    In the sections that
follow, we interpret the results of our modeling.  

It is important to note that \editone{throughout (unless otherwise
specified) all ages correspond to ``light weighted'' ages (see
above) as these are better  constrained because they correspond to the
stellar populations that dominate the light in the galaxies' spectra
(when weighted by their luminosity).}  These are related to the
parametric form of the SFHs through
Equation~\ref{avgage}, and different choices in this can impact the
(light-weighted or mass-weighted) ages by $\sim$0.4~Gyr
\citep[e.g.,][]{carn17}.   We will explore the effects of different
SFHs in a future work.

\subsection{The Ages and Metallicities of Quiescent Galaxy Populations
  from $z\sim 1.1$ to $1.6$} 

Generally, our modeling of the quiescent galaxies from $\zone$ to
$\zsix$  favor (\editone{light-weighted}) ages of $\simeq$2--4 Gyr
(with  age increasing with decreasing redshift)  and near Solar
metallicities. Here, we comment on the values and quality of the model
fits.  

For each redshift subgroup, Figures \ref{lowz} and \ref{hiz} show 
the stacked posterior on the age and
metallicity.  The model fits in the Figures have
metallicity and age equal to the median value from
these stacked posteriors (that is, these are \textit{not} best fit
models to the stack).  Rather, the posteriors are derived by stacking
the posteriors of the individual galaxies in each redshift subgroup
using the method in Section~\ref{section:fit_data}, and the data are the stacks of
the individual galaxies (weighted by their inverse variance).  The agreement between the models and data match is qualitatively quite good.


The \zone\ subgroup has median parameter values of
$Z=1.16 \pm 0.29$~\zsol, $t_L = 3.2 \pm 0.7$~Gyr (Figure~\ref{lowz} 
top panel).  The regions
around \hb\ and Mg$b$ are particularly well reproduced by the model.
This may be expected as the S/N of the data are highest in those
regions and these features are age and
metallicity sensitive.  The model agrees well with the data
at bluer wavelengths, but it does not capture all the apparent
features. For example, the data show possible absorption at H$\delta$
that is not reproduced in the model.  However, the G102 grism has
less sensitivity at these wavelengths, reflected by the larger
uncertainties on the data in this part of the spectrum, and the
spectral regions at bluer wavelengths constrain the models less.
Furthermore, we would have expected stronger H$\delta$ absorption to be
accompanied by stronger H$\gamma$ absorption, which is not seen in
the data (providing additional constraints on the models).  

The \ztwo\ sample has median parameter values of
$Z=1.05 \pm 0.34$~\zsol,  $t_L= 2.2 \pm 0.6$~Gyr (Figure~\ref{lowz} bottom panel).  Qualitatively,
we see excellent agreement between the model and data, especially in
the regions from the 4000~\AA/Balmer break out past \hb, and these
features drive the constraints on the model parameters. 

The \zthree\ sample has median parameter values of $Z=1.00 \pm 0.31$~\zsol\ and
$t_L = 3.1 \pm 0.6$~Gyr (Figure~\ref{hiz} top panel).   
The 68\%-tile confidence intervals on \editone{light-weighted} age and
metallicity are as tight as those at \zone\ and \ztwo\ samples (see
Table \ref{STKTBL}), implying the shape of the continuum
has important constraining power even as  important features (like
H$\beta$ and Mg$b$) shift out of the wavelength coverage.    Most of
the information constraining the models comes from the wavelength
region of the 4000~\AA\ break to the rest-frame $B$-band, where the
S/N is highest. 

The \zsix\ sample has median parameter values of $Z=0.95 \pm 0.39$~\zsol,
and $t_L = 2.0 \pm 0.6$~Gyr (Figure~\ref{hiz} bottom panel).   
The 68\%-tile range on the \editone{light-weighted} age
is nearly as tight as for the lower redshift subgroups. The metallicity
measurements place this group as slightly sub-solar, with a fairly broad
68\%-tile range.

\subsection{On the Star-Formation and Quenching Histories of Quiescent
  Galaxies at $z > 1$,\label{approx}}

\begin{figure*}[t]
\epsscale{1.}
\plotone{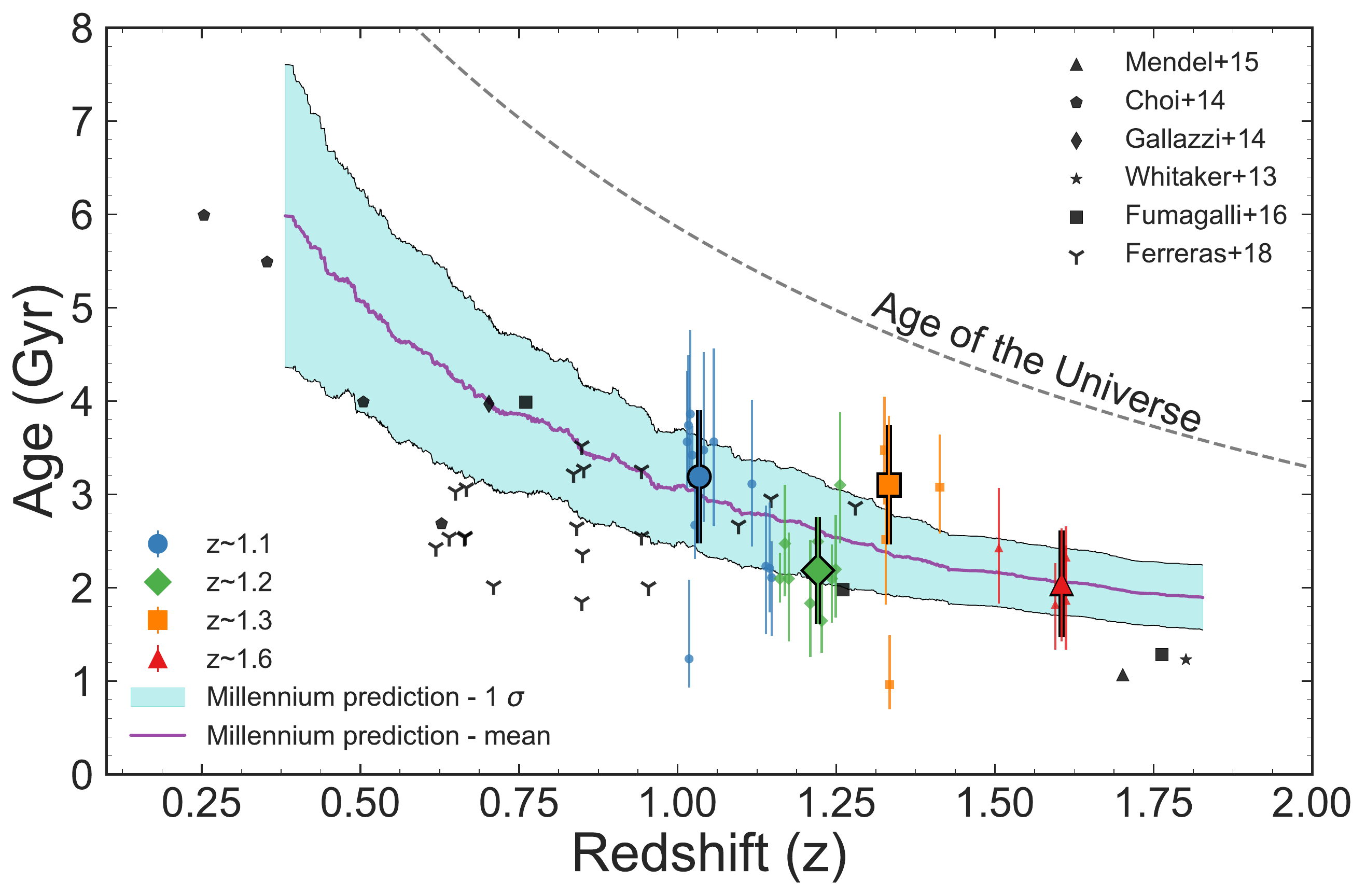}
\caption{The evolution of light-weighted age as a function
of redshift.  The small, colored data points show results for the 
individual $1 < z_{grism} <1.8$ galaxies in our sample.   
Large colored data points correspond to median
values derived from the stacked posteriors for redshift subgroup as labeled.
Error bars show 68\% confidence intervals.  Other small (black) data points
correspond to results from \cite{mend15, choi14,  gall14, whit13, 
fuma16,ferr18}. Generally, quiescent galaxies have younger stellar 
populations at higher redshifts, where their \editone{light-weighted} 
age has nearly a constant offset from the age of the universe.
This agrees with predictions from the Millennium simulation
\citep{henr15}, where the shaded band shows the median and 68\%-tile 
scatter in light-weighted ages of quiescent galaxies in their predictions.
\label{avzfg}}
\end{figure*} 

\begin{figure*}[t]
\epsscale{1.}
\plotone{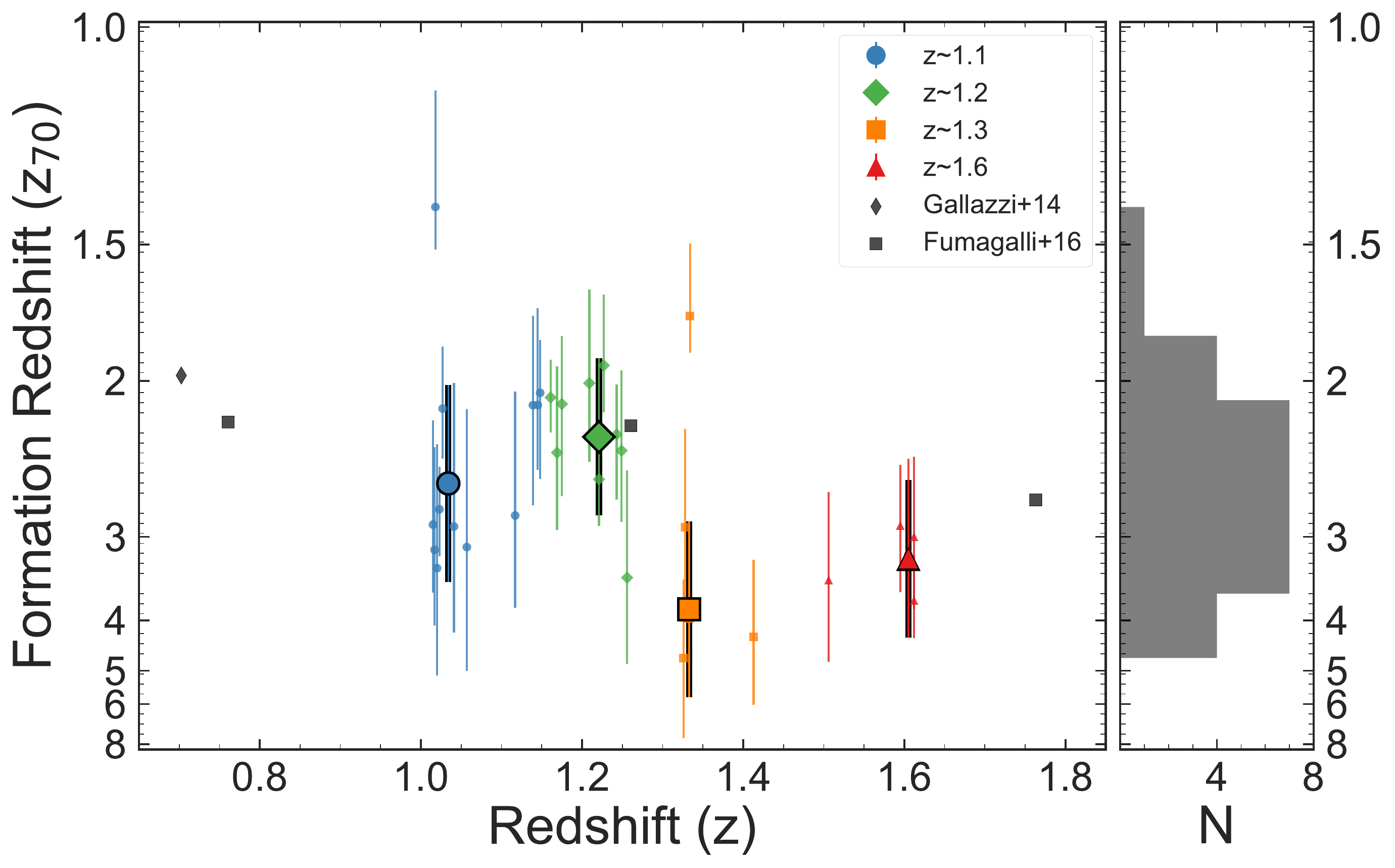}
\caption{Distribution of galaxy formation redshifts. The Left panel shows the
derived formation redshifts, $z_{70}$ of quiescent galaxies as a
function of observed redshift.  The Right panel shows the distribution
of median formation redshifts for our samples.   The formation
redshift, $z_\mathrm{form}$,  corresponds to the redshift where the
galaxies had formed more than 68\% of their stellar mass (see Section
\ref{approx}). Here, we only include age measurements (shown in black)
from studies which measured  light-weighted ages. The quiescent
galaxies at $1 < z_{grism} < 1.8$  in our sample have formation
redshifts $z_\mathrm{form} > 2-3$ nearly independent of observed
galaxy redshift. \label{zform}}
\end{figure*} 

Figure~\ref{avzfg} shows the measured \editone{(light-weighted)} ages
as a function of redshift for the galaxies in our sample.  The Figure
shows the median values and 68$\%$ confidence intervals on the  ages
for each quiescent galaxy in our sample,  as well as the ages derived
from the stacks for each redshift subgroup.   Generally, galaxies in
our sample have  ages $2-4$~Gyr, where there is a trend that  higher
redshift galaxies have younger ages.  
\editone{In Figure~\ref{avzfg} we also compare the (light-weighted)
ages for the galaxies in our study to some other studies in the
literature that also report ages.  Comparing
our derived age measurements to many other  studies is complicated by
the fact that there are different definitions and conventions of
``age''.   There are three widely used definitions of age:  the
instantaneous age of the stellar population model; mass-weighted ages,
and light (luminosity-)weighted ages (we use the latter here).  For a
given (declining) SFH , the instantaneous age is the
oldest as it measures the time since the onset of star-formation.
Mass--weighted and light-weighted ages will be younger than
instantaneous ages as they average over the mass (or light) of stars that
continue to form after the initial burst of star-formation.   As
discussed above, the light-weighted ages are more robust against
uncertainties in the SFH.   For this
reason we only show  other results for light-weighted ages in
Figure \ref{zform} as
they are most directly comparable to the ones here.}   

The ages we derive for the galaxies in our  sample tend to be larger 
than some other values for
galaxies at similar redshift taken from the literature
\citep[cf.][]{whit13,mend15,fuma16}.   This may result from systematic
effects owing to different definitions of  age in previous studies,
and the fact that we treat the metallicity as a free parameter.  As
seen in the right panel of  Figure \ref{galspec}, the
metallicity--age degeneracy shows that lower metallicity solutions
push  the median age higher (consistent with the offsets between our
work and most literature studies). For example, \cite{carn17}  allow
the metallicity to be free and find similar (mass-weighted) ages for
quiescent galaxies in the same redshift range as our sample here. 

Interestingly, the \editone{light-weighted} ages we derive are
consistent with predictions of ($r$--band light-weighted) from the
semi-analytic model (SAM) based on the Millennium simulation from
\cite{henr15}.   Figure~\ref{avzfg} shows the 68\% scatter in
ages from galaxies from this SAM selected to
be quiescent (defined by sSFR$ < 10^{-10}$ yr$^{-1}$).  This implies
that the quenching epochs predicted in the models are consistent with
the results we derive here for our sample. (We plan a more detailed
comparison in a future study.)

Figure \ref{avzfg} also shows that the \editone{light-weighted} ages
of the quiescent galaxies in our sample have a nearly constant offset
from the age of the Universe at the observed redshifts of the galaxies
in our sample.  This implies the galaxies all quenched at
approximately the same time in the past.  Figure~\ref{zform}
illustrates this by showing the formation redshift $z_\mathrm{form}$
for each galaxy in our sample (and showing galaxies from the
literature for comparison).   Indeed, the galaxies in our samples show
a near constant formation redshift, $z_\mathrm{form} > 2.5 $. 

We can relate $z_\mathrm{form}$ to the amount of stellar mass the
galaxies had formed at this formation time. This requires integrating the
SFHs, accounting for the light-weighted ages and
mass losses from stellar evolution.   This is not straight-forward:  for
example there is no analytical solution (to our knowledge) to the
integral in Equation~\ref{avgage}, owing to the dependence of 
$L(t^* -
t)$ on \editone{light-weighted} age, metallicity, and SFH.   
However, we can make an
approximation as the ages of the 
quiescent galaxies in our samples are
all in the regime where $t^* \gg \tau$ (where $t^\ast$ is the
instantaneous age of the model and $\tau$ is the $e$-folding time
constant of the SFR).  In this case we compare $\langle t(t^\ast,
\tau) \rangle_L$ (the light-weighted age) to $\langle t(t^\ast, \tau)
\rangle$ (i.e., the mass-weighted age), which we define as
\begin{equation}\label{eqn:aveage_nolum}
\langle t(t^\ast, \tau) \rangle = \frac {\int _0 ^{t^*}
  \Psi(t,\tau)\  (t^*-t)\ dt} {\int _0 ^{t^*} \Psi(t,\tau)\ dt}
\end{equation}
For the case where $t^* > \tau$ it will always be the case that 
\begin{equation}
\langle t(t^\ast, \tau) \rangle_L < \langle t(t^\ast, \tau) \rangle. 
\end{equation}
For the case $t^* \gg \tau$,
it can be shown that for delayed-$\tau$ SFHs,
equation~\ref{eqn:aveage_nolum} reduces to
\begin{equation}
\langle t(t^\ast \gg \tau, \tau) \rangle \approx t^* - 2 \tau.
\end{equation}
That is, the light-weighted age is always less than the (mass-weighted) age because for our galaxy sample the stellar populations fade monotonically.   
This leads to an upper bound on the
light-weighted age, 
$\langle t(t^\ast, \tau) \rangle_L$, 
\begin{equation}
\langle t(t^\ast, \tau) \rangle_L < t^\ast - 2\tau.
\label{eqn:approx}
\end{equation}
Empirically, we find that Equation~\ref{eqn:approx} holds for $\langle
t(t^\ast, \tau) \rangle_L$ $>$ 1.5~Gyr,  and $t^\ast \geq 4\tau$. 
For younger \editone{light-weighted} ages, there is a bias of
$\lesssim$15\% (so long as $t^* \geq 4\tau$ is satisfied).  

\begin{figure*}[t]
\epsscale{1.}
\plotone{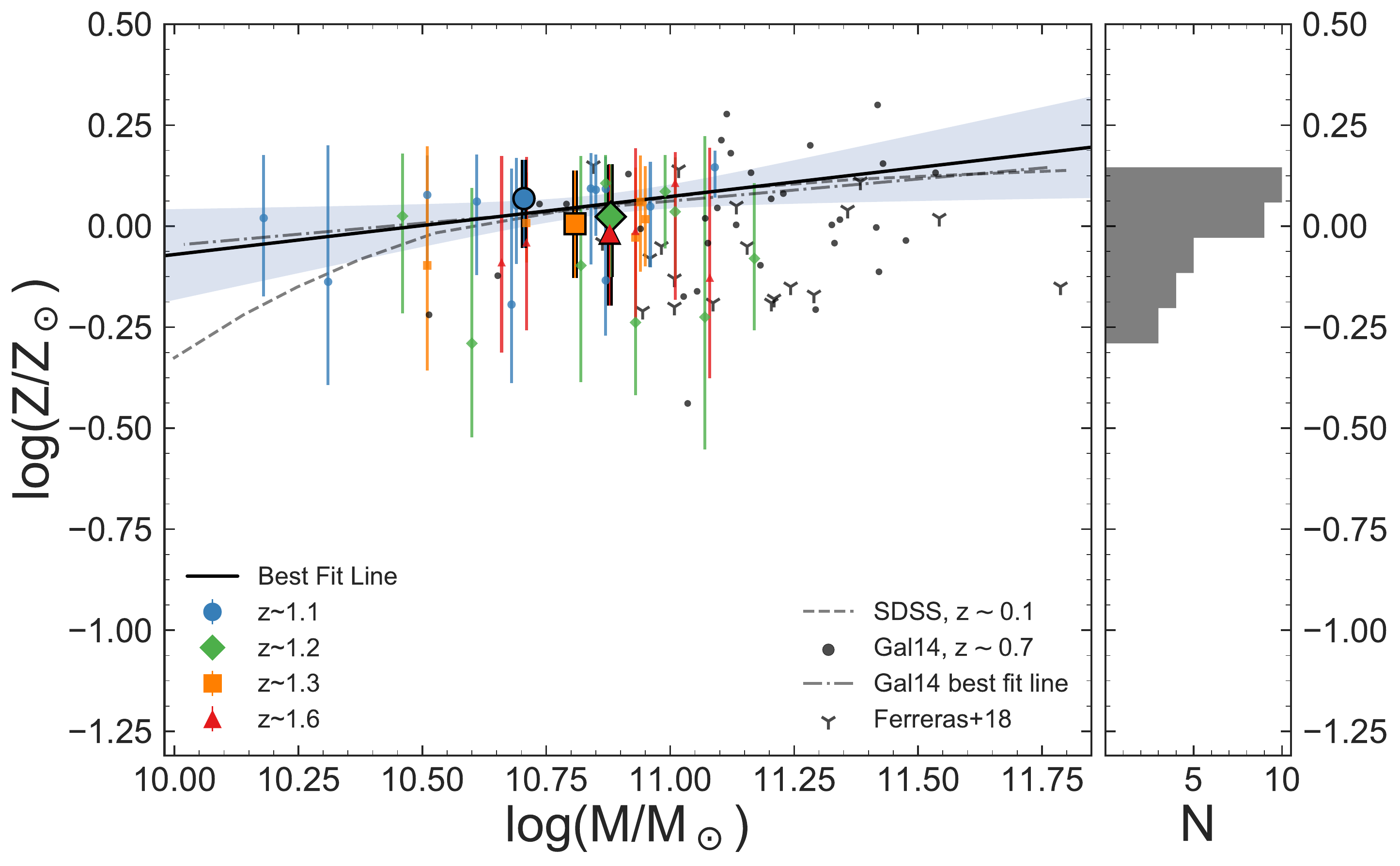}
\caption{Left panel: Mass--metallicity relation for quiescent galaxies
at $1.0 < z_{grism} < 1.7$.   The small, colored data points and error
bars show the median values and 68\%-tile range for the individual
galaxies in our redshift subgroups, as labeled in the figure legend.
The large  colored data points show the metallicities from the stacked
posteriors for each subgroup.  The thick solid line shows a linear fit
to the individual galaxies and the shaded region shows the 68\%--tile
bound.   The dashed and dot--dashed lines show the mass-metallicity
relation for quiescent galaxies from lower redshift samples (from SDSS
at $z < 0.22$, \cite{gall05}; and $z\sim 0.7$ from Gal14). We also
include measurements of individual galaxies from Gal14 and
\cite{ferr18}. Right panel: Histogram of the median  metallicities for
the galaxies in our $1.0 < z_{grism} < 1.7$ sub-groups.  The majority
of the probability density lies around $\approx Z_\odot$. \label{mmfg}}
\end{figure*} 

The stellar--mass formed for a SFH parameterized as
a ``delayed-$\tau$'' model is then 
\begin{equation}
M(t , \tau) = m(t,\tau)\ C\ \int_0^{t}t'\ e^{-t'/\tau}\ dt'
\end{equation}
\noindent where $m(t,\tau)$ is the fractional mass loss function (and
depends on age, SFH and stellar population model).
$C$ is a constant we can solve for by setting $t=t^*$ and $M(t^*,\tau)
= M_{obs}$ (the measured stellar mass). 

Substituting in the value for $C$ leads to
\begin{equation}
M(t\ |\ M_{obs}, t^\ast, \tau) = \frac{m(t,\tau)\ M_{obs}\ \int_0^t\
  t'\ e^{-t'/\tau}\
 dt'}{m(t^*,\tau)\ \tau^2\ (1-e^{-t^*/\tau} [ 1 +t^*/ \tau] )} 
\end{equation}
\noindent Using the approximation for $\langle t(t^\ast, \tau)
\rangle_L  < t^\ast - 2\tau$, we find a lower limit on the stellar
mass formed at $\langle t(t^\ast, \tau)
\rangle_L $ (called the ``quenching mass'', $M_Q$),
\begin{equation}
M_Q >M(t=2\tau) = \frac{ 0.6\ m(2 \tau,\tau)\ M_{obs}}{m(t^*,\tau)\ (
  1-e^{-t^*/\tau} [1+ t^*/\tau])} 
\end{equation}
\noindent Finally, again applying the limit of $t^* \gg \tau$, we can approximate $M_Q$ to be
\begin{equation}
M_Q >0.6\times \frac{m(2 \tau,\tau)}{m(t^*,\tau)} \times  M_{obs}
\end{equation}

\noindent The values of $m$ are determined by the stellar population
model, including the effects of the IMF. For a Salpeter IMF $M_Q > 0.68$ $M_{obs}$, and for a Chabrier IMF  
$M_Q > 0.75$ $M_{obs}$. Therefore, for the assumed SFHs, the galaxies in
our sample would have formed $\gsim$70\% of their stellar mass by the
redshift that corresponds to their light-weighted ages, which we
define as $z_\mathrm{form}$.

Both $z_{70}$ and $M_Q$ are related to the assumed star-formation
history.  To test how different parameterizations for this  affect
our light-weighted ages, we refit our galaxies with both $\tau$ and delayed $\tau$ 
models. The only effect seen on the light-weighted ages was a systematic
shift (delayed $\tau$ models produce lower light-weighted ages).
Using $\tau$-models rather than delayed-$\tau$ models increases the \editone{light-weighted}
ages and shifts the formation redshifts to $z_\mathrm{form} > 4$. 

 Compared to the galaxies in our sample, quiescent galaxies at lower
redshifts have lower formation redshifts \editone{\citep[see Figure~\ref{zform}, and][although a direct comparison is complicated by the different
definitions of ``age'' in some studies]{mend15, choi14,  gall14,
whit13, fuma16,ferr18}.}    This is likely a consequence of progenitor
bias, where the progenitors of quiescent galaxies at $z < 1$ are a mix
of quiescent galaxies and some star-forming galaxies at $z > 1$ (that
quench at later time).  Because some fraction of the population is
star-forming at higher redshift, they will be younger at lower
redshift and shift the mean formation redshift lower.
\editone{Similarly, quiescent galaxies could become ``younger'' if
they accrete enough mass through minor (dry) mergers of (quenched)
lower mass galaxies with younger ages \citep[see discussion
in][]{choi14}. }

Our results imply there should exist a population of massive quiescent
galaxies at redshifts as high as $z > 3$.  Indeed, candidates of such
galaxies have been identified
\citep[e.g.,][]{renz06,guo12,spli14,stra14,merl18},  where recent
spectroscopic confirmation shows \editone{the rest-frame optical
light}  \editone{strong Balmer absorption features (i.e., dominated by
A-type stars), indicative of recent quenching} \citep{glaz17,schr18b}.   

We can gain insight  into the evolution of these objects by comparing
the (comoving) number densities of these massive galaxy candidates at
$z > 2$ to those in our sample.  \editone{ Figure~\ref{zform} shows
that roughly one-third of our sample experienced early quenching, with
$z_{70}$ $>$ 3.  \cite{spli14} measure a number density  when
considering \textit{all} massive galaxies ($\log M_\ast/M_\odot > 10.6$) at
$3< z < 4$ of $5.1 \times 10^{-5}$ Mpc$^{-3}$.   We can  compare this
to number density of quiescent galaxies  in our sample ($1 < z_{grism}
< 1.8$) with  $\log M/M_\odot > 10.85$ (the higher mass here is needed
to account for the $\sim 30 \%$ growth expected as our estimates of
the "formation  redshift", $z_{70}$, correspond to the point where our
galaxies had  formed 70\% of their mass).   Using this mass limit we
find a number density of $1.3 (\pm 0.2) \times 10^{-4}$ Mpc$^{-3}$,
where the uncertainty is purely Poissonian.    Taking one-third of
this number density (to account for the one-third of objects that
quenched at $z_{70} > 3$), this is consistent with the measured number
density from \cite{spli14}, but it requires that $\approx$all of the
massive galaxies at $z > 3$ quench to account for the galaxies with
high $z_{70}$ in our sample.   }

\editone{The rest of the galaxies quench
at lower redshift ($2 < z < 3$, see Figure~\ref{zform}). Integrating
the galaxy stellar mass functions from \cite{tomc14}, we find that
quiescent galaxies with $\log M_\ast/M_\odot > 10.6$ have a number
density of $1.3 \times 10^{-4}$ Mpc$^{-3}$.  These are equal to the
density number we derive for quiescent galaxies in our sample, showing
we can account for all the population of quiescent galaxies we observe at
$1 < z < 1.8$ (not even accounting for uncertainties or from changes
in stellar mass through mergers, but see below). This evidence
supports the $z_{70}$ values we derive. }



\subsection{The Mass--Metallicity Relation for Quiescent Galaxies
  at $z > 1$}

The CLEAR WFC3 G102 data allows us to measure stellar  metallicities
for quiescent galaxies at $z > 1$.   Because our sample is closer
\editone{(in  redshift)} to the epoch when the galaxies quenched, we
can better constrain their enrichment history.  

Figure~\ref{mmfg} shows the mass-metallicity relation for the
quiescent galaxies in our sample at $1 < z_{grism} < 1.7$.  The figure
shows evidence that (1) quiescent galaxies at this redshift  have near
Solar Metallicities, and (2) that a mass--metallicity relationship for
these quiescent galaxies is nearly unchanged from the $z\sim 0.1$ to
$z > 1$. 

The mass-metallicity relation for $z\sim 1-1.7$ quiescent galaxies is
either flat (no dependence) with stellar mass or slightly increasing
with stellar mass. We fit a linear relation to the data of the form,
%
%
\begin{equation}
\log( Z_* / Z_\odot) = A \log (M_* / 10^{11}\ M_\odot) + B
\end{equation}
for all the $1 < z_{grism} < 1.7$ galaxies in our sample.  This
yielded a fit with $A=0.15 \pm 0.13$, $B=0.08\pm 0.03$, with a
covariance, $\sigma_{AB}=0.002$, which is illustrated as the shaded
area in the figure. 

The zero-point and shallow slope of the mass-metallicity relation show
that the galaxies in our sample are consistent with Solar metallicity.
Furthermore,  Figure~\ref{mmfg} shows that the stacked posteriors of
galaxies in each redshift subgroup  are strongly clustered around
Solar metallicities, \editone{and that the distribution (histogram) of median
metallicities for all the galaxies in our sample are likewise peaked
around Solar metallicities.}  Therefore, we interpret this as strong
evidence for Solar-metallicity enrichment in quiescent galaxies at
$z\sim 1-1.7$.

In contrast, there is no evidence for evolution in the slope and
zeropoint of the mass-metallicity relation for quiescent galaxies from
$1 \lsim z \lsim 1.7$ to $z\sim 0.2$.    Figure~\ref{mmfg} shows that
the measurements for galaxies at $z < 0.22$ from SDSS \citep{gall05}
and at $z\sim 0.7$ (Gal14) are consistent with the same
mass-metallicity relation we infer at $z > 1$. \editone{\citep[Similar
results are found by][]{lono15,onod15,ferr18}.}  This is consistent
with the assertion that these galaxies enriched rapidly while they
formed their stars up to Solar metallicities, they have evolved
passively since.  

\editone{At higher redshift, $z > 2$, there is some evidence that
massive ($\log(M_*/M_\odot) > 11.0$) quiescent galaxies include
objects with $\approx$Solar (and super-Solar) metallicities, as well
as objects with less (sub-Solar metallicity) enrichment
\citep{krie16,toft17,mori18}, with (in at least one case) evidence for
super-Solar [$\alpha$/Fe] enrichment \citep[e.g.,][]{krie16}.    This
could imply that at these epochs we are seeing the quenching of the
cores of galaxies, which may evolve to higher metallicity at later
times through mergers (see below, and discussion in Kriek et al.\
2016).    However, current constraints at $z > 2$ are  based on only a
few (four) galaxies (highlighting the difficulty of these
measurements) and larger galaxy samples are required.} 

\editone{Similarly interesting would be to study the evolution of the
metallicity ([Fe/H])  and the $\alpha$-element enrichment
([$\alpha$/Fe]) in our galaxies.   This is clearly an important
problem given that massive quiescent galaxies at low redshifts show
evidence for [$\alpha$/Fe] $>$ 0 \citep[e.g.,][]{thom10,conr14}, with
some recent evidence at high redshift
\citep[e.g.,][]{krie16,stei16}.  This is likely related to their short
SFHs combined with delay times in of  Type Ia
supernovae \citep[e.g.,][]{dahl12,frie18}. Given that the $z\sim
1-1.7$ quiescent galaxies in our sample likely have similar histories
to those at $z\sim 0$ (see discussion below), we may expect them to be
similarly $\alpha$-enriched.   We plan to study this in a future
work. }
  
\subsection{Implications for Enrichment and Quenching of $z > 1$
Quiescent Galaxies}

Star-forming galaxies show a (gas-phase) mass-metallicity relation
that evolves strongly with redshift.  This contrasts strongly with the
(lack of) evolution in the mass-metallicity relation for quiescent
galaxies we find out to $1.0 < z < 1.7$.  For example, the
mass-metallicity relation for galaxies from SDSS at $z\sim 0.1$
derived from  nebular emission lines shows that the gas-phase
metallicity rises quickly with  mass, and is Solar (or exceeds Solar)
for the most massive galaxies  \citep[with stellar masses $\log M_\ast
/ M_\odot > 10.5$,][]{trem04,andr13}.  This evolves with redshift: at
$z\sim  2.3$ the metallicities of star-forming galaxies at
$\log(M_\ast /  M_\odot)$=10.5 are lower by about a factor of $\sim$2
compared to $z\sim 0.1$  \citep[although this depends slightly on
systematics owing to different  metallicity
indicators]{tada13,zahi14,sand17}, and by about a factor of $\sim$3 at
$z\sim 3.3$ \citep{onod16}.   This evolution persists even for
absorption-line  studies, where Gal14 report a similar offset in the
\editone{stellar metallicity} for  star-forming galaxies at $z\sim
0.7$ at fixed stellar mass (in contrast to their  mass-metallicity
relation for quiescent galaxies which shows no evolution to  $z\sim
0.7$). 

What does it mean that the metallicities of quiescent galaxies at $1 <
z < 1.7$ all favor near Solar metallicities?  The lack of evolution in
their mass-metallicity relation means that the progenitors of the
galaxies in our sample presumably also had Solar metallicities at the
time of quenching, $z_{70} \gtrsim 2.$. However, this makes their
star-forming progenitors strong outliers in metallicity, offset from
the  gas--phase mass-metallicity relation observed at $z\sim 2.3$ and
$z\sim 3.3$ for galaxies with stellar mass $\log M_\ast/M_\odot >
10.5$  \citep[see above]{onod16,sand17}.    Observationally,
there are few (if any) star-forming galaxies at these masses and
redshifts with Solar metallicity that are candidates for the
progenitors of the quiescent galaxies  in our sample (this is assuming
metallicity indicators used for quiescent and  star forming galaxies
are consistent, which is unclear).  

%

Insight may be garnered from theory.  In simulations of galaxies at
these redshifts, the slope and scatter of the gas--phase
mass-metallicity relation is a result of the combination of gas
accretion, gas fraction, feedback, and metal-retention efficiency
\citep{oppe12,torr17}.   The slope of the gas-phase mass-metallicity
relation is a result of mass-dependent gas accretion and feedback,
resulting in lower metal retention efficiency for lower mass
galaxies. This reproduces the observed evolution of the gas-phase
mass-metallicity relation \citep[e.g.,][]{torr17}. 

The simulations show that at $z\sim 2$ the mean metallicity of massive
galaxies is sub-Solar.  However, the distribution extends to higher
values and some galaxies have roughly Solar metallicity
\citep{torr17}.  One explanation is that gas accretion has ceased in
these galaxies (or slowed down, possibly as the galaxies are in the
act of quenching, see below), and/or their metal retention
efficiencies are higher (related to the fact that their escape
velocities are larger), making it easier for them to enrich to higher
metallicities faster.  Interestingly, the lack of evolution in the
observed mass-metallicity relation for quiescent galaxies out to $z >
1$ could mean that these processes of galaxy quenching are redshift
independent, which would set a requirement on simulations. 

Why is it then that quiescent galaxies observed at nearly all
redshifts have Solar metallicity?    It could be that obtaining Solar
metallicity is a symptom of the  processes that lead to quenching and
be related to the observational  differences between quiescent and
star-forming galaxies.    One way to correlate the build-up of metals
to quenching is by connecting the metal retention  efficiency to
galaxy feedback and galaxy size \citep{mura15,chri16}.    Quiescent
galaxies are known to have smaller effective  radii compared to
mass-matched star-forming galaxies \citep{vanw11}, and this
observation has led to theories such as the process of
``compactification'', which relates sizes sizes to quenching
\citep[e.g.,][]{deke09,barr13}.  Such scenarios  could also lead to a
natural connection between the high Solar metallicity and quenching in
galaxies:  as galaxies become more compact, the increase in  their
mass- and SFR-surface densities leads to stronger feedback, likely
pushing the galaxy beyond the quenching point \citep{thom05a,barr13};
the increase in their escape velocities and SFR leads to higher metal
production and retention \citep{elli08,torr17}.   This is a plausible
explanation for our observations, but  it remains to be seen if
simulations can reproduce the detailed values and  evolution.  For
example, one prediction if the metal enrichment occurs prior  to
quenching is that more compact star-forming galaxies should likewise
have higher metallicity \citep[consistent with some results for
galaxies in  the SDSS, see][]{elli08}.  We plan to test this at higher
redshift in a future  study. 

Regardless, if the Solar metallicities in our sample are correct, then their 
progenitors must exist \textit{somewhere}.  Some may be related to the compact 
star-forming galaxies discussed above \citep{barr13}.  \editone{It is also possible 
that the time for galaxies to enrich from $\sim$1/3--1/2 $Z_\odot$ to 
$Z_\odot$ is very short-lived, in which case they may be very rare in
surveys.    For example, if quenching is tied to the process of 
``compaction'', the timescales could be $\ll 1$~Gyr \citep{zolo15}, such that 
finding metal-rich star-forming galaxies in this stage is rare.  This
``rapid'' phase would also be consistent with high [$\alpha$/Fe]
abundances \citep[e.g.,][and we plan to test for this in future
work]{krie16}.} However, as 
surveys of $z > 2$ galaxies become larger \citep{krie15}, we expect them to 
include examples of this population.   

Another possible explanation for the lack of metal-rich star-forming
progenitors is that current samples of spectroscopically studied
galaxies at  $z\sim 2$ and 3 contain components of metal-enriched gas,
but only in regions  with heavy dust obscuration.  In this case, the
(rest-frame optical) spectra  only probe the lower-metallicity gas
(possibly diluted by recent inflows of pristine  gas), yielding a
mass-metallicity relation biased low.   Alternatively, it may be that
current spectroscopic studies of star-forming  galaxies lack the
progenitors of the quiescent galaxies in our samples  entirely because
the former are \textit{completely} attenuated by dust.      Massive,
dusty star-forming galaxies are known to exist \citep[e.g.,][]
{papo06,case13,spil14}, with the space densities that connect them to
quiescent galaxies \citep{spli14,toft14}.   Some dusty star-forming
galaxies are observed to have  compact gas reservoirs
\citep{simp15,tada15}, which likewise make them  candidate progenitors
to quiescent galaxies at lower redshifts.   \editone{Future  observations at
at (rest-frame) near-IR to far-IR wavelengths (such as with the \textit{James Webb Space
Telescope} [\jwst] and the \textit{Atacama Large Millimeter Array} [ALMA])} may
identify this galaxy population (and/or the  higher- metallicity gas
components in galaxies at these redshifts, see previous  paragraph).
It may also be possible to resolve these regions of high  metallicity
through higher angular resolution studies (e.g., with the Giant
Segmented Mirror Telescopes [GSMTs] combined with adaptive optics).

\section{Summary}

In this work we used deep \hst\ spectroscopy to  constrain the
metallicities and \editone{light-weighted} ages of massive ($\log
M_\ast/M_\odot\gtrsim10$), quiescent galaxies at $1.0<z<1.8$.   We
selected 32 galaxies from existing \hst\ and multi-wavelength imaging
in the GOODS-N and -S fields also covered by our deep, 12--orbit
WFC3/G102 grism pointings from CLEAR.   For redshifts $1.0<z<1.8$ the
data cover important stellar population features in the rest-frame
optical, including the Ca HK feature, 4000~\AA\ break, Balmer lines,
Mg$b$, several other Lick indices and \ion{Fe}{1} Iron lines, which
are sensitive to stellar population age and metallicity.  All galaxies
are selected to have rest-frame optical colors ($U-V$ and $V-J$)
indicative of quiescent stellar populations, so that we can use the
\hst\ grism data to study the relation between galaxy star-formation
and enrichment histories and galaxy quenching.

We developed a method to forward model stellar population models,
combining the grism dispersion model with the galaxy morphology to model
accurately the galaxy spectra and contamination from spectra of nearby
sources.  We extracted 1D spectra from these models so they match the
1D spectra measured from the data.  We validated this method by
fitting these models to simulated data from which we are able to recover
stellar population ages and metallicities.  We showed that there are
four redshift subgroups with median redshifts, \zone, \ztwo, \zthree,
\zsix, which contain different spectral features in the G102
wavelength coverage, and we discussed the systematic differences and
statistical uncertainties in the results for stellar population
parameters for galaxies that fall in these redshift subgroups. 

We then fit the suite of stellar population models to the 1D G102
grism data for each galaxy in our sample, taking into account the
galaxies' morphologies to correctly model the grism resolution.  We
first re-fit for galaxy redshifts using the grism data, and find small
corrections from the photometric redshifts derived from broad-band
photometry.  The grism redshifts yield an accuracy of $\sigma_z
\approx 0.005$ compared to the typical photometric redshift from
broad-band photometry, $\sigma_z \approx 0.02-0.10$.   Using the grism
redshifts, we then fit the models to derive posterior likelihood
distributions for metallicity and \editone{light-weighted} age for
each galaxy.    We considered two sets of stellar population models:
FSPS and BC03, but we argue the FSPS models are a better
representation for these galaxies based on evidence from Bayes Factors
derived from the fitting these models to all the galaxies.    We then
stack the posteriors for \editone{light-weighted} ages and
metallicities from the fits of stellar population models to each
galaxy in each of the four redshift subgroups to derive constraints on
the stellar population parameters for the galaxy populations.  

Because we are observing the quiescent galaxies closer to their
quenching epoch, our results place tighter constraints on their
formation history than has generally been possible.  Considering the
full range of SFHs in our models, we derive
\editone{light-weighted} ages for the galaxies at $1.0 < z < 1.8$ that
correspond to a  ``formation'' redshift \editone{of $z_{70} > 2$, with
approximately one-third of the sample showing $z_{70} > 3$}.    We show
that for the SFHs, these formation redshifts
correspond to the epoch when the galaxies had formed  formed $\gtrsim
70$\% of their stellar mass.  The implication is that quiescent
galaxies formed the bulk of  their stellar mass early.  This connects
them to recently identified quiescent galaxies at redshifts as high as
$z \sim 3-4$ \citep[e.g.,][]{glaz17,schr18b}.

We derive constraints on the metallicities of the quiescent galaxies
at $1.0 < z < 1.8$, which show that these massive galaxies had
enriched  rapidly to approximately Solar metallicities as early as
$z_{70} \sim 3$. We also show that a mass-metallicity relation exists
for $1 < z_{grism} < 1.7$ quiescent galaxies, and that this is
consistent with \textit{no} evolution from $z\sim 1.7$ to $z < 0.1$.

Logically, the star-forming progenitors of these galaxies must have
been similarly enriched to approximately Solar metallicities prior to
quenching. Because there are few galaxies (at any mass) at $z\sim 2-3$
with Solar gas-phase metallicities, the progenitors of the quiescent
galaxies would be outliers in existing samples.  This means there is
something special about galaxies on the verge of quenching in the high
redshift Universe:  \editone{either they are more compact with higher metal
retention efficiencies (possibly these are the blue-compact galaxies
seen at $z > 2$), very short lived phases ($\ll1$~Gyr),  or their
progenitors are deeply attenuated by dust
which current surveys (that select in the rest-frame optical, i.e.,
$H$-band or $K$-band) are biased against.  It also may be that some
combination of these effects are in play.}  The key to understanding
the quenching of quiescent galaxies observed at $1 \lsim z \lsim 1.7$
will be to identify their Solar-metallicity star-forming progenitors
at $z > 2-3$.   Such studies will be possible in the future with
\jwst, ALMA, and the GSMTs.

\acknowledgements \editone{We wish to thank our colleagues in the CANDELS,
3DHST, and CLEAR collaborations for their intense work to make these
datasets usable.  The authors thank the anonymous referee for their
constructive report, which improved the quality and clarity of this
paper.     We also thank them (and others) for productive
conversations, comments, suggestions and information, including
Charlie Conroy, Romeel Dav\'e, Sandy Faber,  Kristian Finlator, Rob
Kennicutt, Marcin Sawicki, Corentin Schreiber, and Sarah Wellons.  VEC
acknowledges generous support from the Hagler Institute for Advanced
Study at Texas A\&M University.  This work benefited from generous
support from George P. and Cynthia Woods Mitchell Institute for
Fundamental Physics and Astronomy at Texas A\&M University.}  This work
is based on data obtained from the Hubble Space Telescope through
program number GO-14227.  Support for Program number GO-14227 was
provided by NASA through a grant from the Space Telescope Science
Institute, which is operated by the Association of Universities for
Research in Astronomy, Incorporated, under NASA contract NAS5-26555.  

\appendix
\section{\editone{Results of fitting stellar population models to all sample galaxies}}\label{section:appendix_allplots}

\begin{figure}[t]
\epsscale{1.23}
\plotone{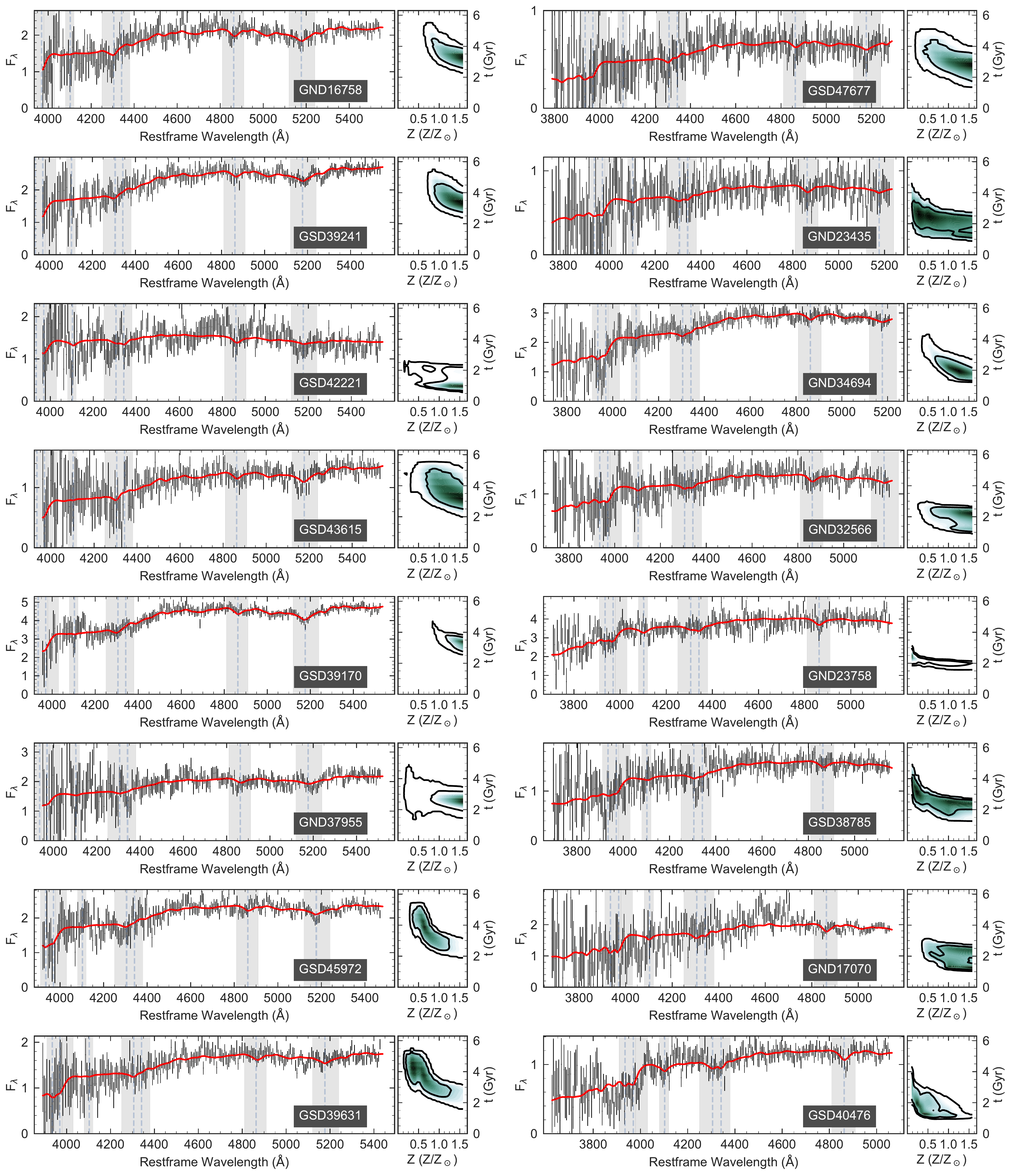}
\caption{\editone{The data and model fits for the first 16 of the 31
    galaxies in our sample.  In the left hand panel of each subplot,
    the gray data points show the measured   spectra (and
    uncertainties).  The red lines show the model fits using median
    values for the parameters.   The shaded regions correspond to the
    metallicity--age spectral  features. The right hand panel of each subplot
    shows the metallicity and (light-weighted) age joint
    likelihoods.  The legend shows the galaxy ID, and
    Table~\ref{optable} gives the derived
    values for each model parameter for each galaxy.}
\label{allspec1}}
\end{figure}

\begin{figure}[t]
\epsscale{1.23}
\plotone{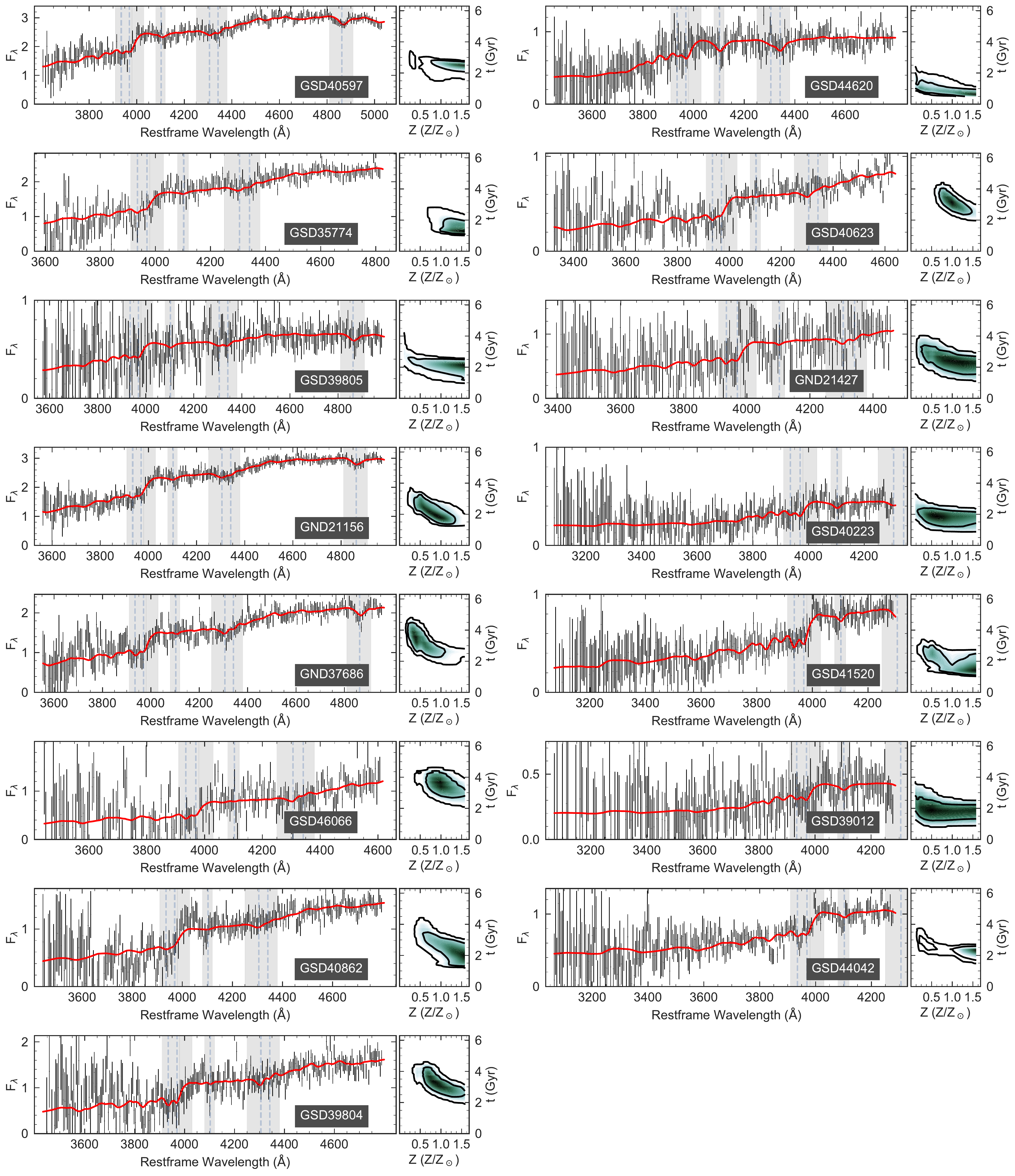}
\caption{\editone{Same as Figure \ref{allspec1} showing the remaining
    15 of 31 galaxies in our dataset.}\label{allspec2}}
\end{figure}

\editone{In this appendix we show the spectra for each galaxy along with the model fit from our analysis.   In Figure~\ref{allspec1} and \ref{allspec2} we show the 1D G102 grism data for all the galaxies in our sample.  The red line in each figure shows the ``median" model fit, the stellar population model with parameter values equal to the median values in Table~\ref{optable}.   For each galaxy, the figures also show the joint posterior on the (light-weighted) age and metallicity from the stellar population model fitting. }

\section{\editone{Template Error Function} \label{app-tmperr}}

\begin{figure}[t]
\epsscale{0.85}
\plotone{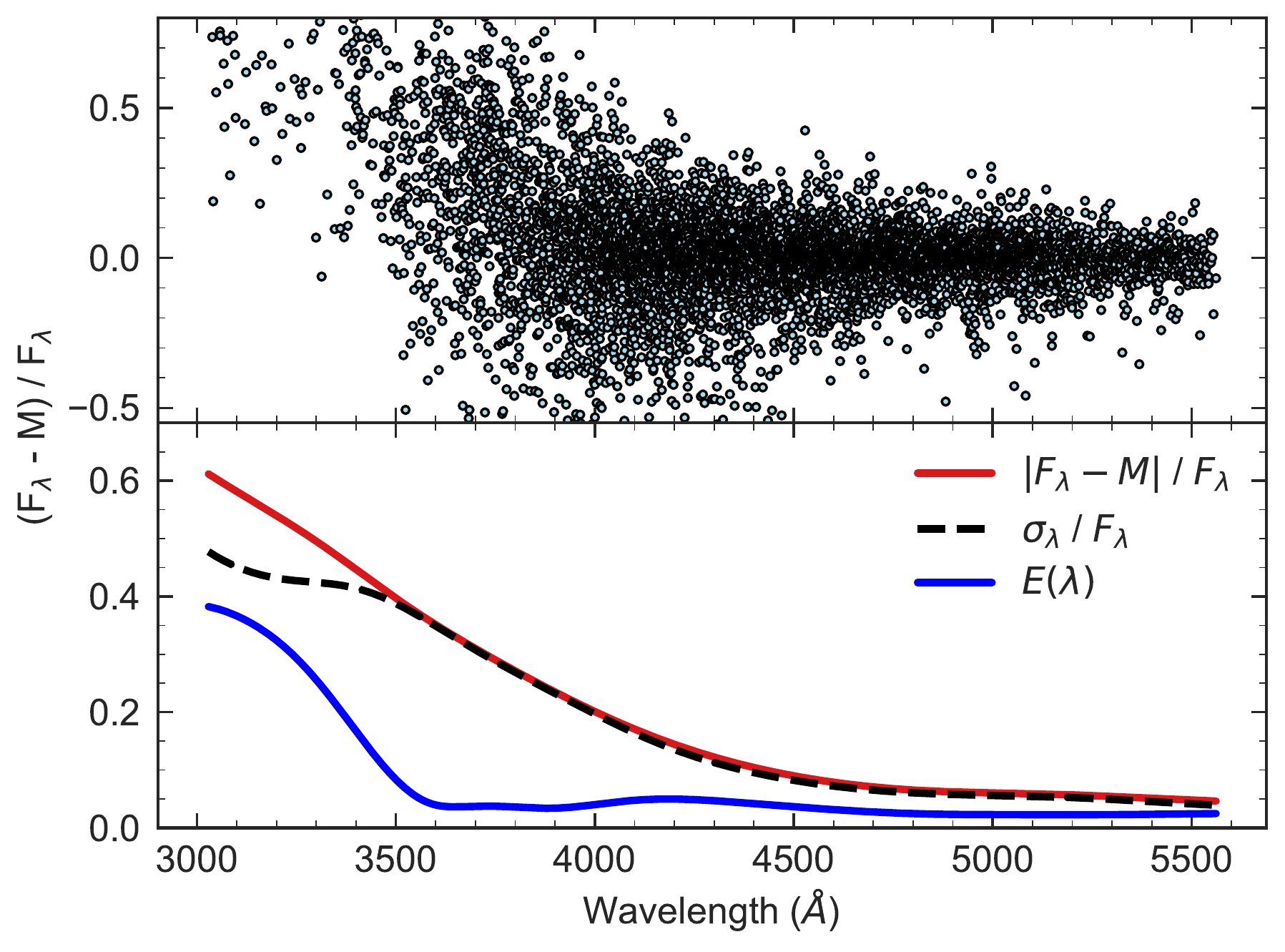}
\caption{\editone{Distribution of the normalized residuals and the
    template error function. The top panel shows ($(F_\lambda - M)$ /
    $F_\lambda$) as a function of wavelength for all galaxies in the
    rest-frame, where $F_\lambda$ are the observed data (e.g., the
    G102 spectra) for each galaxy, $M$ is the best-fit model for each galaxy.  
The bottom panel shows the smoothed absolute normalized residuals (red, solid line; derived from the data in the top plot)
along with the smoothed relative error (($\sigma_\lambda$ / $F_\lambda$), black dashed line, where $\sigma_\lambda$ are the measured uncertainties on the data)
and the derived template error function ($E(\lambda)$, blue solid line) as described in the text.  }
\label{tmp_err}}
\end{figure} 

\editone{
To calculate our template error function we first find the
best-fit model to the data for each galaxy (see
Section~\ref{section:fitData}).    
We then calculate a likelihood, $\mathcal{L}$, of the form
\begin{equation}
\mathcal{L} \propto \frac{1}{\sqrt{2 \pi} s} \exp{(-x^2 / (2s^2))} 
\end{equation}
where $x = F_\lambda - M$ the difference between the data
and model, and  $s = \sqrt{\sigma_{F_\lambda}^2 +
  (E(\lambda)F_\lambda)^2}$ is the sum (in quadrature) of the errors
on the data ($\sigma_{F_\lambda}$) and the ``template error
function'', $E(\lambda)$.  We then take $d\mathcal{L} / dE(\lambda) =
0$, and solve to find
\begin{equation}
E(\lambda) = \sqrt{\left(\frac{F_\lambda-M}{F_\lambda}\right)^2  - \left(\frac{\sigma_{F_\lambda}}{F_\lambda}\right)^2 }
\end{equation}
where the first term is the normalized residuals and
the second term is the normalized error.  We then follow
the steps outlined in \cite{bram08} to derive $E(\lambda)$. }

\editone{The top panel in Figure \ref{tmp_err} shows the distribution
of $(F_\lambda-M) $ / $F_\lambda$ for all of our spectra.   The bottom
panel shows absolute median of this distribution as a function of
wavelength, and shows the contributions from photometric
uncertainties, and the error function, $E(\lambda)$.   For wavelengths
longward of $\gtrsim$3500~\AA, there is only a small template error
function ($\approx$5\%).  At shorter wavelengths, below
$\sim$3500~\AA, the template error function increases to as high as
30\%, and may result of model uncertainties in the rest-frame UV.  We
include this error function in our fits, but in practice this has
little effect on our results (as the data for rest-frame
$\lesssim$3500~\AA\ typically have lower S/N, so contribute little to
the probably density). }
\section{Bayesian Evidence Between the Stellar Population Models\label{fsps-v-bc03}}

\begin{figure}[t]
\epsscale{.7}
\plotone{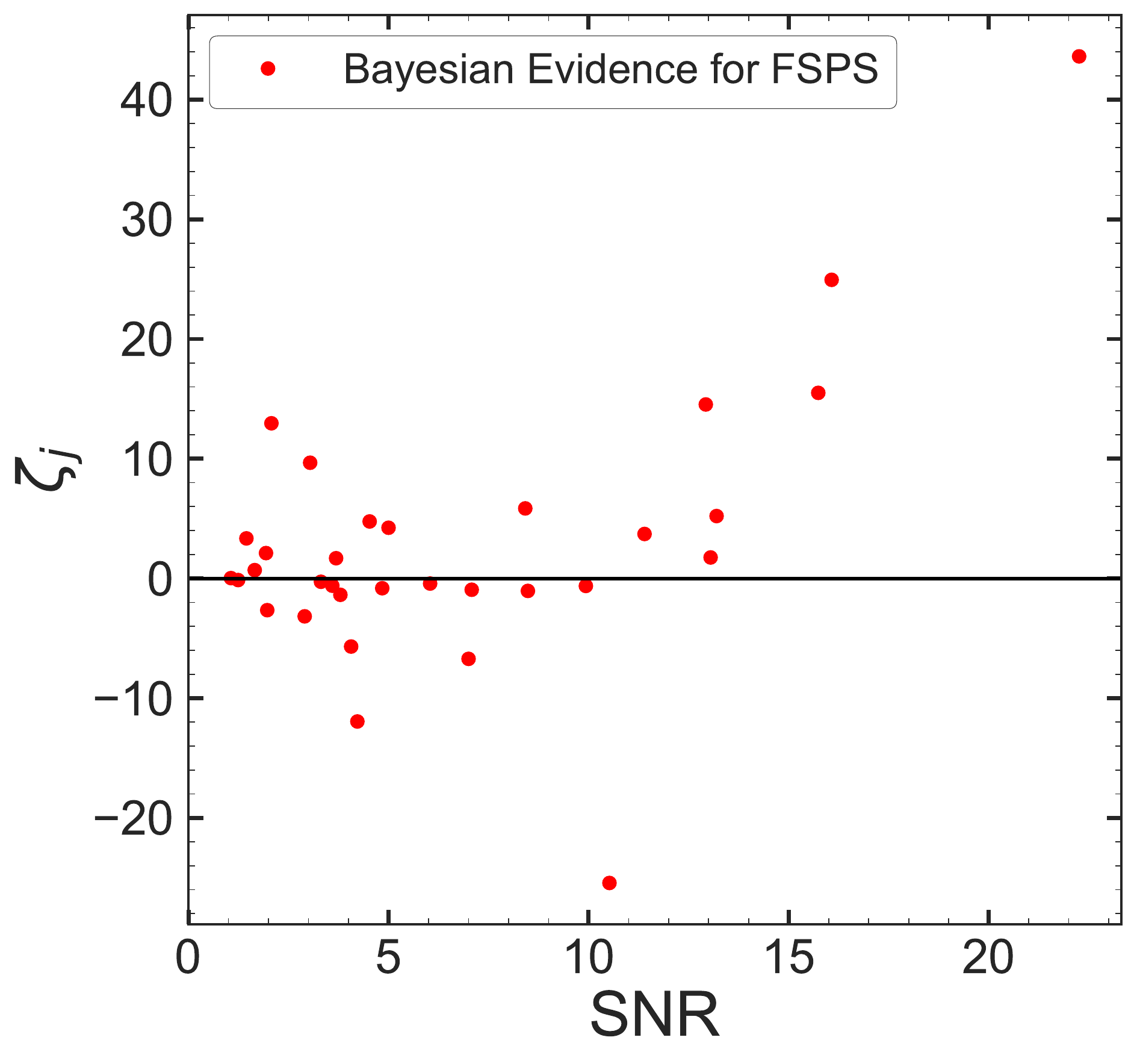}
\caption{Distribution of Bayes-factor evidence ($\zeta_j$), for  each
galaxy $j$ in our sample, as a function of SNR.  Positive (negative)
values of $zeta_j$ denote galaxies with evidence against (in favor of)
the BC03 models compared to the FSPS models. \label{bVs}}
\end{figure} 

In this study we performed model fits using two sets of composite
stellar population synthesis models, those from FSPS and BC03
(see Section~\ref{section:models}).    In our analysis we favored results from
FSPS, for the reason that it better fits the rest-frame wavelength
region probed by the G102 data for our sample, approximately, $3000 <
\lambda_\mathrm{rest}/\hbox{\AA} < 5800$.   Similar results are found
by Fum16 for quiescent galaxies at different redshifts with similar rest-frame wavelength coverage.  
Here, we use Bayesian evidence to support our use of the FSPS models in our conclusions.    

We followed the description of the model selection process describe in
\cite[see also Kass \& Raftery 1995]{salm16}.  We used the results from the fit of each set of models
(BC03 and FSPS) to each galaxy in our sample using the same range of
age and SFH parameters. For FSPS we use the same metallicity griding 
we use in our fitting, while for BC03 we use several of the metallicities made available 
($Z = $[0.02, 0.2, 0.4, 1.0, 2.5] \Zsol) .
To compare the models, we consider all 
parameters, $\Theta$, as nuisance
parameters.  Marginalizing over the full parameter space includes
probability contributions from all possible combinations of parameters
for a given model, and the ratio of these then contains information
about the relative probability for one model compared to the other. 
We may use this to derive the ``odds'' that favor one model over the
other using the ratio of the model posteriors: 
\begin{equation}\label{eqn:odds}
\frac{ P(  m_1 | D) } {P( m_2 | D)} = \frac{ P(D | m_1)}{ P(D | m_2)} \times \frac{P(m_1)}
{P(m_2)}
\end{equation}
for the different stellar population models, $m_1$ and $m_2$, where
$m_i$ is the i-th model (e.g., 1=FSPS, 2=BC03).  In
Equation~\ref{eqn:odds}, the term on the left-hand side is the ratio of posteriors (``Posterior Odds'').  On the right-hand side, the last term is the ratio of
priors (``Prior Odds''), and the first term is the so-called
``Bayes Factor'' \citep[e.g.,][]{jeff35,kass95}, defined as
\begin{equation}
\mathrm{Bayes\ Factor} \equiv B_{12} = \frac{P(D|m_1)}{P(D|m_2)},
\end{equation}
where the values $P(D | m_{1,2})$ are the normalizations of the
posteriors for each model, derived by
marginalizing the posteriors over the full set of parameters,
\begin{equation}
%
P(D | m_{1,2})= \int_\Theta\ P(D|\Theta, 
m_{1,2})\ P(\Theta | m_{1,2})\  d\Theta. 
\end{equation}

Conceptually, the Bayes factor is the ratio of the normalizations
of the posteriors.  A model that produces an overall better fit (over the
full range of parameters, $\Theta$) will yield a higher normalization,
and the ratio of these normalizations then indicates whether one model
generally fits the data better, and is thus favored.  

We quantified this using the definition of the Bayes-factor evidence ($\zeta$) \citep{kass95},
which is related to the Bayes factor by, 
$\zeta_j = 2 \ln B_{12,j}. $  Here the subscript $j$ corresponds to each galaxy in our sample.
We then sum over all $\zeta_j$ to measure the significance of the
Bayes factor evidence.    If both models ($m_{1,2}$) fit the data 
equally well then $\Sigma \zeta_j=0$.   A positive sum of the Bayes factor evidence ($\Sigma \zeta\ >\ 0 $) shows evidence in favor of model 1 ($m_1$), while a negative sum of the Bayes factor evidence ($\Sigma \zeta\ <\ 0$) shows evidence towards model 2 ($m_2$).
\cite{kass95} provide significance statements based on value of the
Bayes factor evidence, which we adopt here.

We computed the Bayes factor for the FSPS and BC03 models for each
galaxy in our sample, and then derived the overall Bayes factor
evidence (the distribution of values are shown in Figure \ref{bVs}).    
Summing over $\zeta_j$ for all $j$ galaxies, we derived
$\Sigma\ \zeta\ =\ 164$.  This corresponds to ``very strong'' evidence (defined as $\Sigma\ \zeta > 10$) against
BC03 in favor of FSPS for the galaxies in our sample.
We note that the grid of FSPS models is much finer in metallicity than
BC03, and this could be a driving factor. We therefore recomputed the
Bayes factor evidence using a set of FSPS models with the same
metallicity griding as for BC03.  In this case we derive $\zeta =
96$, which is still ``very strong'' evidence for FSPS compared to
BC03, and argues that for the galaxies in our sample the FSPS models provide a better overall fit to the data.   For these reasons we adopt the FSPS for the conclusions in this study. 

\section{Testing the ``Stack--Smooth--Iterate'' Method to Combine
  Parameter Likelihoods \label{stacking}}

\begin{figure*}[t]
\epsscale{1.1}
\plotone{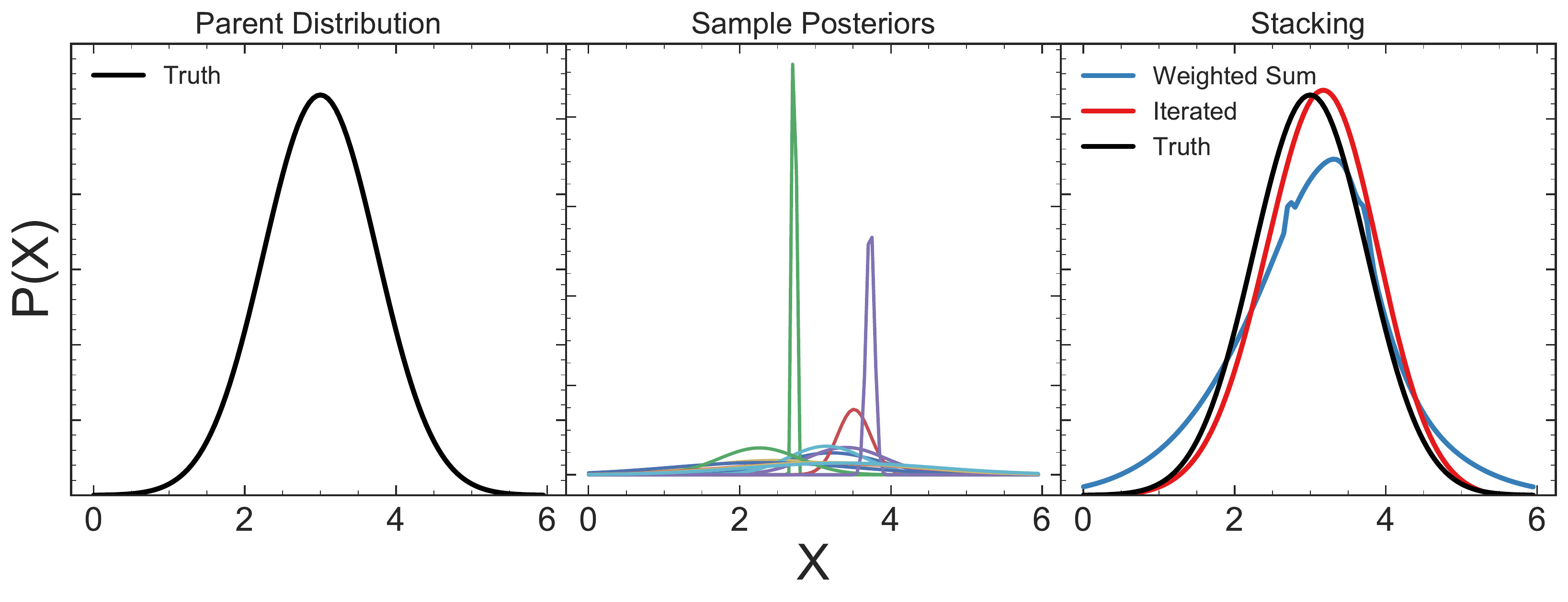}
\caption{Test of our stacking method to recover a parent distribution. Panel 1 shows the true distribution, panel 2 show the randomly selected sample
from distribution, and panel 3 displays a weighted sum of the sample posteriors along with the fully
processed stacked posterior.\label{stx1}}
\end{figure*} 

\begin{figure*}[t]
\epsscale{1.1}
\plotone{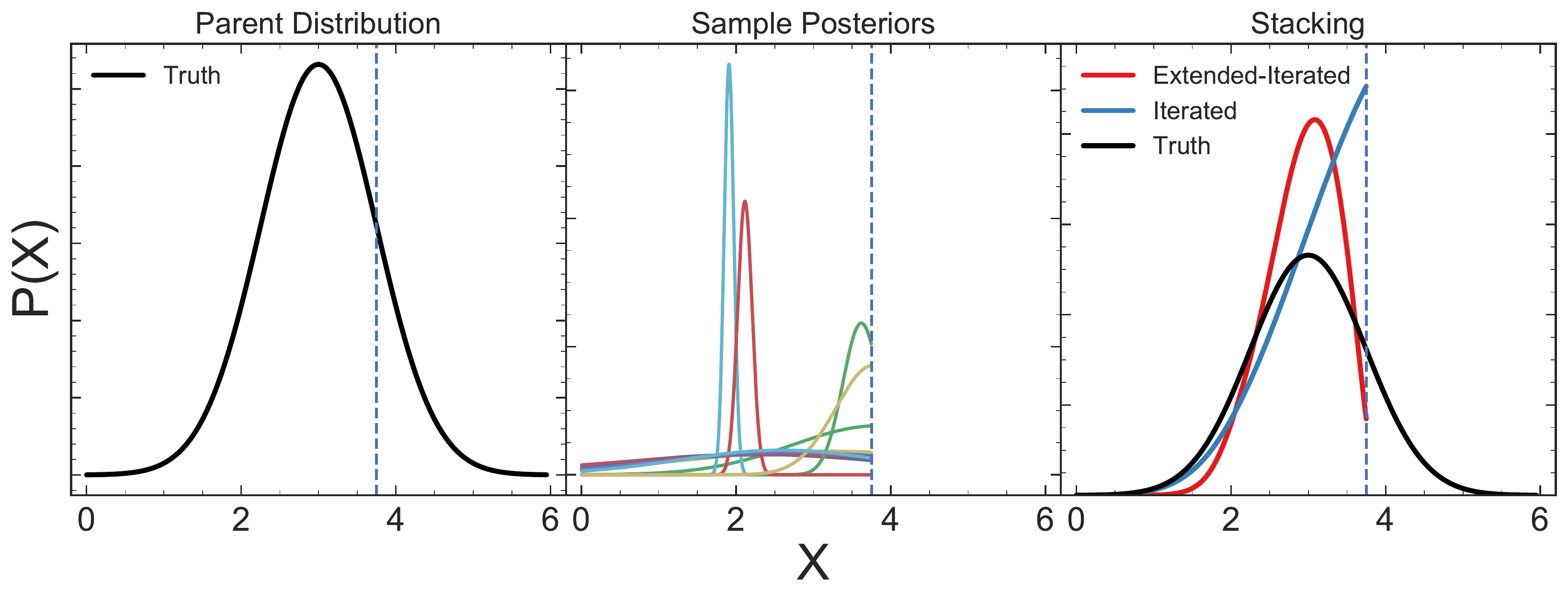}
\caption{Test of our stacking method to recover a parent distribution when our parameter space does not cover the entirety of the parent distribution. Panel 1 shows the true distribution, panel 2 show the randomly selected sample
from distribution, and panel 3 displays the fully
processed stacked posterior with and without extending the parameter space.\label{stxex}}
\end{figure*} 

\begin{figure}[h]
\epsscale{.9}
\plotone{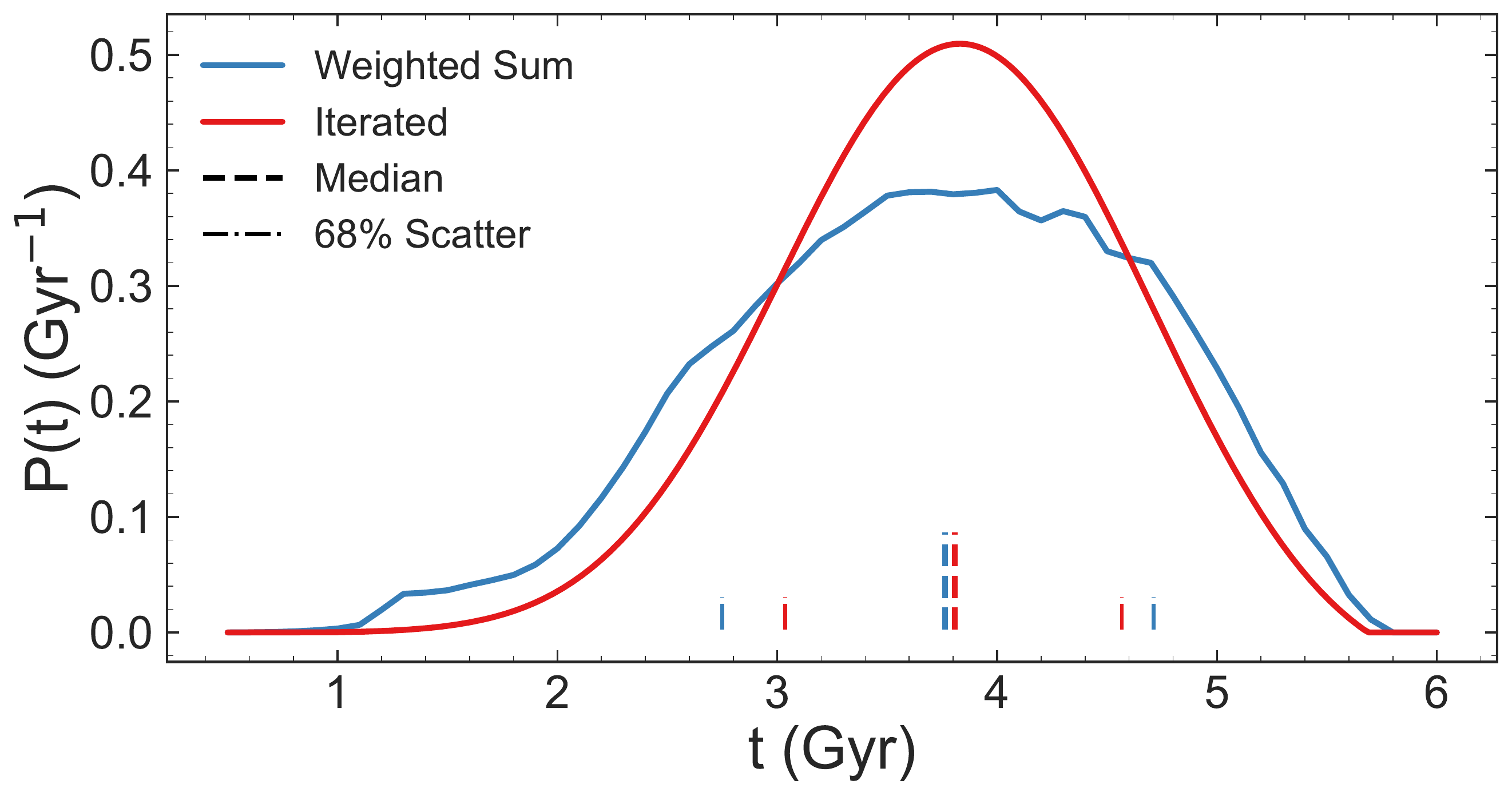}
\caption{An example of the stacking method applied to the \editone{light-weighted} age posteriors of our \zone\ redshift
group.   The solid blue line shows the direct weighted sum (the result
of direct stacking in the first step of the interaction using
Equation~\ref{stack}).   This yields the median and 68\% confidence
range illustrated by the blue dashed and dot-dashed lines.    The
solid red line shows the results after iterating.  Because the
iterations smooth over objects with sharply peaked  likelihoods it
yields a slightly tighter (and smoother) final likelihood.   The red
dashed and dot-dashed lines show the change in the median and 68\%
confidence range, respectively.  We use the
``stack-smooth-iterate'' method to derive likelihoods from the
galaxies in each of our redshift subgroups. \label{stx2}}
\end{figure} 

Here we test the stacking method described in Section \ref{section:fit_data}
to see how well it reproduces a known true parent distribution.
This process is illustrated in Figure \ref{stx1}. The test is done as follows:
(1) We take a  parent (Gaussian) distribution (shown in the first
panel of Figure \ref{stx1}). (2) We randomly draw from it  several
data points (here we use 12 samplings to approximate a sample size
similar to what we have in our \zone\ redshift subgroup) with random
errors (this is similar to a situation where some galaxies constrain
their parameters better than others):  this is illustrated in the
middle panel of Figure~\ref{stx1}.   (3) Then we iteratively stack our
sample posteriors to recover an approximation of the parent
distribution following the method of 
Section~\ref{section:fit_data}.  This is
illustrated in the right panel of Figure~\ref{stx1}.  In that panel,
the blue curve shows the recovered distribution after performing the
weighted  summation (the first step in the stacking process, which is
identical to a weighted stack of the likelihoods).  After the first
step, the weighted distribution contains sharp peaks from objects with
highly constrained likelihood functions. If these sharp peaks are not
removed the iteration process will tend to follow the peaks, and the final
product will be highly skewed. The
red curve shown in panel 3 is the final product of the iterative
stacking process. This included 20 iterations, although even a few
iterations ($\sim$5) produce a good approximation of the correct distribution (and
subsequent iterations approach an estimation of the truth distribution asymptotically).
With a fixed sample size, n=12 in this case, the algorithm may converge to some estimate, 
but this will not be the true distribution. Convergence to the true 
distribution is only expected as the sample size gets larger.

\editone{One of the limitations of the stacking process described above is that
it can fail in cases where the parameter space does not fully encompass the parent
distribution ($ \lsim 90\%$ of the probability). When this occurs much of the 
probability mass will lie at the edge
of the parameter space, and when the iterative stacking is applied the resulting
distribution is driven to the edge (as seen in panel 3 of Figure \ref{stxex}).
In order to resolve this we extend the parameter space artificially during the
stacking process. This allows the distribution to fold over during the stacking process,
preventing the distribution from sticking to the edge. This extended-iterative stack
better reproduces the median value (though it often underestimates the confidence
intervals).}

Figure \ref{stx2} shows the outcome of this ``stack-smooth-iterate''
process applied to our data for the galaxies in our \zone\ sample.   
Here we see the differences between
the weighted sum and iterated stack. There is a very slight shift in
the median, $\sim$ 5\% for metallicity and $\sim$ 1\% for \editone{light-weighted} age.
The 68\% confidence ranges are slightly tighter by $\sim$ 40\% for 
metallicity and $\sim$ 10\% for \editone{light-weighted} age. The reason that the iterated
distribution is ``tighter'' is that the presence of sharply peaked
likelihoods from individual galaxies are smoothed in successive
iterations as the method converges.  We apply this method to all the
galaxies in each of our redshift groups (\zone, \ztwo, \zthree, \zsix)
to derive probability density functions parameters for each
population.

\bibliography{Vince_Quiescent_galaxies}{}

\end{document}